\documentclass[
  aps,
  pre,
  twocolumn,
	superscriptaddress,
	showpacs,
  preprintnumbers,
  amsmath,
  amssymb,
  nofootinbib, 
	longbibliography
	]{revtex4-2}
\usepackage{braket,slashed,url,multirow,amsmath,amsfonts,xcolor}
\usepackage[T1]{fontenc} 
\allowdisplaybreaks[1]

\definecolor{def_blue}{rgb}{0.33,0.30,0.85}
\definecolor{def_red}{rgb}{1.02,0,0.05}

\usepackage[
colorlinks=true, 
pdfstartview=FitV, 
linkcolor=blue, 
citecolor=red,
urlcolor=blue, 
bookmarks=true,
bookmarksnumbered=true,
pdftitle={},
pdfauthor={}
]{hyperref}

\usepackage{graphicx,color}
\usepackage[normalem]{ulem}

\newcommand{\diff}{\mathrm{d}} 
 
\newcommand{\rme}{\mathrm{e}}

\begin{document}

\title{Effects of local minima and bifurcation delay on combinatorial optimization with continuous variables}

\author{Shintaro Sato}
\affiliation{NTT Computer and Data Science Laboratories, NTT Corporation, Musashino 180-8585, Japan}

\date{\today}

\begin{abstract}
Combinatorial optimization problems can be mapped onto Ising models, and their ground state is generally difficult to find. 
A lot of heuristics for these problems have been proposed, and one promising approach is to use continuous variables. 
In recent years, one such algorithm has been implemented by using parametric oscillators known as coherent Ising machines.
Although these algorithms have been confirmed to have high performance through many experiments, unlike other familiar algorithms such as simulated annealing, their computational ability has not been fully investigated.
In this paper, we propose a simple heuristic based on continuous variables whose static and dynamical properties are easy to investigate. 
Through the analyses of the proposed algorithm, we find that many local minima in the early stage of the optimization and bifurcation delay reduce its performance in a certain class of Ising models.
\end{abstract}

\maketitle

\section{Introduction}~\label{sec: intro}
Combinatorial optimization problems are generally hard to solve, and a lot of algorithms to tackle them have been proposed.
Such computationally difficult problems can be mapped onto the ground state search problem for the Ising models \cite{lucas2014ising}.
Algorithms based on this correspondence have also been proposed, and many of them are inspired by physical phenomena.
Some use discrete spin variables, which appear in Ising models.
Simulated Annealing (SA) is known as a representative example \cite{kirkpatrick1983optimization},
and its implementation and development have been actively pursued.
Moreover, many different types of algorithms based on continuous variables such as soft spins have also been proposed \cite{hopfield1985neural,wang2013coherent,goto2016bifurcation,kalinin2018global,molnar2018continuous,goto2019combinatorial,doi:10.7566/JPSJ.88.061015,Kalinin2020}.

The coherent Ising machine (CIM) is one of the heuristics that use continuous variables.
It basically consists of degenerate optical parametric oscillators (DOPOs) and their phases are interpreted as Ising spins \cite{wang2013coherent}.
It is devised in the expectation that by controlling the interactions among DOPOs corresponding to the Ising coupling matrix,
the system stabilizes to the ground state configuration of the Ising models.
The high performance of such methods in optimization has actually been reported in both numerical simulations \cite{wang2013coherent,ng2022efficient} and experiments with optical devices \cite{takata201616,mcmahon2016fully,inagaki2016coherent,honjo2021100}.
To describe the dynamics of the CIM, theoretical models have been discussed \cite{PhysRevA.96.043850,PhysRevA.96.053834,doi:10.1063/5.0016140}. 
In addition, various methods have been devised to improve the performance \cite{PhysRevE.95.022118,Leleu2019,doi:10.1063/5.0016140,https://doi.org/10.1002/qute.202000045}.

Despite such experimental, numerical, and theoretical developments,
the principle of optimization in the CIM itself has not been studied extensively.
Unlike other heuristics such as SA \cite{geman1984stochastic}, no firm theoretical basis has been found to guarantee its high performances.
The CIM is a type of gradient based optimization. 
It consists of bosons interacting via the Ising coupling matrix,
and through the optimization, the state evolves under the time dependent external fields.
Even within the mean field approximation, it is a nonlinear dynamical system controlled by the time dependent parameter.
Although its nonlinearities play an important role for the optimizations~\cite{PhysRevLett.126.143901}, it is generally difficult to analyze~\cite{wang2013coherent}.
Researchers have attempted to investigating the property by researching the energy landscape or its steady state through the optimizations \cite{wang2013coherent,PhysRevA.98.053839,ito2018bifurcation}. 
However, these analyses only deal with the small and simple Ising models, or the case of a large degrees of freedom limit.
In applying the CIMs to more complex problems, their potentials need to be clarified in more general cases from the theoretical point of view.

In this paper, we propose a simple model, which facilitates analyses of its computational property by introducing time dependent Lagrange multipliers to the mean field CIM.
In contrast to the usual mean field CIM, positions of the fixed points in the landscape are analytically obtained in our model,
so the linear stability around them is also easy to discuss.
In addition to its static aspect, we also discuss the existence of the bifurcation delay~\cite{shishkova1973examination,PhysRevLett.53.1818,PhysRevLett.78.1691,10.1143/PTPS.139.325}. 
It can also be observed in the usual CIM at least at the mean field level.
These properties will affect optimization ability of the proposed model in a kind of gradient-like algorithm. 
Finally, we numerically investigate these effects on the optimizations by using random matrices as the Ising models.
To verify that our model is a good toy model for investigating the property of the CIM,
we also discuss the relation between our model and mean field CIM without noise.

The paper is organized as follows.
In Sec.~\ref{sec: algorithm}, we introduce a simple heuristic using continuous variables like the CIM.
In Sec.~\ref{sec: property},  we analyze the linear stability around its fixed points and discuss the bifurcation delay as a dynamical property. 
To see the relation between local minima and the performance, we numerically simulate the dynamics for a certain class of Ising models in Sec. \ref{sec: numerical}.
Our conclusions are presented in Sec.~\ref{sec: Conclusion}.

\section{Model}~\label{sec: algorithm}
We discuss the property of the algorithm show later for solving Ising problems without magnetic fields.
The Ising problem is defined as finding the ground state of the following Ising Hamiltonian:
\begin{align}\label{ising_prob}
  H(\mathbf{S})=\frac{1}{2}\sum_{i\neq j}J_{ij}S_iS_j,
\end{align}
where $\mathbf{S}$ is a $N$ dimensional Ising spin vector with $S_i=\pm1\ (i,j=1,\ldots,N)$ in the elements.
An $N\times N$ dimensional matrix $J_{ij}$ represents the Ising coupling matrix, which is a symmetric $J_{ij}=J_{ji}$ and $J_{ii}=0$. 
For later discussion, the eigenvalues of $J_{ij}$ are represented as $l_i$ and the smallest ones are written as $l^{\text{min}}$.

We propose the mean field approximated CIM algorithm using continuous variables $x_i(t)$ with auxiliary variables $y_i(t)$ that obey the following equations:
 \begin{align}\label{system_def1}
    \frac{\diff x_i}{\diff t}&=-\left(x_i^3+m(t)x_i+\beta\sum_{j}J_{ij}x_j\right)+2\kappa\Theta(t-t_c)y_i x_i,
\end{align}
\begin{align}\label{system_def2}
    \frac{\diff y_i}{\diff t}&=-\kappa\Theta(t-t_c)(-\delta_m(t)+x^2_i),
\end{align}
where $\Theta(t)$ is the Heaviside step function and $\beta>0, \kappa\geq0, t_c, m_c$ are time independent parameters. 
Time dependent parameters $m(t)$ and $\delta_m(t)$ can be arbitral functions of time $t$, and  
in the later analysis, they are taken as linear functions.

This system can be regarded as a gain-dissipative system with auxiliary variables or a kind of the Augmented Lagrange method (see Ref.~\cite{bertsekas1997nonlinear,boyd2004convex}). 
Some variations have been proposed in the context of the Ising solver ~\cite{Leleu2019,kalinin2018global,doi:10.1073/pnas.2015192117,vadlamani2022equivalence}.
We note that in particular for $\kappa=0$ our model reduces to the usual mean field CIM \cite{wang2013coherent,doi:10.1063/5.0016140} without noise terms, and continuous variables $x_i$ correspond to amplitudes of DOPO signals.
In solving the Ising problem, we set the matrix $J_{ij}$ as the Ising coupling matrix and evolve the system from initial time $t=0$ to final time $t=T$. 
Finally the ground state of the Hamiltonian (\ref{ising_prob}) is calculated on the basis of the final state obtained as $S_i=\text{sgn}( x_i(T))$.  
In this paper, initial conditions for $x_i(0)$ are chosen uniformly at random from the interval $[-r, r], r\in\mathbb{R}$ and those for the auxiliary variables are $y_i(0)=0$.  
\section{Model property}~\label{sec: property}
In this section, we analyze the static and dynamical aspects of the systems especially related to the searching ability of the solutions for the Ising problem.
Our algorithm is based on the gradient method, and we consider the situation where the time dependent parameters slowly change in time for sufficiently large $T$.
Then we assume that its algorithmic property can be captured by analyzing the fixed points and their linear stability. 
Due to the presence of auxiliary variables, fixed point analyses are much easier than the usual mean field CIM. 
In the following discussion, the forms of the time dependent parameter are chosen as $m(t)=-\epsilon t+m_c$ and $\delta_m(t)=-m(t-t_c)+m_c$ with $\epsilon>0$ and $m_c=\epsilon t_c-\beta l^{\text{min}}$.

\subsection{Static property}~\label{sec: static}
Let us consider here the static property of this system that includes locations of the fixed points and their linear stability. 
In this section, we focus on a particular situation where the time dependent parameters have a certain fixed value $m(t)=m,\ \delta_m(t)=\delta_m$. 

When $t\leq t_c$, there is only a trivial fixed point $x_i=0$, and its linear stability can be obtained by considering the corresponding Jacobian matrix for $x_i$ defined $\mathrm{J}_0\equiv -m-\beta J_{ij}$. 
Because of the value of $m_c$, it is linearly stable as long as $t<t_c$, and at $t=t_c$ it has zero eigenvalues. 

On the other hand $t>t_c$, non-trivial fixed points are given by the solutions of the simultaneous equations calculated from Eqs.(\ref{system_def1}) and (\ref{system_def2}) as 
\begin{align}
    x_i&=\pm\sqrt{\delta_m},\\
    y_i&=\frac{1}{2\kappa}\left(m+\delta_m+\beta\sigma_i\sum_jJ_{ij}\sigma_j\right),
\end{align}
where $\sigma_i\equiv{\text{sgn}}(x_i)$. 
It is found that the bifurcation occurs through the changing parameters.
We note that these $2^N$ fixed points correspond to all configurations that $\mathbf{S}$ can take, so the ground state of the Hamiltonian (\ref{ising_prob}) is given as one of those.  

Let us focus on the latter situation in which the non-trivial fixed points appear.
Before going to the discussion for the case of the non-trivial fixed points, we note that the origin is not a fixed point for $\kappa\neq0$. 
The linear stability analysis for these non-trivial fixed points can be performed by considering the eigenvalues of the following $2N\times2N$ dimensional matrix:
\begin{align}
    \mathrm{J}=\left[
                \begin{array}{cc}
                \mathrm{J_{xx}} & \mathrm{J_{xy}}\\
                \mathrm{J_{yx}} & \mathrm{J_{yy}}
                \end{array}
                \right],
\end{align}
where each $N\times N$ dimensional matrices are defined as $\mathrm{J_{xx}}\equiv\text{diag}(-2\delta_m+\beta\sigma_i\sum_jJ_{ij}\sigma_j)-\beta J_{ij},\ \mathrm{J_{xy}}=-\mathrm{J_{yx}}\equiv2\kappa\sqrt{\delta_m}\text{diag}(\sigma_i),\ $and $\mathrm{J_{yy}}\equiv0$.
Here $\text{diag}(v_i)$ means $N\times N$ dimensional diagonal matrix with the elements of some $N$ dimensional vector $v_i$.
As reported in \cite{Leleu2019}, the eigenvalues of this type of Jacobian matrix $\lambda_i^{\pm}(\mathbf{\sigma},\delta_m)$ can be expressed as follows:
\begin{align}\label{eig vals}
    \lambda_i^{\pm}(\mathbf{\sigma},\delta_m)=\frac{1}{2}\left(\mu_i(\mathbf{\sigma})\pm\sqrt{\mu^2_i(\mathbf{\sigma})-16\kappa^2\delta_m}\right),
\end{align}
by using $\mu_i(\mathbf{\sigma})$ which is defined as the $i$th eigenvalues of $\mathrm{J_{xx}}$. 
Since the inside of the square root is not necessarily positive, these eigenvalues are complex numbers in general. 
These $2N$ eigenvalues consist of $N$ pairs of $\lambda_i^{+}(\mathbf{\sigma},\delta_m)$ and $\lambda_i^{-}(\mathbf{\sigma},\delta_m)$ for each index $i$. 
We note that their signs are completely determined by $\mu_i(\mathbf{\sigma})$: $\text{sgn}(\text{Re}(\lambda_i^{\pm}(\mathbf{\sigma},\delta_m)))=\text{sgn}(\mu_i(\mathbf{\sigma}))$, so it is sufficient to examine the signs of $\mu_i(\mathbf{\sigma})$ to discuss the stability.
In addition, if $\delta_m$ is sufficiently large (i.e., near the final time of the evolution), $\mu_i(\mathbf{\sigma})$ will be dominated by the influence of the $-2\delta_m$ term in the diagonal component of $\mathrm{J_{xx}}$ and all eigenvalues of the Jacobian $\mathrm{J}$ become negative. 
As a result, all fixed points will be stable.
By using these results, we can discuss the number of local minima in the optimizations.

\subsection{Dynamical property}~\label{sec: dynamical}
Now let us consider the effect of the time dependent parameters. 
In general, the time evolution of the system is difficult to obtain analytically, but within a particular time interval, their dynamics can be captured by linearized equations. 
In this section, we discuss the dynamical aspect of our model from the viewpoint of the bifurcation delay~\cite{shishkova1973examination,PhysRevLett.53.1818,PhysRevLett.78.1691,10.1143/PTPS.139.325}.  
We will see that this phenomenon delays the actual time when the bifurcation occurs due to the time dependence of the parameter $m(t)$.

To improve the perspective of the following analysis, it is convenient to rewrite Eqs.~(\ref{system_def1}) and (\ref{system_def2}) by using new variables $a_i(t)\equiv y_i+\kappa\int_{0}^t\diff t^\prime \Theta(t^\prime-t_c)\ \delta_m(t)$ and 
$M(t;t_c)\equiv m(t)+2\kappa^2\Theta(t-t_c)\int_{t_c}^t\diff t^\prime \delta_m(t^\prime)$.
We set initial conditions for new introduced variables as $a_i(0)=y_i(0)=0$.
From here, we consider the following linear approximation around the origin like 
\begin{align}
  \frac{\diff x_i}{\diff t}&=-\sum_j\left(M(t;t_c)\delta_{ij}+\beta J_{ij}\right)x_j,\\
  \frac{\diff a_i}{\diff t}&=0,  
\end{align}
where $\delta_{ij}$ is the Kronecker delta.
These linear equations can describe the dynamics if the effect of the nonlinear terms is negligible and such situations are expected to occur from $t=0$ to around $t_c$.

Auxiliary variables $a_i(t)$ do not evolve in time, and we only focus on the dynamics of $x_i(t)$ whose explicit forms are given by
\begin{align}
    x_i(t)=\sum_j\rme^{-U_{ij}(t;t_c)} x_j(0),
\end{align} 
where we introduce the time dependent matrix $U_{ij}(t;t_c)\equiv\int_{0}^t\diff t^{\prime\prime}\left(M(t^{\prime\prime};t_c)\delta_{ij}+\beta J_{ij}\right)$.
To discuss the bifurcation delay, its instantaneous eigenvalues play an important role.
They are represented as
\begin{align}
    \Lambda_i(t;t_c)=U(t;t_c)+\beta tl_i,
\end{align}
and we here introduce $U(t;t_c)\equiv\int_{0}^t\diff t^{\prime\prime}M(t^{\prime\prime};t_c)$ related to $m(t)$.
They can take both positive and negative values. 

\begin{figure}[t]
    \centering
    \includegraphics[width=1.1\linewidth]{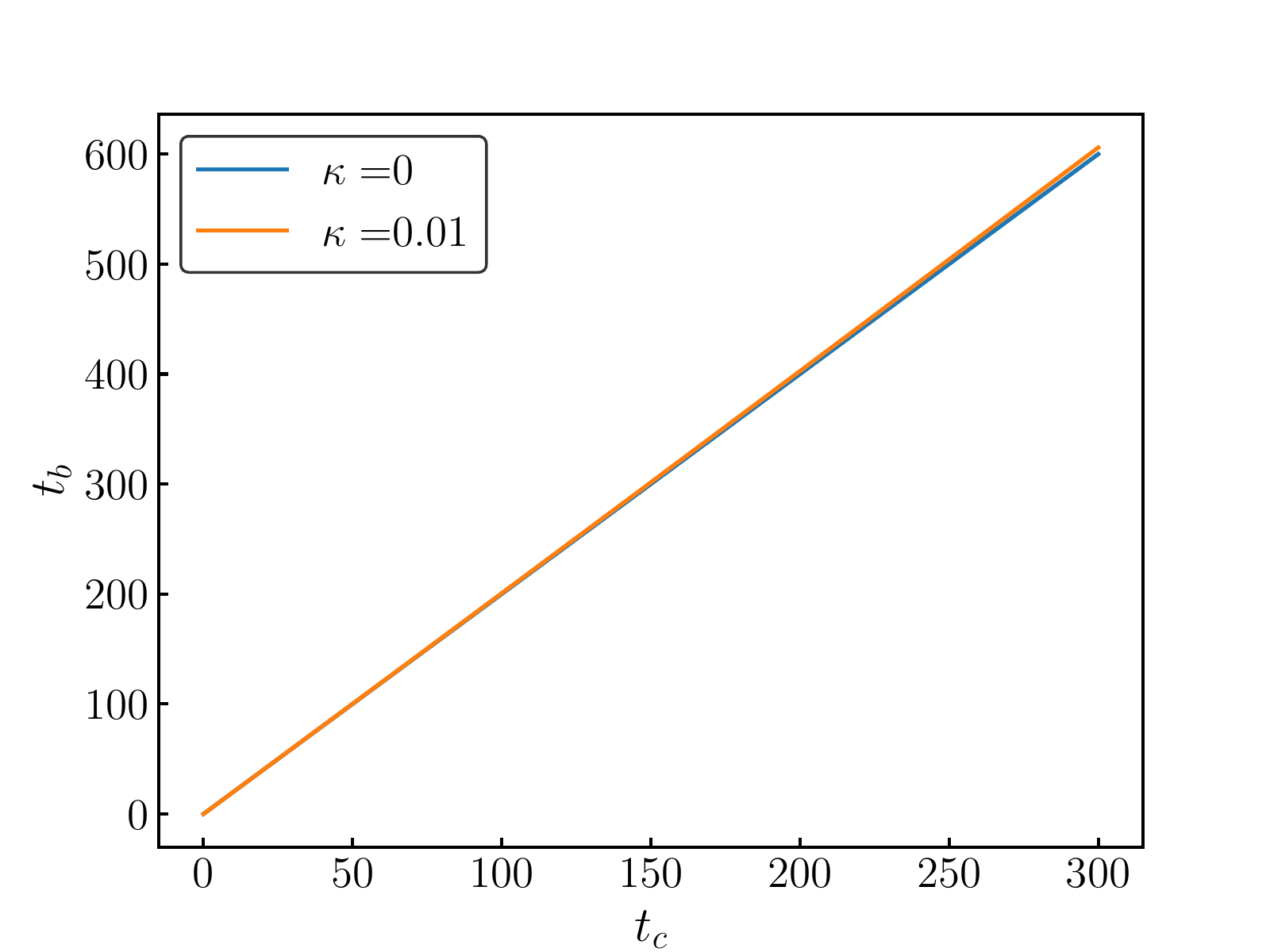}
    \caption{The bifurcation delay time $t_b$ as a function of $t_c$ for $\kappa=0$ (Eq.~(\ref{tb1})) and $0.01$ (Eq.~(\ref{tb2})).}
    \label{fig:tb-tc}    
\end{figure}

While all $\Lambda_i(t;t_c)$ are positive, $x_i(t)$ approaches the origin heading in $\mathbf{v}^{\text{min}}$, which is the eigenvector of $J_{ij}$ corresponding to $l^{\text{min}}$.
Although the origin is no longer a linearly stable fixed point at $t=t_c$, as long as all $\Lambda_i(t;t_c)$ remain positive, $x_i(t)$ is still trapped in the neighborhood of the origin and remains proportionally toward $\mathbf{v}^{\text{min}}$.  
This is known as the bifurcation delay, and its time of trapping can be characterized by the bifurcation delay time $t_b$ that satisfies the following equation:
\begin{align}~\label{BD eq}
    \Lambda^{\text{min}}(t_b;t_c)=0,
\end{align} 
where $\Lambda^{\text{min}}(t_b;t_c)=U(t_b;t_c)+\beta t_bl^{\text{min}}$ is the smallest eigenvalues of $U_{ij}$.
Because of the linear time dependence of parameters $m(t)$ and $\delta_m(t)$, we can obtain the explicit form of $t_b$ as a function of $t_c$ and $\kappa$.
For $\kappa=0$, the solution of the above equation is
\begin{align}~\label{tb1}
    t_b=2t_c,
\end{align}
and for $\kappa\neq0$, it is 
\begin{align}~\label{tb2}
    t_b=\frac{1}{2\kappa^2}\left(\sqrt{4\kappa^2t_c+1}\left(\sqrt{3}\sin{\frac{\theta}{3}}+\cos{\frac{\theta}{3}}\right)+2\kappa^2t_c-1\right),
\end{align}  
where $\frac{\pi}{2}\leq \theta\leq\pi$ satisfies the following relation using $z\equiv\kappa^2t_c$ as $\theta=\tan^{-1}\left(2z\sqrt{16z^4+12z^2+3}\right)$.
Due to our definition of $m_c$, the solution does not depend on $\beta$.
We plot the relations between $t_b$ and $t_c$ in Fig.~\ref{fig:tb-tc}.
From these results, the effective optimization is expected to occur in $T-t_b$.
In addition, $t_b$ exhibits similar behavior in both cases $\kappa=0$ and $0.01$.
The important thing is that $t_b$ is independent of $\epsilon$.
No matter how small $\epsilon$ is, it takes a finite value and may affect the results of the Ising optimizations, which we will see in the numerical experiments.

\section{Numerical Results}~\label{sec: numerical}
In this section, we discuss the optimization properties in terms of the linear stability and the bifurcation delay through numerical calculations.
First, we consider the relations between these quantities and the success probability for obtaining the ground state in our proposed model.
In addition, the comparison between our model and mean field CIM without noises is also numerically investigated from the viewpoint of success probability. 
Throughout this paper, we consider the relation for fixed energy scale $\beta=0.3$ with normalized Ising matrices that is defined in the following subsection.
For other $\beta$, we also numerically investigate these relations reported in Appendix \ref{various_beta}.
Here the system size is set to $N=20$.
The results of their time evolutions obeying Eqs.~(\ref{system_def1}) and (\ref{system_def2}) are numerically calculated by using the fourth order Runge-Kutta method with time step $\Delta t=0.01$ and final time $T=500$.

\subsection{Ising Models}~\label{sec: ising model}
For numerical experiments, we use the so-called Wishart planted ensembles (WPE) proposed by Hamze et al. \cite{PhysRevE.101.052102} as Ising models.
Their algorithmic hardness for the ground state search can be tuned via the parameter $\alpha$ introduced later.
The detailed properties of the model are given in the reference, and here we only give the definition of the model.
When we specify a solution (spin vector) $\mathbf{S}^{\text{(GS)}}$, the corresponding WPE is defined by using the following quantities:
\begin{align}
    \tilde{J}_{ij}\equiv\frac{1}{N}\sum_{\mu=1}^M{\mathbf{w}}^\mu\otimes{\mathbf{w}}^\mu-\text{Diag}\left(\frac{1}{N}\sum_{\mu=1}^M{\mathbf{w}}^\mu\otimes{\mathbf{w}}^\mu\right),
\end{align}
where $\text{Diag}(A)$ denotes a matrix with the diagonal component of some matrix $A$ as its diagonal elements and $M$ column vectors in $N$ dimensions $\mathbf{w}^\mu \ (\mu=1,\ldots,M)$ are taken to follow the Gaussian distribution $\mathcal{N}(0,\Sigma)$ whose covariance is defined as
\begin{align}
    \Sigma\equiv\frac{N}{N-1}\left({\mathbf{I}}-\frac{1}{N}\mathbf{S}^{\text{(GS)}}{\mathbf{S}^{\text{(GS)}}}^{\text{T}}\right),
\end{align}
with the $N$ dimensional identity matrix $\mathbf{I}$. 
The WPE can be expected to be a more difficult problem in which to find the solution for small $\alpha\equiv \frac{M}{N}$ as discussed in Ref.~\cite{PhysRevE.101.052102}.
To fix energy scales between different $\alpha$, we use the following normalized WPE in the numerical simulations:
\begin{align}
  J_{ij}=\frac{1}{C}\tilde{J}_{ij},
\end{align}
with a normalization constant $C\equiv \text{max}(|\tilde{J}_{ij}|)$.
We use $\mathbf{S}^{\text{(GS)}}=(1,\ldots,1)$ and fixed random seed for generating the distribution $\mathcal{N}(0,\Sigma)$.
We report the results for the following numerical experiments of different random seeds in Appendix \ref{other_seeds}.

\subsection{Calculation of the linear stable fixed points}~\label{sec: calc lin stab points}
To investigate the relation between the property for optimal solution search and static properties of the system, the number of stable fixed points $\mathcal{N}_{\text{SP}}$ is computed as a function of $\delta_m$. 
The definition of $\mathcal{N}_{\text{SP}}$ is a number of the fixed points whose eigenvalues of corresponding Jacobian $\mathrm{J}$ are all negative for $\delta_m$.

For the normalized WPE models, the vector $\mathbf{w}^\mu$ is perpendicular to $\mathbf{S}^{\text{(GS)}}$ by construction,
so $\mathrm{J_{xx}}$ for $\mathbf{S}^{\text{(GS)}}$ is obtained as
\begin{align}
    \mathrm{J_{xx}}=\text{diag}\left(-2\delta_m\right)-\frac{\beta}{N C}\sum_{\mu=1}^M{\mathbf{w}}^\mu\otimes{\mathbf{w}}^\mu.
\end{align}
Since $\sum_{\mu=1}^M{\mathbf{w}}^\mu\otimes{\mathbf{w}}^\mu$ is a positive semi-definite matrix and its eigenvalues are non-negative.
From these property, we see that the fixed point corresponding to $\mathbf{S}^{\text{(GS)}}$ is stable for any positive $\delta_m$.

\begin{figure}[t]
\centering
\includegraphics[width=1.1\linewidth]{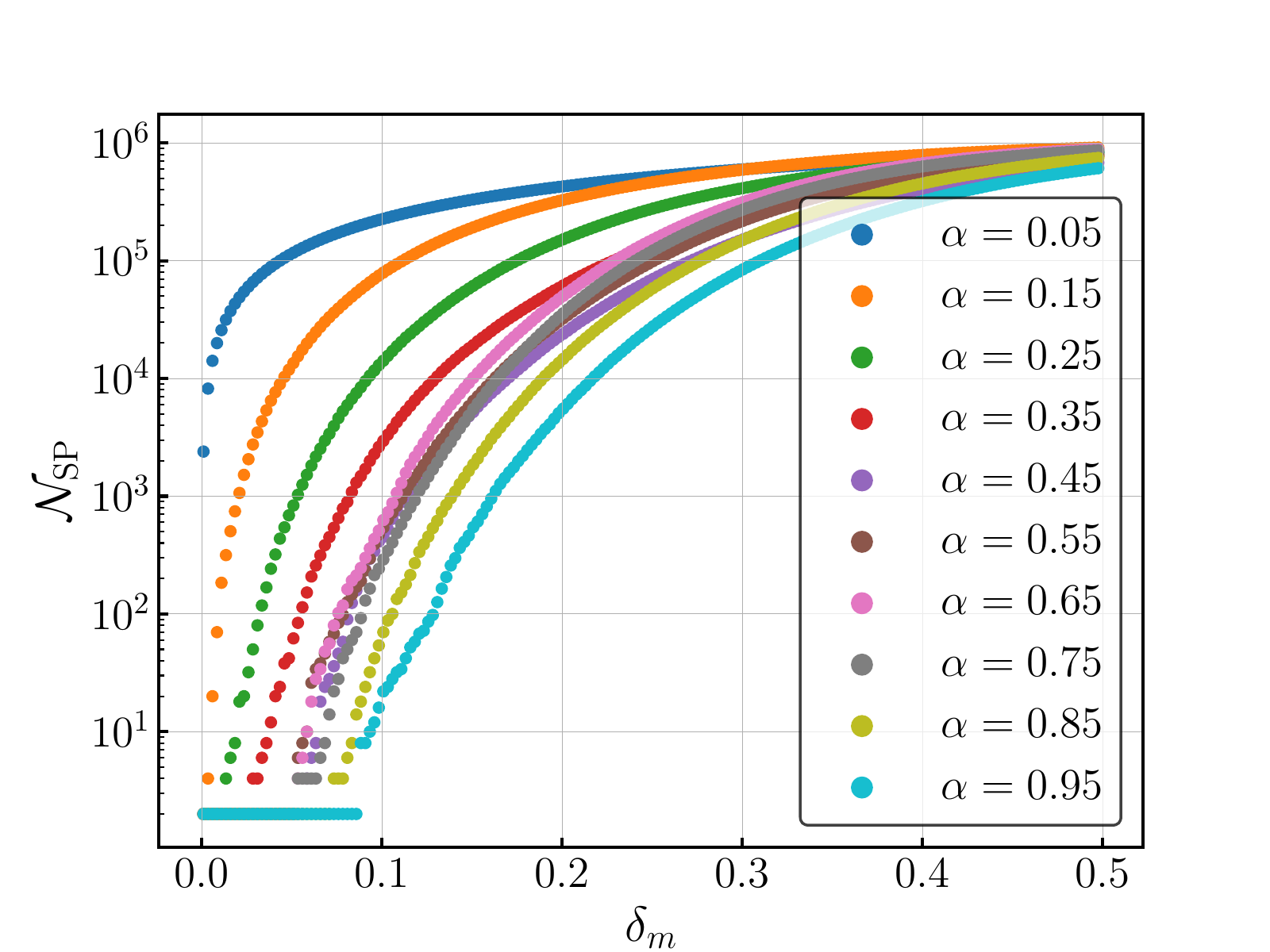}
\caption{Number of stable fixed points during the optimization for $\beta=0.3$.}   
\label{fig:nsp} 
\end{figure}

The numbers of stable fixed points for $\beta=0.3$ are shown in Fig.~\ref{fig:nsp} that represents the relations between $\delta_m$ and $\mathcal{N}_{\text{SP}}$. 
In each $\alpha$, we take the value of $\delta_m$ from $0.001$ to $0.5$ divided by $200$.
The dependence of $\delta_m$ on $\mathcal{N}_{\text{SP}}$ differs from instance to instance.
In particular for small $\alpha$, many fixed points appear between $\delta_m=0$ and $0.1$.
It is expected that their differences probably affect the success probability, which we will numerically observe in a later section.

\begin{figure}[t]
    \centering
    \includegraphics[width=1.1\linewidth]{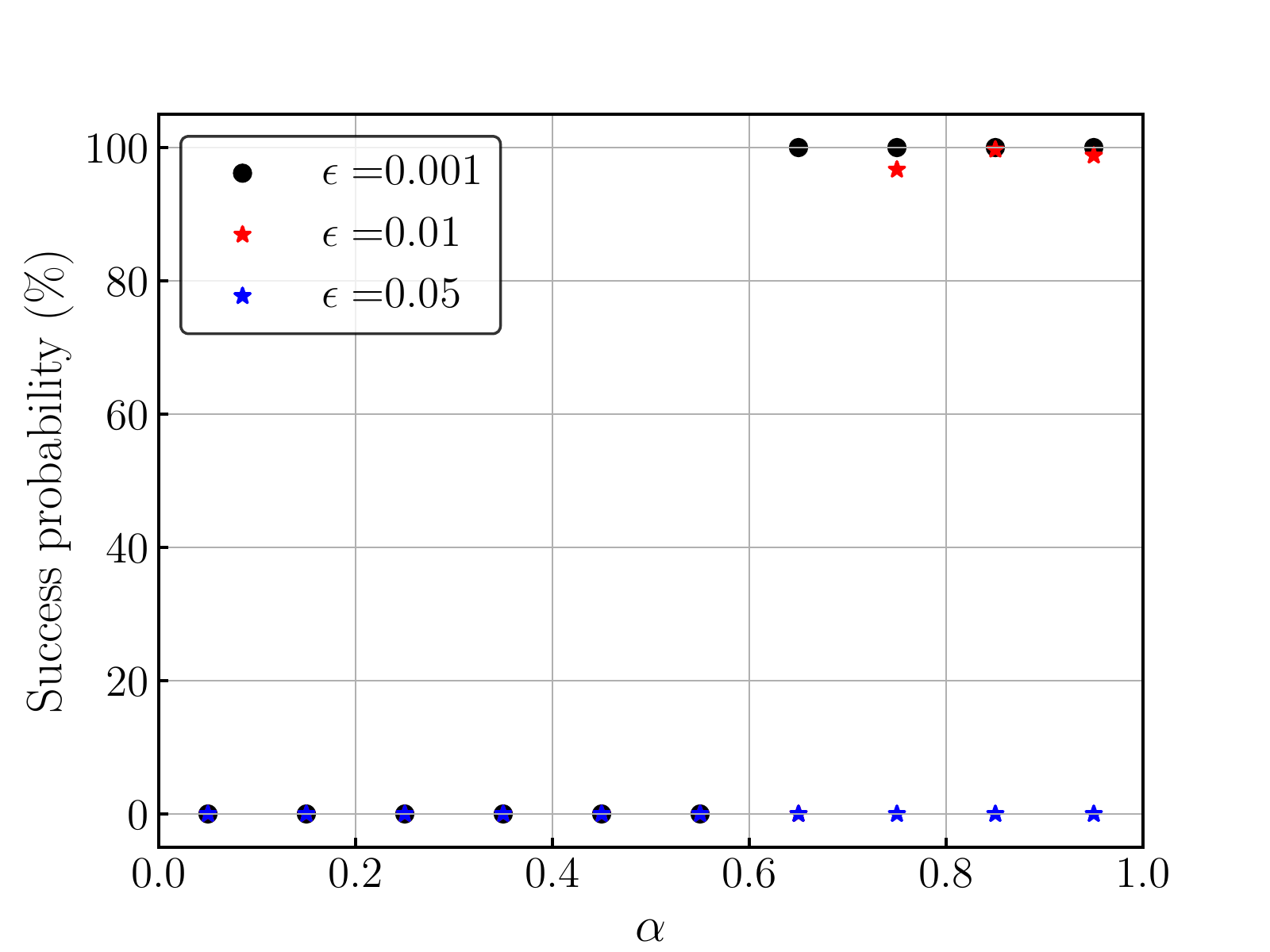}
    \caption{Success probability of the optimizations for $\alpha=0.05,0.15,\ldots,0.95$.}
    \label{fig:lin}    
\end{figure}

\subsection{The effect of the number of stable fixed points on the Success probability}~\label{sec: Success prob}
In this subsection, we report the numerical results that show the relations between $\mathcal{N}_{\text{SP}}$ and the success probability. 
The success probability means the percentage of trials in which the ground state is obtained among multiple trials from randomly generated initial conditions. 
For each Ising model and each parameter, 1000 trials were performed.

We here take $\epsilon=0.05, 0.01, 0.001,\ t_c=20$, and initial conditions are taken from the interval $[-0.1,0.1]$. 
We show the results of the optimizations for $\kappa=0.01$ in Fig.~\ref{fig:lin}.
The result shows that small $\alpha$ instances are difficult in all $\epsilon$,
 and it is found that the smaller the $\epsilon$ , the higher the accuracy. 
Considered together with the results of small $\alpha$ instances in Fig.~\ref{fig:nsp}, many local minima that appear in large numbers early in the calculation interfere with the Ising optimization.
We note that this can also be found to occur in other random seeds (see Appendix \ref{other_seeds}).
In the normalized WPE, the fixed point corresponding to the ground state stabilizes before any other fixed points.
To increase the success probability, it is better to find the correct fixed points before many local minima stabilize.
Therefore, the slow changing of $\delta_m(t)$ in time is needed for the optimizations. 

\subsection{Bifurcation delay effect on the Success probability}~\label{sec: BD_effect}
\begin{figure}[t]
   \begin{center}
    \includegraphics[width=1.1\linewidth]{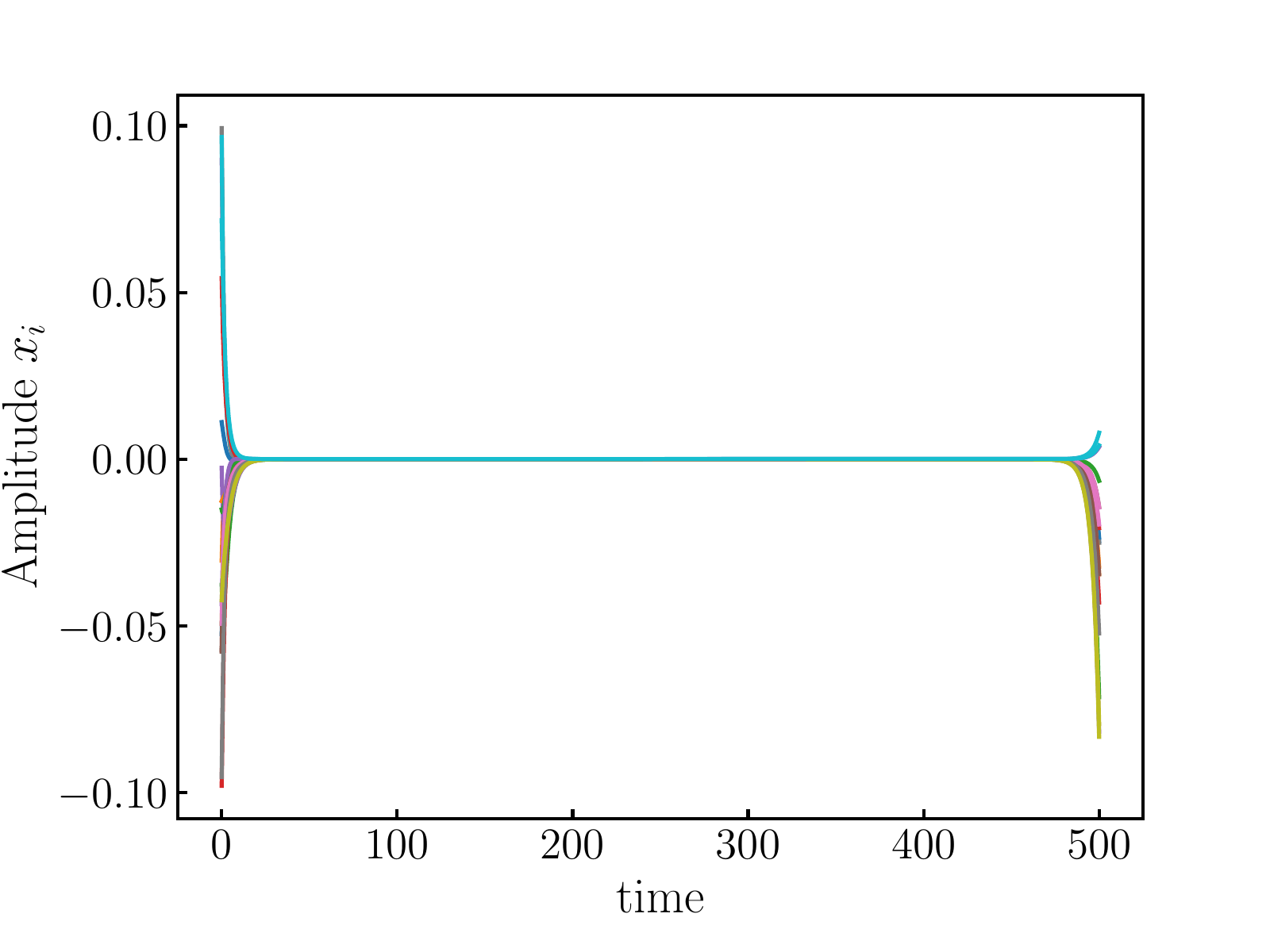}
    \caption{Time evolution of $x_i(t)$ with $t_c=250$ in $\alpha=0.95$. Each colored line shows amplitude $x_i(t)$ for each index $i$.}
    \label{fig:tb_effects}
    \end{center}    
\end{figure}

\begin{figure}[t]
    \centering
    \includegraphics[width=1.1\linewidth]{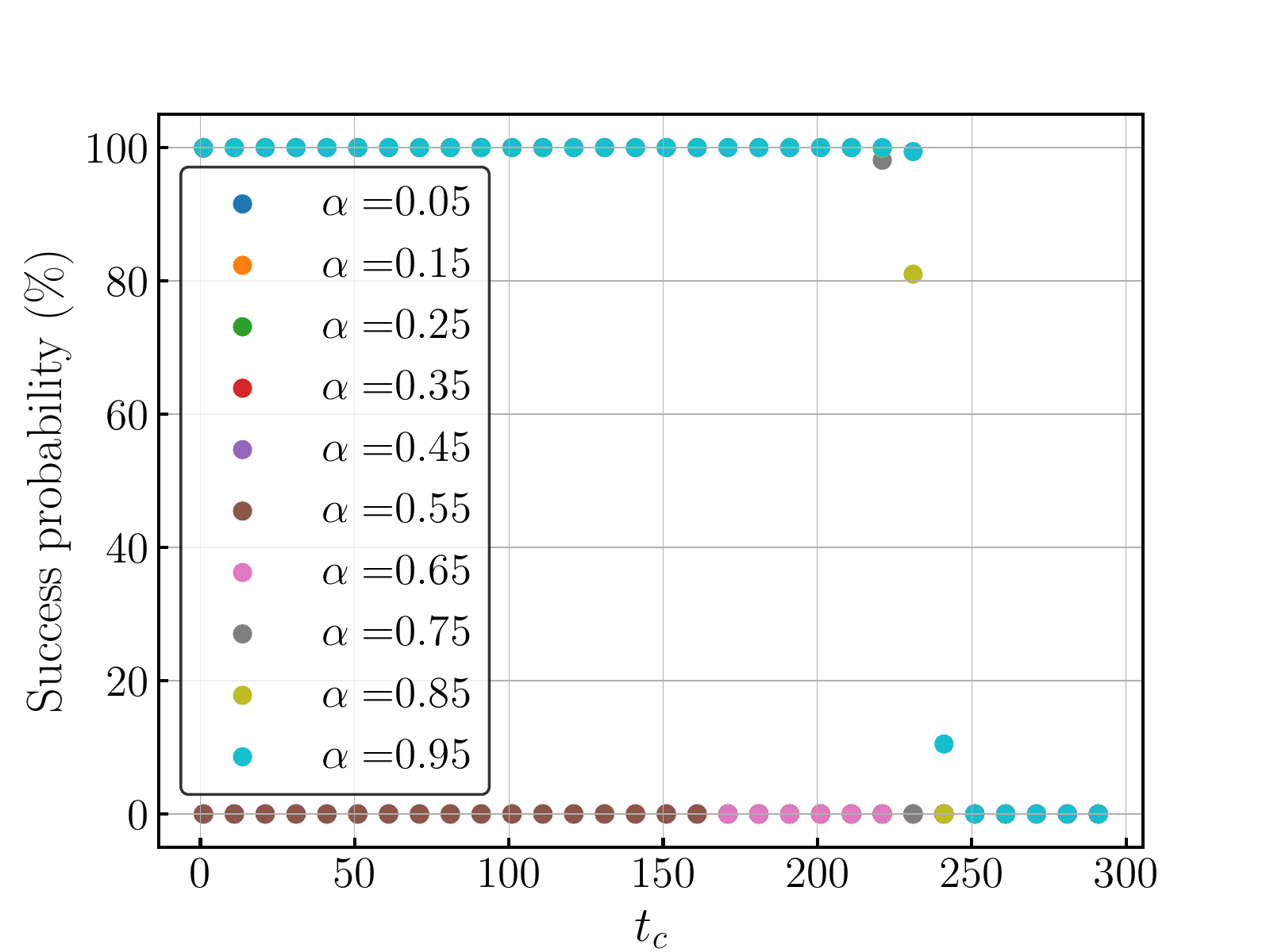}
    \caption{Success probability of the optimizations with $t_c=1,11,\ldots,291$ and $\kappa=0.01$ in all $\alpha$.}
    \label{fig:tc}    
\end{figure}

We report the dynamical effect on the searching ability of our model in terms of the bifurcation delay. 
Noting that here we also fix the calculation time $T=500$ for all experiments.
As discussed in Sec.~\ref{sec: dynamical}, when $t_c$ is $250$, bifurcation delay time $t_b$ equals approximately $500(=T)$ from Eq.~(\ref{tb2}),
 and this can also be checked by the numerical results in Fig.~\ref{fig:tb_effects}.
The results of the optimizations in $t_c=1,11,21,\ldots,291$ are shown in Fig.~\ref{fig:tc}.
From these results, we see significant decreases in the success probability for all $\alpha$ in $t_c$ larger than around $250$ as expected. 
We now conclude the bifurcation delay affects the optimizations, and see that the value of $t_c$ (or $m(t)$ and $\delta_m(t)$) should be properly chosen for the optimizations.

\subsection{Comparison between proposed model and mean field CIM without noises}
\begin{figure}[t]
    \begin{center}
     \includegraphics[width=1.1\linewidth]{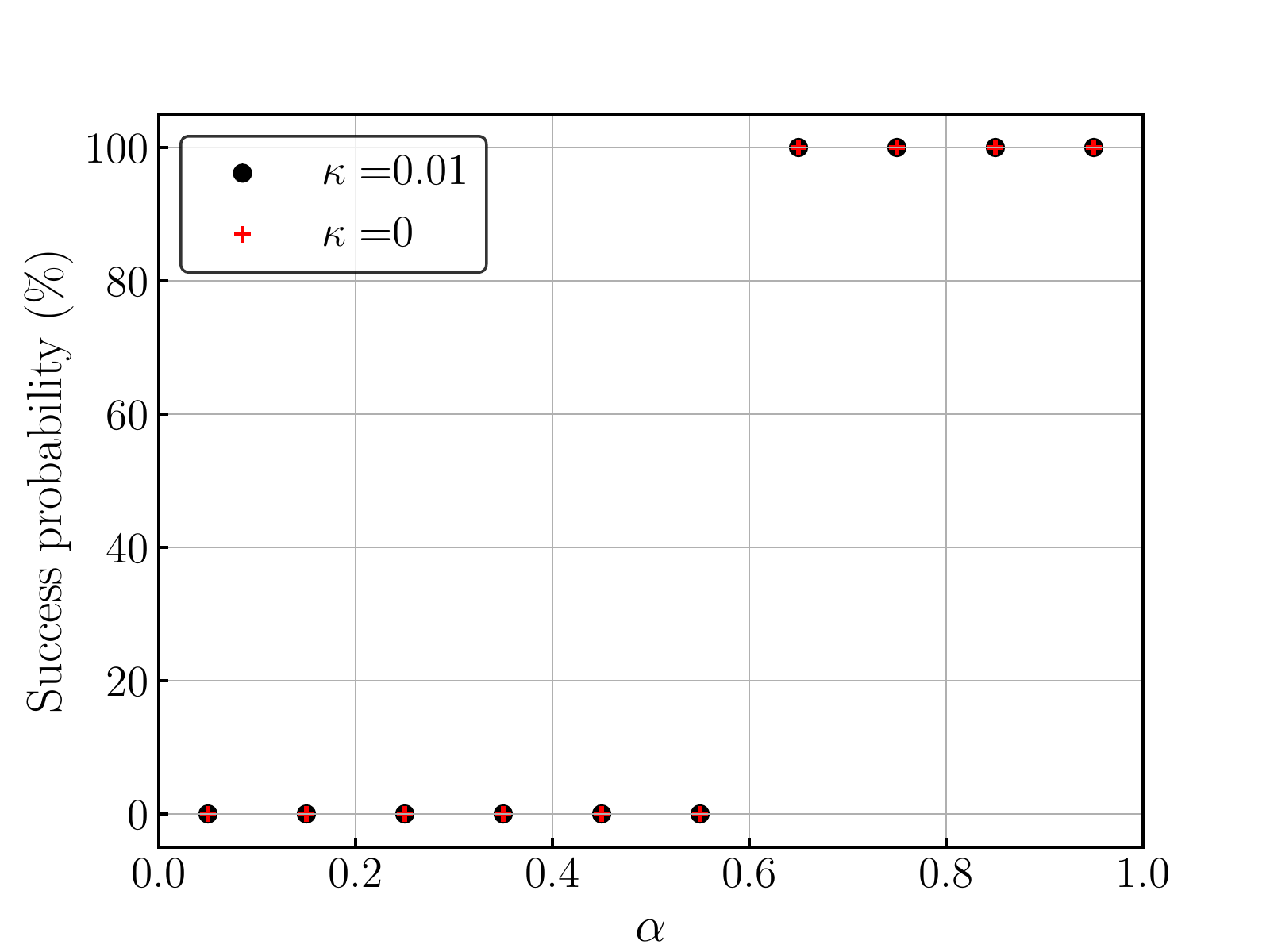}
     \caption{Success probability of the optimizations in $\alpha=0.05,0.15,\ldots,0.95$ for $\epsilon=0.001$ with $\kappa=0$ and $0.01$.}
     \label{fig:sp_normal}
     \end{center}    
 \end{figure}

We also discuss the optimization property of our model and mean field CIM without noises.
Here the latter is defined as $\kappa=0$ in Eq.~(\ref{system_def1}). 
First, the results of the optimizations are shown in Fig.~\ref{fig:sp_normal}.
We take the same parameters used in Sec.~\ref{sec: Success prob}.
Although fixed point analysis discussed in Sec.~\ref{sec: static} cannot be applied for $\kappa=0$, we can see that both results are in good quantitative agreement at least for these problems.
For other WPE with different random seeds, these tendencies are also confirmed, which we discuss in detail in Appendix \ref{other_seeds}. 

\begin{figure}[t]
    \begin{center}
     \includegraphics[width=1.1\linewidth]{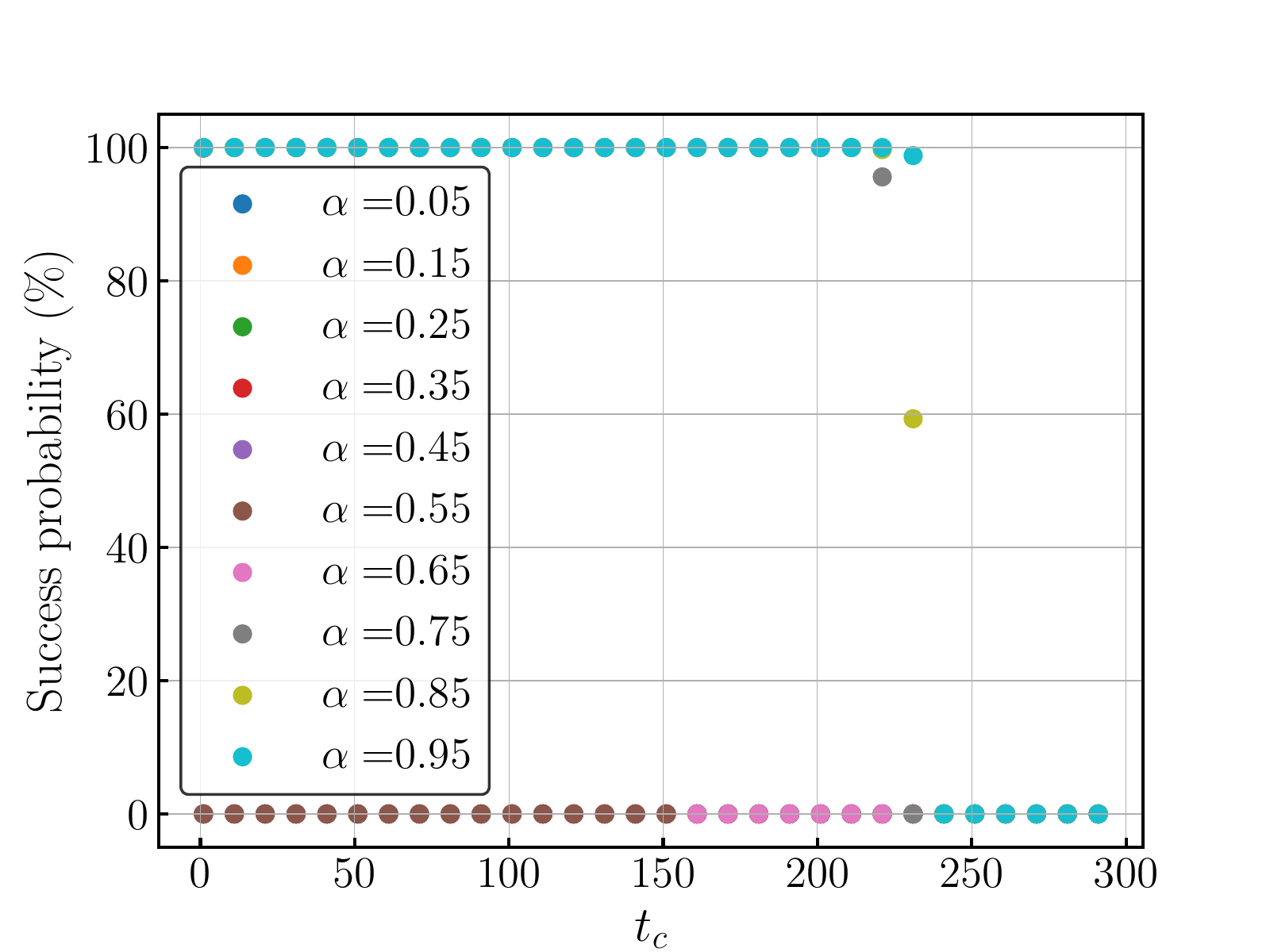}
     \caption{Success probability of the optimizations with $t_c=1,11,\ldots,291$ and $\kappa=0$ in all $\alpha$.}
     \label{fig:tb_effects_normal}
     \end{center}    
\end{figure}

In addition, we numerically check the effect of bifurcation delay for $\kappa=0$.
Here we also take the same parameters used in Sec.~\ref{sec: BD_effect}.
Figure \ref{fig:tb_effects_normal} shows the results which show that the bifurcation delay also obstructs the optimization in $\kappa=0$.
In this case, decline in the success probability around $t_c=250$ is consistent with the result of Eq.~(\ref{tb1}).

\section{Conclusion}~\label{sec: Conclusion}
In this paper, we proposed a coherent Ising machine (CIM)-like heuristic whose properties are easier to investigate by time dependent Lagrange multipliers
and studied the effects of local minima and the bifurcation delay on the optimizations.
To see these effects clearly, we used the normalized Wishart planted ensembles (WPE) for the experiments.
From the results (Figs.~\ref{fig:nsp} and \ref{fig:lin}), 
we have numerically observed that the ground state search is hindered by a lot of local minima in the early stage of the optimizations.
This information is easily obtained from the linear stability analyses.
It should be noted that the existence of many local minima is merely a sufficient condition for problems to be difficult. 
In fact, there is no significant difference in Fig.~\ref{fig:nsp} for $\alpha=0.55$ and $0.65$,
but, the difference of the success probability between them is obvious.
It is expected that the factors of this difference are other than the number of local minima.
We have also pointed out the effects of the bifurcation delay on the optimizations (Fig.~\ref{fig:tb_effects}).
Due to the parameter varying in time, the dynamics are changed from what is expected from the bifurcation theory for static property.
Moreover, the optimization results of our model quantitatively agree with those of mean field CIM without noises (Fig.~\ref{fig:sp_normal}).
The effects of the bifurcation delay on the optimizations have also seen in mean field CIM without noises (Fig.~\ref{fig:tb_effects_normal}).
These results suggest that our model not only has easy-to-investigate  property but also is useful to analyze the performance of other CIM-like algorithms.

Let us comment on applicability of our results such as the relationship between local minima and the success probability for other Ising models.
Our model and analyses discussed in Sec.~\ref{sec: property} can be applied to any Ising models.
However, the fixed points corresponding to the ground state are not stable in the first stage of the calculation in general.
Therefore, as well as the ratio of increase in the number of local minima, the timing of the stabilization of the fixed points corresponding to the ground state could also affect the optimizations.
Investigating these factors in detail is important future work.

We also note applications of our results for using algorithms like CIMs.
In general, the Ising Hamiltonian has several hyper-parameters such as coefficients of the penalty terms.
To solve the problems efficiently, these hyper-parameters need to be tuned properly.
For example, in Quantum Annealing, because of the adiabatic theorem, the energy gap between the ground state and first excited state during the annealing is a good indicator in tunings.
In contrast, there has been no general characteristic of difficulties in optimizations for CIMs so far.
It would be an interesting direction to investigate the relation between the property of local minima and the appropriate values of hyper-parameters in our model and CIMs.

\acknowledgements
The author thanks Kazuhiro Miyahara, Yasuhiro Yamada, Kentaro Ohno and Rion Shimazu for useful discussions and comments.

\bibliography{draft.bib} 

\providecommand{\noopsort}[1]{}\providecommand{\singleletter}[1]{#1}%
\begin{thebibliography}{34}%
\makeatletter
\providecommand \@ifxundefined [1]{%
 \@ifx{#1\undefined}
}%
\providecommand \@ifnum [1]{%
 \ifnum #1\expandafter \@firstoftwo
 \else \expandafter \@secondoftwo
 \fi
}%
\providecommand \@ifx [1]{%
 \ifx #1\expandafter \@firstoftwo
 \else \expandafter \@secondoftwo
 \fi
}%
\providecommand \natexlab [1]{#1}%
\providecommand \enquote  [1]{``#1''}%
\providecommand \bibnamefont  [1]{#1}%
\providecommand \bibfnamefont [1]{#1}%
\providecommand \citenamefont [1]{#1}%
\providecommand \href@noop [0]{\@secondoftwo}%
\providecommand \href [0]{\begingroup \@sanitize@url \@href}%
\providecommand \@href[1]{\@@startlink{#1}\@@href}%
\providecommand \@@href[1]{\endgroup#1\@@endlink}%
\providecommand \@sanitize@url [0]{\catcode `\\12\catcode `\$12\catcode
  `\&12\catcode `\#12\catcode `\^12\catcode `\_12\catcode `\%12\relax}%
\providecommand \@@startlink[1]{}%
\providecommand \@@endlink[0]{}%
\providecommand \url  [0]{\begingroup\@sanitize@url \@url }%
\providecommand \@url [1]{\endgroup\@href {#1}{\urlprefix }}%
\providecommand \urlprefix  [0]{URL }%
\providecommand \Eprint [0]{\href }%
\providecommand \doibase [0]{https://doi.org/}%
\providecommand \selectlanguage [0]{\@gobble}%
\providecommand \bibinfo  [0]{\@secondoftwo}%
\providecommand \bibfield  [0]{\@secondoftwo}%
\providecommand \translation [1]{[#1]}%
\providecommand \BibitemOpen [0]{}%
\providecommand \bibitemStop [0]{}%
\providecommand \bibitemNoStop [0]{.\EOS\space}%
\providecommand \EOS [0]{\spacefactor3000\relax}%
\providecommand \BibitemShut  [1]{\csname bibitem#1\endcsname}%
\let\auto@bib@innerbib\@empty
\bibitem [{\citenamefont {Lucas}(2014)}]{lucas2014ising}%
  \BibitemOpen
  \bibfield  {author} {\bibinfo {author} {\bibfnamefont {A.}~\bibnamefont
  {Lucas}},\ }\bibfield  {title} {\bibinfo {title} {Ising formulations of many
  np problems},\ }\href@noop {} {\bibfield  {journal} {\bibinfo  {journal}
  {Frontiers in physics}\ ,\ \bibinfo {pages} {5}} (\bibinfo {year}
  {2014})}\BibitemShut {NoStop}%
\bibitem [{\citenamefont {Kirkpatrick}\ \emph {et~al.}(1983)\citenamefont
  {Kirkpatrick}, \citenamefont {Gelatt~Jr},\ and\ \citenamefont
  {Vecchi}}]{kirkpatrick1983optimization}%
  \BibitemOpen
  \bibfield  {author} {\bibinfo {author} {\bibfnamefont {S.}~\bibnamefont
  {Kirkpatrick}}, \bibinfo {author} {\bibfnamefont {C.~D.}\ \bibnamefont
  {Gelatt~Jr}},\ and\ \bibinfo {author} {\bibfnamefont {M.~P.}\ \bibnamefont
  {Vecchi}},\ }\bibfield  {title} {\bibinfo {title} {Optimization by simulated
  annealing},\ }\href@noop {} {\bibfield  {journal} {\bibinfo  {journal}
  {science}\ }\textbf {\bibinfo {volume} {220}},\ \bibinfo {pages} {671}
  (\bibinfo {year} {1983})}\BibitemShut {NoStop}%
\bibitem [{\citenamefont {Hopfield}\ and\ \citenamefont
  {Tank}(1985)}]{hopfield1985neural}%
  \BibitemOpen
  \bibfield  {author} {\bibinfo {author} {\bibfnamefont {J.~J.}\ \bibnamefont
  {Hopfield}}\ and\ \bibinfo {author} {\bibfnamefont {D.~W.}\ \bibnamefont
  {Tank}},\ }\bibfield  {title} {\bibinfo {title} {“neural” computation of
  decisions in optimization problems},\ }\href@noop {} {\bibfield  {journal}
  {\bibinfo  {journal} {Biological cybernetics}\ }\textbf {\bibinfo {volume}
  {52}},\ \bibinfo {pages} {141} (\bibinfo {year} {1985})}\BibitemShut
  {NoStop}%
\bibitem [{\citenamefont {Wang}\ \emph {et~al.}(2013)\citenamefont {Wang},
  \citenamefont {Marandi}, \citenamefont {Wen}, \citenamefont {Byer},\ and\
  \citenamefont {Yamamoto}}]{wang2013coherent}%
  \BibitemOpen
  \bibfield  {author} {\bibinfo {author} {\bibfnamefont {Z.}~\bibnamefont
  {Wang}}, \bibinfo {author} {\bibfnamefont {A.}~\bibnamefont {Marandi}},
  \bibinfo {author} {\bibfnamefont {K.}~\bibnamefont {Wen}}, \bibinfo {author}
  {\bibfnamefont {R.~L.}\ \bibnamefont {Byer}},\ and\ \bibinfo {author}
  {\bibfnamefont {Y.}~\bibnamefont {Yamamoto}},\ }\bibfield  {title} {\bibinfo
  {title} {Coherent ising machine based on degenerate optical parametric
  oscillators},\ }\href {https://doi.org/10.1103/PhysRevA.88.063853} {\bibfield
   {journal} {\bibinfo  {journal} {Phys. Rev. A}\ }\textbf {\bibinfo {volume}
  {88}},\ \bibinfo {pages} {063853} (\bibinfo {year} {2013})}\BibitemShut
  {NoStop}%
\bibitem [{\citenamefont {Goto}(2016)}]{goto2016bifurcation}%
  \BibitemOpen
  \bibfield  {author} {\bibinfo {author} {\bibfnamefont {H.}~\bibnamefont
  {Goto}},\ }\bibfield  {title} {\bibinfo {title} {Bifurcation-based adiabatic
  quantum computation with a nonlinear oscillator network},\ }\href@noop {}
  {\bibfield  {journal} {\bibinfo  {journal} {Scientific reports}\ }\textbf
  {\bibinfo {volume} {6}},\ \bibinfo {pages} {1} (\bibinfo {year}
  {2016})}\BibitemShut {NoStop}%
\bibitem [{\citenamefont {Kalinin}\ and\ \citenamefont
  {Berloff}(2018)}]{kalinin2018global}%
  \BibitemOpen
  \bibfield  {author} {\bibinfo {author} {\bibfnamefont {K.~P.}\ \bibnamefont
  {Kalinin}}\ and\ \bibinfo {author} {\bibfnamefont {N.~G.}\ \bibnamefont
  {Berloff}},\ }\bibfield  {title} {\bibinfo {title} {Global optimization of
  spin hamiltonians with gain-dissipative systems},\ }\href
  {https://doi.org/10.1038/s41598-018-35416-1} {\bibfield  {journal} {\bibinfo
  {journal} {Scientific reports}\ }\textbf {\bibinfo {volume} {8}},\ \bibinfo
  {pages} {1} (\bibinfo {year} {2018})}\BibitemShut {NoStop}%
\bibitem [{\citenamefont {Moln{\'a}r}\ \emph {et~al.}(2018)\citenamefont
  {Moln{\'a}r}, \citenamefont {Moln{\'a}r}, \citenamefont {Varga},
  \citenamefont {Toroczkai},\ and\ \citenamefont
  {Ercsey-Ravasz}}]{molnar2018continuous}%
  \BibitemOpen
  \bibfield  {author} {\bibinfo {author} {\bibfnamefont {B.}~\bibnamefont
  {Moln{\'a}r}}, \bibinfo {author} {\bibfnamefont {F.}~\bibnamefont
  {Moln{\'a}r}}, \bibinfo {author} {\bibfnamefont {M.}~\bibnamefont {Varga}},
  \bibinfo {author} {\bibfnamefont {Z.}~\bibnamefont {Toroczkai}},\ and\
  \bibinfo {author} {\bibfnamefont {M.}~\bibnamefont {Ercsey-Ravasz}},\
  }\bibfield  {title} {\bibinfo {title} {A continuous-time maxsat solver with
  high analog performance},\ }\href@noop {} {\bibfield  {journal} {\bibinfo
  {journal} {Nature communications}\ }\textbf {\bibinfo {volume} {9}},\
  \bibinfo {pages} {1} (\bibinfo {year} {2018})}\BibitemShut {NoStop}%
\bibitem [{\citenamefont {Goto}\ \emph {et~al.}(2019)\citenamefont {Goto},
  \citenamefont {Tatsumura},\ and\ \citenamefont
  {Dixon}}]{goto2019combinatorial}%
  \BibitemOpen
  \bibfield  {author} {\bibinfo {author} {\bibfnamefont {H.}~\bibnamefont
  {Goto}}, \bibinfo {author} {\bibfnamefont {K.}~\bibnamefont {Tatsumura}},\
  and\ \bibinfo {author} {\bibfnamefont {A.~R.}\ \bibnamefont {Dixon}},\
  }\bibfield  {title} {\bibinfo {title} {Combinatorial optimization by
  simulating adiabatic bifurcations in nonlinear hamiltonian systems},\
  }\href@noop {} {\bibfield  {journal} {\bibinfo  {journal} {Science advances}\
  }\textbf {\bibinfo {volume} {5}},\ \bibinfo {pages} {eaav2372} (\bibinfo
  {year} {2019})}\BibitemShut {NoStop}%
\bibitem [{\citenamefont {Goto}(2019)}]{doi:10.7566/JPSJ.88.061015}%
  \BibitemOpen
  \bibfield  {author} {\bibinfo {author} {\bibfnamefont {H.}~\bibnamefont
  {Goto}},\ }\bibfield  {title} {\bibinfo {title} {Quantum computation based on
  quantum adiabatic bifurcations of kerr-nonlinear parametric oscillators},\
  }\href {https://doi.org/10.7566/JPSJ.88.061015} {\bibfield  {journal}
  {\bibinfo  {journal} {Journal of the Physical Society of Japan}\ }\textbf
  {\bibinfo {volume} {88}},\ \bibinfo {pages} {061015} (\bibinfo {year}
  {2019})}\BibitemShut {NoStop}%
\bibitem [{\citenamefont {Kalinin}\ and\ \citenamefont
  {Berloff}(2020)}]{Kalinin2020}%
  \BibitemOpen
  \bibfield  {author} {\bibinfo {author} {\bibfnamefont {K.~P.}\ \bibnamefont
  {Kalinin}}\ and\ \bibinfo {author} {\bibfnamefont {N.~G.}\ \bibnamefont
  {Berloff}},\ }\bibinfo {title} {Nonlinear systems for unconventional
  computing},\ in\ \href {https://doi.org/10.1007/978-3-030-44992-6_15} {\emph
  {\bibinfo {booktitle} {Emerging Frontiers in Nonlinear Science}}},\ \bibinfo
  {editor} {edited by\ \bibinfo {editor} {\bibfnamefont {P.~G.}\ \bibnamefont
  {Kevrekidis}}, \bibinfo {editor} {\bibfnamefont {J.}~\bibnamefont
  {Cuevas-Maraver}},\ and\ \bibinfo {editor} {\bibfnamefont {A.}~\bibnamefont
  {Saxena}}}\ (\bibinfo  {publisher} {Springer International Publishing},\
  \bibinfo {address} {Cham},\ \bibinfo {year} {2020})\ pp.\ \bibinfo {pages}
  {345--369}\BibitemShut {NoStop}%
\bibitem [{\citenamefont {Ng}\ \emph {et~al.}(2022)\citenamefont {Ng},
  \citenamefont {Onodera}, \citenamefont {Kako}, \citenamefont {McMahon},
  \citenamefont {Mabuchi},\ and\ \citenamefont {Yamamoto}}]{ng2022efficient}%
  \BibitemOpen
  \bibfield  {author} {\bibinfo {author} {\bibfnamefont {E.}~\bibnamefont
  {Ng}}, \bibinfo {author} {\bibfnamefont {T.}~\bibnamefont {Onodera}},
  \bibinfo {author} {\bibfnamefont {S.}~\bibnamefont {Kako}}, \bibinfo {author}
  {\bibfnamefont {P.~L.}\ \bibnamefont {McMahon}}, \bibinfo {author}
  {\bibfnamefont {H.}~\bibnamefont {Mabuchi}},\ and\ \bibinfo {author}
  {\bibfnamefont {Y.}~\bibnamefont {Yamamoto}},\ }\bibfield  {title} {\bibinfo
  {title} {Efficient sampling of ground and low-energy ising spin
  configurations with a coherent ising machine},\ }\href
  {https://doi.org/10.1103/PhysRevResearch.4.013009} {\bibfield  {journal}
  {\bibinfo  {journal} {Phys. Rev. Research}\ }\textbf {\bibinfo {volume}
  {4}},\ \bibinfo {pages} {013009} (\bibinfo {year} {2022})}\BibitemShut
  {NoStop}%
\bibitem [{\citenamefont {Takata}\ \emph {et~al.}(2016)\citenamefont {Takata},
  \citenamefont {Marandi}, \citenamefont {Hamerly}, \citenamefont {Haribara},
  \citenamefont {Maruo}, \citenamefont {Tamate}, \citenamefont {Sakaguchi},
  \citenamefont {Utsunomiya},\ and\ \citenamefont {Yamamoto}}]{takata201616}%
  \BibitemOpen
  \bibfield  {author} {\bibinfo {author} {\bibfnamefont {K.}~\bibnamefont
  {Takata}}, \bibinfo {author} {\bibfnamefont {A.}~\bibnamefont {Marandi}},
  \bibinfo {author} {\bibfnamefont {R.}~\bibnamefont {Hamerly}}, \bibinfo
  {author} {\bibfnamefont {Y.}~\bibnamefont {Haribara}}, \bibinfo {author}
  {\bibfnamefont {D.}~\bibnamefont {Maruo}}, \bibinfo {author} {\bibfnamefont
  {S.}~\bibnamefont {Tamate}}, \bibinfo {author} {\bibfnamefont
  {H.}~\bibnamefont {Sakaguchi}}, \bibinfo {author} {\bibfnamefont
  {S.}~\bibnamefont {Utsunomiya}},\ and\ \bibinfo {author} {\bibfnamefont
  {Y.}~\bibnamefont {Yamamoto}},\ }\bibfield  {title} {\bibinfo {title} {A
  16-bit coherent ising machine for one-dimensional ring and cubic graph
  problems},\ }\href@noop {} {\bibfield  {journal} {\bibinfo  {journal}
  {Scientific reports}\ }\textbf {\bibinfo {volume} {6}},\ \bibinfo {pages} {1}
  (\bibinfo {year} {2016})}\BibitemShut {NoStop}%
\bibitem [{\citenamefont {McMahon}\ \emph {et~al.}(2016)\citenamefont
  {McMahon}, \citenamefont {Marandi}, \citenamefont {Haribara}, \citenamefont
  {Hamerly}, \citenamefont {Langrock}, \citenamefont {Tamate}, \citenamefont
  {Inagaki}, \citenamefont {Takesue}, \citenamefont {Utsunomiya}, \citenamefont
  {Aihara} \emph {et~al.}}]{mcmahon2016fully}%
  \BibitemOpen
  \bibfield  {author} {\bibinfo {author} {\bibfnamefont {P.~L.}\ \bibnamefont
  {McMahon}}, \bibinfo {author} {\bibfnamefont {A.}~\bibnamefont {Marandi}},
  \bibinfo {author} {\bibfnamefont {Y.}~\bibnamefont {Haribara}}, \bibinfo
  {author} {\bibfnamefont {R.}~\bibnamefont {Hamerly}}, \bibinfo {author}
  {\bibfnamefont {C.}~\bibnamefont {Langrock}}, \bibinfo {author}
  {\bibfnamefont {S.}~\bibnamefont {Tamate}}, \bibinfo {author} {\bibfnamefont
  {T.}~\bibnamefont {Inagaki}}, \bibinfo {author} {\bibfnamefont
  {H.}~\bibnamefont {Takesue}}, \bibinfo {author} {\bibfnamefont
  {S.}~\bibnamefont {Utsunomiya}}, \bibinfo {author} {\bibfnamefont
  {K.}~\bibnamefont {Aihara}}, \emph {et~al.},\ }\bibfield  {title} {\bibinfo
  {title} {A fully programmable 100-spin coherent ising machine with all-to-all
  connections},\ }\href@noop {} {\bibfield  {journal} {\bibinfo  {journal}
  {Science}\ }\textbf {\bibinfo {volume} {354}},\ \bibinfo {pages} {614}
  (\bibinfo {year} {2016})}\BibitemShut {NoStop}%
\bibitem [{\citenamefont {Inagaki}\ \emph {et~al.}(2016)\citenamefont
  {Inagaki}, \citenamefont {Haribara}, \citenamefont {Igarashi}, \citenamefont
  {Sonobe}, \citenamefont {Tamate}, \citenamefont {Honjo}, \citenamefont
  {Marandi}, \citenamefont {McMahon}, \citenamefont {Umeki}, \citenamefont
  {Enbutsu}, \citenamefont {Tadanaga}, \citenamefont {Takenouchi},
  \citenamefont {Aihara}, \citenamefont {ichi Kawarabayashi}, \citenamefont
  {Inoue}, \citenamefont {Utsunomiya},\ and\ \citenamefont
  {Takesue}}]{inagaki2016coherent}%
  \BibitemOpen
  \bibfield  {author} {\bibinfo {author} {\bibfnamefont {T.}~\bibnamefont
  {Inagaki}}, \bibinfo {author} {\bibfnamefont {Y.}~\bibnamefont {Haribara}},
  \bibinfo {author} {\bibfnamefont {K.}~\bibnamefont {Igarashi}}, \bibinfo
  {author} {\bibfnamefont {T.}~\bibnamefont {Sonobe}}, \bibinfo {author}
  {\bibfnamefont {S.}~\bibnamefont {Tamate}}, \bibinfo {author} {\bibfnamefont
  {T.}~\bibnamefont {Honjo}}, \bibinfo {author} {\bibfnamefont
  {A.}~\bibnamefont {Marandi}}, \bibinfo {author} {\bibfnamefont {P.~L.}\
  \bibnamefont {McMahon}}, \bibinfo {author} {\bibfnamefont {T.}~\bibnamefont
  {Umeki}}, \bibinfo {author} {\bibfnamefont {K.}~\bibnamefont {Enbutsu}},
  \bibinfo {author} {\bibfnamefont {O.}~\bibnamefont {Tadanaga}}, \bibinfo
  {author} {\bibfnamefont {H.}~\bibnamefont {Takenouchi}}, \bibinfo {author}
  {\bibfnamefont {K.}~\bibnamefont {Aihara}}, \bibinfo {author} {\bibfnamefont
  {K.}~\bibnamefont {ichi Kawarabayashi}}, \bibinfo {author} {\bibfnamefont
  {K.}~\bibnamefont {Inoue}}, \bibinfo {author} {\bibfnamefont
  {S.}~\bibnamefont {Utsunomiya}},\ and\ \bibinfo {author} {\bibfnamefont
  {H.}~\bibnamefont {Takesue}},\ }\bibfield  {title} {\bibinfo {title} {A
  coherent ising machine for 2000-node optimization problems},\ }\href
  {https://doi.org/10.1126/science.aah4243} {\bibfield  {journal} {\bibinfo
  {journal} {Science}\ }\textbf {\bibinfo {volume} {354}},\ \bibinfo {pages}
  {603} (\bibinfo {year} {2016})}\BibitemShut {NoStop}%
\bibitem [{\citenamefont {Honjo}\ \emph {et~al.}(2021)\citenamefont {Honjo},
  \citenamefont {Sonobe}, \citenamefont {Inaba}, \citenamefont {Inagaki},
  \citenamefont {Ikuta}, \citenamefont {Yamada}, \citenamefont {Kazama},
  \citenamefont {Enbutsu}, \citenamefont {Umeki}, \citenamefont {Kasahara}
  \emph {et~al.}}]{honjo2021100}%
  \BibitemOpen
  \bibfield  {author} {\bibinfo {author} {\bibfnamefont {T.}~\bibnamefont
  {Honjo}}, \bibinfo {author} {\bibfnamefont {T.}~\bibnamefont {Sonobe}},
  \bibinfo {author} {\bibfnamefont {K.}~\bibnamefont {Inaba}}, \bibinfo
  {author} {\bibfnamefont {T.}~\bibnamefont {Inagaki}}, \bibinfo {author}
  {\bibfnamefont {T.}~\bibnamefont {Ikuta}}, \bibinfo {author} {\bibfnamefont
  {Y.}~\bibnamefont {Yamada}}, \bibinfo {author} {\bibfnamefont
  {T.}~\bibnamefont {Kazama}}, \bibinfo {author} {\bibfnamefont
  {K.}~\bibnamefont {Enbutsu}}, \bibinfo {author} {\bibfnamefont
  {T.}~\bibnamefont {Umeki}}, \bibinfo {author} {\bibfnamefont
  {R.}~\bibnamefont {Kasahara}}, \emph {et~al.},\ }\bibfield  {title} {\bibinfo
  {title} {100,000-spin coherent ising machine},\ }\href@noop {} {\bibfield
  {journal} {\bibinfo  {journal} {Science advances}\ }\textbf {\bibinfo
  {volume} {7}},\ \bibinfo {pages} {eabh0952} (\bibinfo {year}
  {2021})}\BibitemShut {NoStop}%
\bibitem [{\citenamefont {Clements}\ \emph {et~al.}(2017)\citenamefont
  {Clements}, \citenamefont {Renema}, \citenamefont {Wen}, \citenamefont
  {Chrzanowski}, \citenamefont {Kolthammer},\ and\ \citenamefont
  {Walmsley}}]{PhysRevA.96.043850}%
  \BibitemOpen
  \bibfield  {author} {\bibinfo {author} {\bibfnamefont {W.~R.}\ \bibnamefont
  {Clements}}, \bibinfo {author} {\bibfnamefont {J.~J.}\ \bibnamefont
  {Renema}}, \bibinfo {author} {\bibfnamefont {Y.~H.}\ \bibnamefont {Wen}},
  \bibinfo {author} {\bibfnamefont {H.~M.}\ \bibnamefont {Chrzanowski}},
  \bibinfo {author} {\bibfnamefont {W.~S.}\ \bibnamefont {Kolthammer}},\ and\
  \bibinfo {author} {\bibfnamefont {I.~A.}\ \bibnamefont {Walmsley}},\
  }\bibfield  {title} {\bibinfo {title} {Gaussian optical ising machines},\
  }\href {https://doi.org/10.1103/PhysRevA.96.043850} {\bibfield  {journal}
  {\bibinfo  {journal} {Phys. Rev. A}\ }\textbf {\bibinfo {volume} {96}},\
  \bibinfo {pages} {043850} (\bibinfo {year} {2017})}\BibitemShut {NoStop}%
\bibitem [{\citenamefont {Yamamura}\ \emph {et~al.}(2017)\citenamefont
  {Yamamura}, \citenamefont {Aihara},\ and\ \citenamefont
  {Yamamoto}}]{PhysRevA.96.053834}%
  \BibitemOpen
  \bibfield  {author} {\bibinfo {author} {\bibfnamefont {A.}~\bibnamefont
  {Yamamura}}, \bibinfo {author} {\bibfnamefont {K.}~\bibnamefont {Aihara}},\
  and\ \bibinfo {author} {\bibfnamefont {Y.}~\bibnamefont {Yamamoto}},\
  }\bibfield  {title} {\bibinfo {title} {Quantum model for coherent ising
  machines: Discrete-time measurement feedback formulation},\ }\href
  {https://doi.org/10.1103/PhysRevA.96.053834} {\bibfield  {journal} {\bibinfo
  {journal} {Phys. Rev. A}\ }\textbf {\bibinfo {volume} {96}},\ \bibinfo
  {pages} {053834} (\bibinfo {year} {2017})}\BibitemShut {NoStop}%
\bibitem [{\citenamefont {Yamamoto}\ \emph {et~al.}(2020)\citenamefont
  {Yamamoto}, \citenamefont {Leleu}, \citenamefont {Ganguli},\ and\
  \citenamefont {Mabuchi}}]{doi:10.1063/5.0016140}%
  \BibitemOpen
  \bibfield  {author} {\bibinfo {author} {\bibfnamefont {Y.}~\bibnamefont
  {Yamamoto}}, \bibinfo {author} {\bibfnamefont {T.}~\bibnamefont {Leleu}},
  \bibinfo {author} {\bibfnamefont {S.}~\bibnamefont {Ganguli}},\ and\ \bibinfo
  {author} {\bibfnamefont {H.}~\bibnamefont {Mabuchi}},\ }\bibfield  {title}
  {\bibinfo {title} {Coherent ising machines―quantum optics and neural network
  perspectives},\ }\href {https://doi.org/10.1063/5.0016140} {\bibfield
  {journal} {\bibinfo  {journal} {Applied Physics Letters}\ }\textbf {\bibinfo
  {volume} {117}},\ \bibinfo {pages} {160501} (\bibinfo {year} {2020})},\
  \Eprint {https://arxiv.org/abs/https://doi.org/10.1063/5.0016140}
  {https://doi.org/10.1063/5.0016140} \BibitemShut {NoStop}%
\bibitem [{\citenamefont {Leleu}\ \emph {et~al.}(2017)\citenamefont {Leleu},
  \citenamefont {Yamamoto}, \citenamefont {Utsunomiya},\ and\ \citenamefont
  {Aihara}}]{PhysRevE.95.022118}%
  \BibitemOpen
  \bibfield  {author} {\bibinfo {author} {\bibfnamefont {T.}~\bibnamefont
  {Leleu}}, \bibinfo {author} {\bibfnamefont {Y.}~\bibnamefont {Yamamoto}},
  \bibinfo {author} {\bibfnamefont {S.}~\bibnamefont {Utsunomiya}},\ and\
  \bibinfo {author} {\bibfnamefont {K.}~\bibnamefont {Aihara}},\ }\bibfield
  {title} {\bibinfo {title} {Combinatorial optimization using dynamical phase
  transitions in driven-dissipative systems},\ }\href
  {https://doi.org/10.1103/PhysRevE.95.022118} {\bibfield  {journal} {\bibinfo
  {journal} {Phys. Rev. E}\ }\textbf {\bibinfo {volume} {95}},\ \bibinfo
  {pages} {022118} (\bibinfo {year} {2017})}\BibitemShut {NoStop}%
\bibitem [{\citenamefont {Leleu}\ \emph {et~al.}(2019)\citenamefont {Leleu},
  \citenamefont {Yamamoto}, \citenamefont {McMahon},\ and\ \citenamefont
  {Aihara}}]{Leleu2019}%
  \BibitemOpen
  \bibfield  {author} {\bibinfo {author} {\bibfnamefont {T.}~\bibnamefont
  {Leleu}}, \bibinfo {author} {\bibfnamefont {Y.}~\bibnamefont {Yamamoto}},
  \bibinfo {author} {\bibfnamefont {P.~L.}\ \bibnamefont {McMahon}},\ and\
  \bibinfo {author} {\bibfnamefont {K.}~\bibnamefont {Aihara}},\ }\bibfield
  {title} {\bibinfo {title} {Destabilization of local minima in analog spin
  systems by correction of amplitude heterogeneity},\ }\href
  {https://doi.org/10.1103/PhysRevLett.122.040607} {\bibfield  {journal}
  {\bibinfo  {journal} {Phys. Rev. Lett.}\ }\textbf {\bibinfo {volume} {122}},\
  \bibinfo {pages} {040607} (\bibinfo {year} {2019})}\BibitemShut {NoStop}%
\bibitem [{\citenamefont {Kako}\ \emph {et~al.}(2020)\citenamefont {Kako},
  \citenamefont {Leleu}, \citenamefont {Inui}, \citenamefont {Khoyratee},
  \citenamefont {Reifenstein},\ and\ \citenamefont
  {Yamamoto}}]{https://doi.org/10.1002/qute.202000045}%
  \BibitemOpen
  \bibfield  {author} {\bibinfo {author} {\bibfnamefont {S.}~\bibnamefont
  {Kako}}, \bibinfo {author} {\bibfnamefont {T.}~\bibnamefont {Leleu}},
  \bibinfo {author} {\bibfnamefont {Y.}~\bibnamefont {Inui}}, \bibinfo {author}
  {\bibfnamefont {F.}~\bibnamefont {Khoyratee}}, \bibinfo {author}
  {\bibfnamefont {S.}~\bibnamefont {Reifenstein}},\ and\ \bibinfo {author}
  {\bibfnamefont {Y.}~\bibnamefont {Yamamoto}},\ }\bibfield  {title} {\bibinfo
  {title} {Coherent ising machines with error correction feedback},\ }\href
  {https://doi.org/https://doi.org/10.1002/qute.202000045} {\bibfield
  {journal} {\bibinfo  {journal} {Advanced Quantum Technologies}\ }\textbf
  {\bibinfo {volume} {3}},\ \bibinfo {pages} {2000045} (\bibinfo {year}
  {2020})}\BibitemShut {NoStop}%
\bibitem [{\citenamefont {Geman}\ and\ \citenamefont
  {Geman}(1984)}]{geman1984stochastic}%
  \BibitemOpen
  \bibfield  {author} {\bibinfo {author} {\bibfnamefont {S.}~\bibnamefont
  {Geman}}\ and\ \bibinfo {author} {\bibfnamefont {D.}~\bibnamefont {Geman}},\
  }\bibfield  {title} {\bibinfo {title} {Stochastic relaxation, gibbs
  distributions, and the bayesian restoration of images},\ }\href@noop {}
  {\bibfield  {journal} {\bibinfo  {journal} {IEEE Transactions on pattern
  analysis and machine intelligence}\ ,\ \bibinfo {pages} {721}} (\bibinfo
  {year} {1984})}\BibitemShut {NoStop}%
\bibitem [{\citenamefont {Calvanese~Strinati}\ \emph
  {et~al.}(2021)\citenamefont {Calvanese~Strinati}, \citenamefont {Bello},
  \citenamefont {Dalla~Torre},\ and\ \citenamefont
  {Pe'er}}]{PhysRevLett.126.143901}%
  \BibitemOpen
  \bibfield  {author} {\bibinfo {author} {\bibfnamefont {M.}~\bibnamefont
  {Calvanese~Strinati}}, \bibinfo {author} {\bibfnamefont {L.}~\bibnamefont
  {Bello}}, \bibinfo {author} {\bibfnamefont {E.~G.}\ \bibnamefont
  {Dalla~Torre}},\ and\ \bibinfo {author} {\bibfnamefont {A.}~\bibnamefont
  {Pe'er}},\ }\bibfield  {title} {\bibinfo {title} {Can nonlinear parametric
  oscillators solve random ising models?},\ }\href
  {https://doi.org/10.1103/PhysRevLett.126.143901} {\bibfield  {journal}
  {\bibinfo  {journal} {Phys. Rev. Lett.}\ }\textbf {\bibinfo {volume} {126}},\
  \bibinfo {pages} {143901} (\bibinfo {year} {2021})}\BibitemShut {NoStop}%
\bibitem [{\citenamefont {Miyazaki}\ and\ \citenamefont
  {Ohzeki}(2018)}]{PhysRevA.98.053839}%
  \BibitemOpen
  \bibfield  {author} {\bibinfo {author} {\bibfnamefont {R.}~\bibnamefont
  {Miyazaki}}\ and\ \bibinfo {author} {\bibfnamefont {M.}~\bibnamefont
  {Ohzeki}},\ }\bibfield  {title} {\bibinfo {title} {Distributions of steady
  states in a network of degenerate optical parametric oscillators in solving
  combinatorial optimization problems},\ }\href
  {https://doi.org/10.1103/PhysRevA.98.053839} {\bibfield  {journal} {\bibinfo
  {journal} {Phys. Rev. A}\ }\textbf {\bibinfo {volume} {98}},\ \bibinfo
  {pages} {053839} (\bibinfo {year} {2018})}\BibitemShut {NoStop}%
\bibitem [{\citenamefont {Ito}\ \emph {et~al.}(2018)\citenamefont {Ito},
  \citenamefont {Ueta},\ and\ \citenamefont {Aihara}}]{ito2018bifurcation}%
  \BibitemOpen
  \bibfield  {author} {\bibinfo {author} {\bibfnamefont {D.}~\bibnamefont
  {Ito}}, \bibinfo {author} {\bibfnamefont {T.}~\bibnamefont {Ueta}},\ and\
  \bibinfo {author} {\bibfnamefont {K.}~\bibnamefont {Aihara}},\ }\bibfield
  {title} {\bibinfo {title} {Bifurcation analysis of eight coupled degenerate
  optical parametric oscillators},\ }\href@noop {} {\bibfield  {journal}
  {\bibinfo  {journal} {Physica D: Nonlinear Phenomena}\ }\textbf {\bibinfo
  {volume} {372}},\ \bibinfo {pages} {22} (\bibinfo {year} {2018})}\BibitemShut
  {NoStop}%
\bibitem [{\citenamefont {Shishkova}(1973)}]{shishkova1973examination}%
  \BibitemOpen
  \bibfield  {author} {\bibinfo {author} {\bibfnamefont {M.~A.}\ \bibnamefont
  {Shishkova}},\ }\bibfield  {title} {\bibinfo {title} {Examination of a system
  of differential equations with a small parameter in the highest
  derivatives},\ }in\ \href@noop {} {\emph {\bibinfo {booktitle} {Doklady
  Akademii Nauk}}},\ Vol.\ \bibinfo {volume} {209}\ (\bibinfo {organization}
  {Russian Academy of Sciences},\ \bibinfo {year} {1973})\ pp.\ \bibinfo
  {pages} {576--579}\BibitemShut {NoStop}%
\bibitem [{\citenamefont {Mandel}\ and\ \citenamefont
  {Erneux}(1984)}]{PhysRevLett.53.1818}%
  \BibitemOpen
  \bibfield  {author} {\bibinfo {author} {\bibfnamefont {P.}~\bibnamefont
  {Mandel}}\ and\ \bibinfo {author} {\bibfnamefont {T.}~\bibnamefont
  {Erneux}},\ }\bibfield  {title} {\bibinfo {title} {Laser lorenz equations
  with a time-dependent parameter},\ }\href
  {https://doi.org/10.1103/PhysRevLett.53.1818} {\bibfield  {journal} {\bibinfo
   {journal} {Phys. Rev. Lett.}\ }\textbf {\bibinfo {volume} {53}},\ \bibinfo
  {pages} {1818} (\bibinfo {year} {1984})}\BibitemShut {NoStop}%
\bibitem [{\citenamefont {Berglund}\ and\ \citenamefont
  {Kunz}(1997)}]{PhysRevLett.78.1691}%
  \BibitemOpen
  \bibfield  {author} {\bibinfo {author} {\bibfnamefont {N.}~\bibnamefont
  {Berglund}}\ and\ \bibinfo {author} {\bibfnamefont {H.}~\bibnamefont
  {Kunz}},\ }\bibfield  {title} {\bibinfo {title} {Chaotic hysteresis in an
  adiabatically oscillating double well},\ }\href
  {https://doi.org/10.1103/PhysRevLett.78.1691} {\bibfield  {journal} {\bibinfo
   {journal} {Phys. Rev. Lett.}\ }\textbf {\bibinfo {volume} {78}},\ \bibinfo
  {pages} {1691} (\bibinfo {year} {1997})}\BibitemShut {NoStop}%
\bibitem [{\citenamefont {Berglund}(2000)}]{10.1143/PTPS.139.325}%
  \BibitemOpen
  \bibfield  {author} {\bibinfo {author} {\bibfnamefont {N.}~\bibnamefont
  {Berglund}},\ }\bibfield  {title} {\bibinfo {title} {{Dynamic Bifurcations:
  Hysteresis, Scaling Laws and Feedback Control}},\ }\href
  {https://doi.org/10.1143/PTPS.139.325} {\bibfield  {journal} {\bibinfo
  {journal} {Progress of Theoretical Physics Supplement}\ }\textbf {\bibinfo
  {volume} {139}},\ \bibinfo {pages} {325} (\bibinfo {year}
  {2000})}\BibitemShut {NoStop}%
\bibitem [{\citenamefont {Bertsekas}(1997)}]{bertsekas1997nonlinear}%
  \BibitemOpen
  \bibfield  {author} {\bibinfo {author} {\bibfnamefont {D.~P.}\ \bibnamefont
  {Bertsekas}},\ }\bibfield  {title} {\bibinfo {title} {Nonlinear
  programming},\ }\href@noop {} {\bibfield  {journal} {\bibinfo  {journal}
  {Journal of the Operational Research Society}\ }\textbf {\bibinfo {volume}
  {48}},\ \bibinfo {pages} {334} (\bibinfo {year} {1997})}\BibitemShut
  {NoStop}%
\bibitem [{\citenamefont {Boyd}\ \emph {et~al.}(2004)\citenamefont {Boyd},
  \citenamefont {Boyd},\ and\ \citenamefont {Vandenberghe}}]{boyd2004convex}%
  \BibitemOpen
  \bibfield  {author} {\bibinfo {author} {\bibfnamefont {S.}~\bibnamefont
  {Boyd}}, \bibinfo {author} {\bibfnamefont {S.~P.}\ \bibnamefont {Boyd}},\
  and\ \bibinfo {author} {\bibfnamefont {L.}~\bibnamefont {Vandenberghe}},\
  }\href@noop {} {\emph {\bibinfo {title} {Convex optimization}}}\ (\bibinfo
  {publisher} {Cambridge university press},\ \bibinfo {year}
  {2004})\BibitemShut {NoStop}%
\bibitem [{\citenamefont {Vadlamani}\ \emph {et~al.}(2020)\citenamefont
  {Vadlamani}, \citenamefont {Xiao},\ and\ \citenamefont
  {Yablonovitch}}]{doi:10.1073/pnas.2015192117}%
  \BibitemOpen
  \bibfield  {author} {\bibinfo {author} {\bibfnamefont {S.~K.}\ \bibnamefont
  {Vadlamani}}, \bibinfo {author} {\bibfnamefont {T.~P.}\ \bibnamefont
  {Xiao}},\ and\ \bibinfo {author} {\bibfnamefont {E.}~\bibnamefont
  {Yablonovitch}},\ }\bibfield  {title} {\bibinfo {title} {Physics successfully
  implements lagrange multiplier optimization},\ }\href
  {https://doi.org/10.1073/pnas.2015192117} {\bibfield  {journal} {\bibinfo
  {journal} {Proceedings of the National Academy of Sciences}\ }\textbf
  {\bibinfo {volume} {117}},\ \bibinfo {pages} {26639} (\bibinfo {year}
  {2020})}\BibitemShut {NoStop}%
\bibitem [{\citenamefont {Vadlamani}\ \emph {et~al.}(2022)\citenamefont
  {Vadlamani}, \citenamefont {Xiao},\ and\ \citenamefont
  {Yablonovitch}}]{vadlamani2022equivalence}%
  \BibitemOpen
  \bibfield  {author} {\bibinfo {author} {\bibfnamefont {S.~K.}\ \bibnamefont
  {Vadlamani}}, \bibinfo {author} {\bibfnamefont {T.~P.}\ \bibnamefont
  {Xiao}},\ and\ \bibinfo {author} {\bibfnamefont {E.}~\bibnamefont
  {Yablonovitch}},\ }\bibfield  {title} {\bibinfo {title} {Equivalence of
  coupled parametric oscillator dynamics to lagrange multiplier primal-dual
  optimization},\ }\href@noop {} {\bibfield  {journal} {\bibinfo  {journal}
  {arXiv preprint arXiv:2204.02472}\ } (\bibinfo {year} {2022})}\BibitemShut
  {NoStop}%
\bibitem [{\citenamefont {Hamze}\ \emph {et~al.}(2020)\citenamefont {Hamze},
  \citenamefont {Raymond}, \citenamefont {Pattison}, \citenamefont {Biswas},\
  and\ \citenamefont {Katzgraber}}]{PhysRevE.101.052102}%
  \BibitemOpen
  \bibfield  {author} {\bibinfo {author} {\bibfnamefont {F.}~\bibnamefont
  {Hamze}}, \bibinfo {author} {\bibfnamefont {J.}~\bibnamefont {Raymond}},
  \bibinfo {author} {\bibfnamefont {C.~A.}\ \bibnamefont {Pattison}}, \bibinfo
  {author} {\bibfnamefont {K.}~\bibnamefont {Biswas}},\ and\ \bibinfo {author}
  {\bibfnamefont {H.~G.}\ \bibnamefont {Katzgraber}},\ }\bibfield  {title}
  {\bibinfo {title} {Wishart planted ensemble: A tunably rugged pairwise ising
  model with a first-order phase transition},\ }\href
  {https://doi.org/10.1103/PhysRevE.101.052102} {\bibfield  {journal} {\bibinfo
   {journal} {Phys. Rev. E}\ }\textbf {\bibinfo {volume} {101}},\ \bibinfo
  {pages} {052102} (\bibinfo {year} {2020})}\BibitemShut {NoStop}%
\end{thebibliography}%

\clearpage
\appendix

\begin{widetext}
\section{The energy scale dependence on the dynamics and the success probability}\label{various_beta}
\setcounter{figure}{0}
\renewcommand{\thefigure}{\Alph{section}.\arabic{figure}}

In the main text, we focus on the computational property only for fixed energy scale $\beta=0.3$.
However, the $\beta$ dependence on the dynamics needs to be considered, in particular for practical situations.
Here we report the bifurcation dynamics in $\beta=0.03, 0.1, 0.3$, and $1$ for $\alpha=0.05, 0.55$, and $0.95$ as examples.
These results are shown in Fig.~\ref{fig:various_beta}.
For large $\beta$, the bifurcation phenomenon does not completely occur at $t=500$, which is employed for the final time $T$ of the optimizations in the main text.
These situations are out of the scope of our analyses in this paper.

\begin{figure}[h]
  \begin{tabular}{lll}
      \hspace{-45pt}
      \begin{minipage}[t]{0.45\hsize}
        \centering
        \includegraphics[keepaspectratio, scale=0.4]{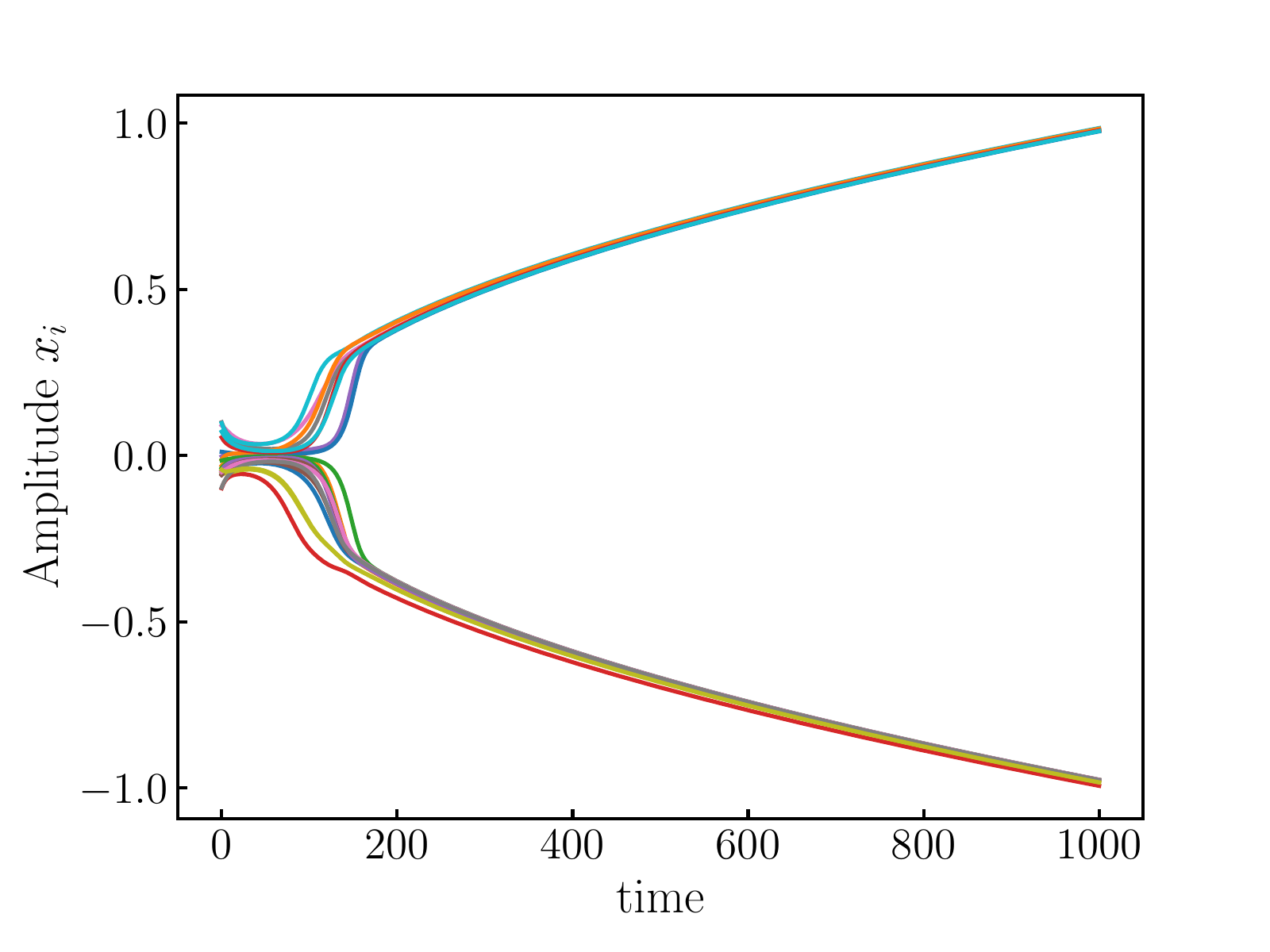}
      \end{minipage} &
      \hspace{-60pt}
      \begin{minipage}[t]{0.45\hsize}
        \centering
        \includegraphics[keepaspectratio, scale=0.4]{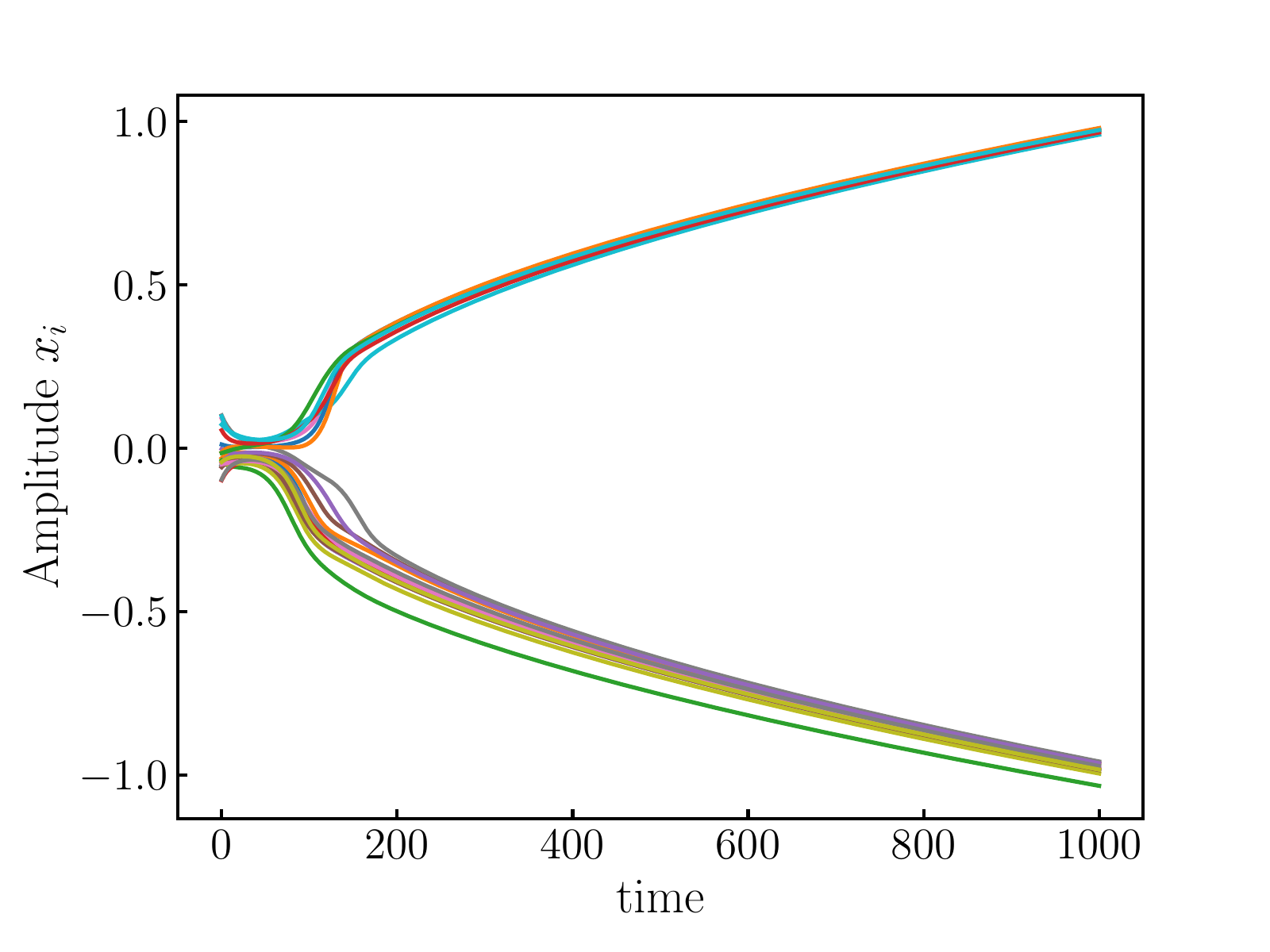}
      \end{minipage} &
      \hspace{-60pt}
      \begin{minipage}[t]{0.45\hsize}
        \centering
        \includegraphics[keepaspectratio, scale=0.4]{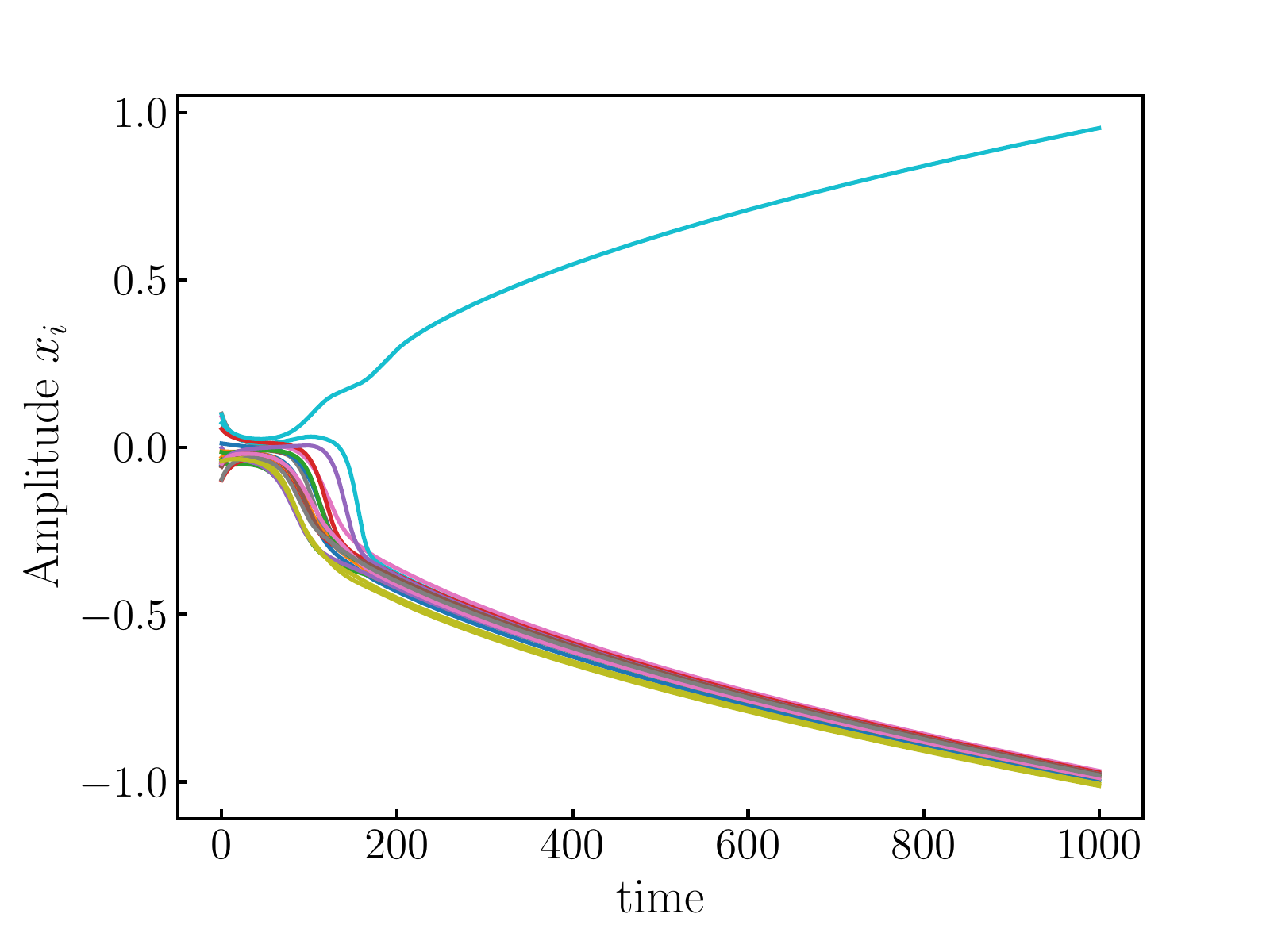}
      \end{minipage} \\    
    \hspace{-45pt}
    \begin{minipage}[t]{0.45\hsize}
      \centering
      \includegraphics[keepaspectratio, scale=0.4]{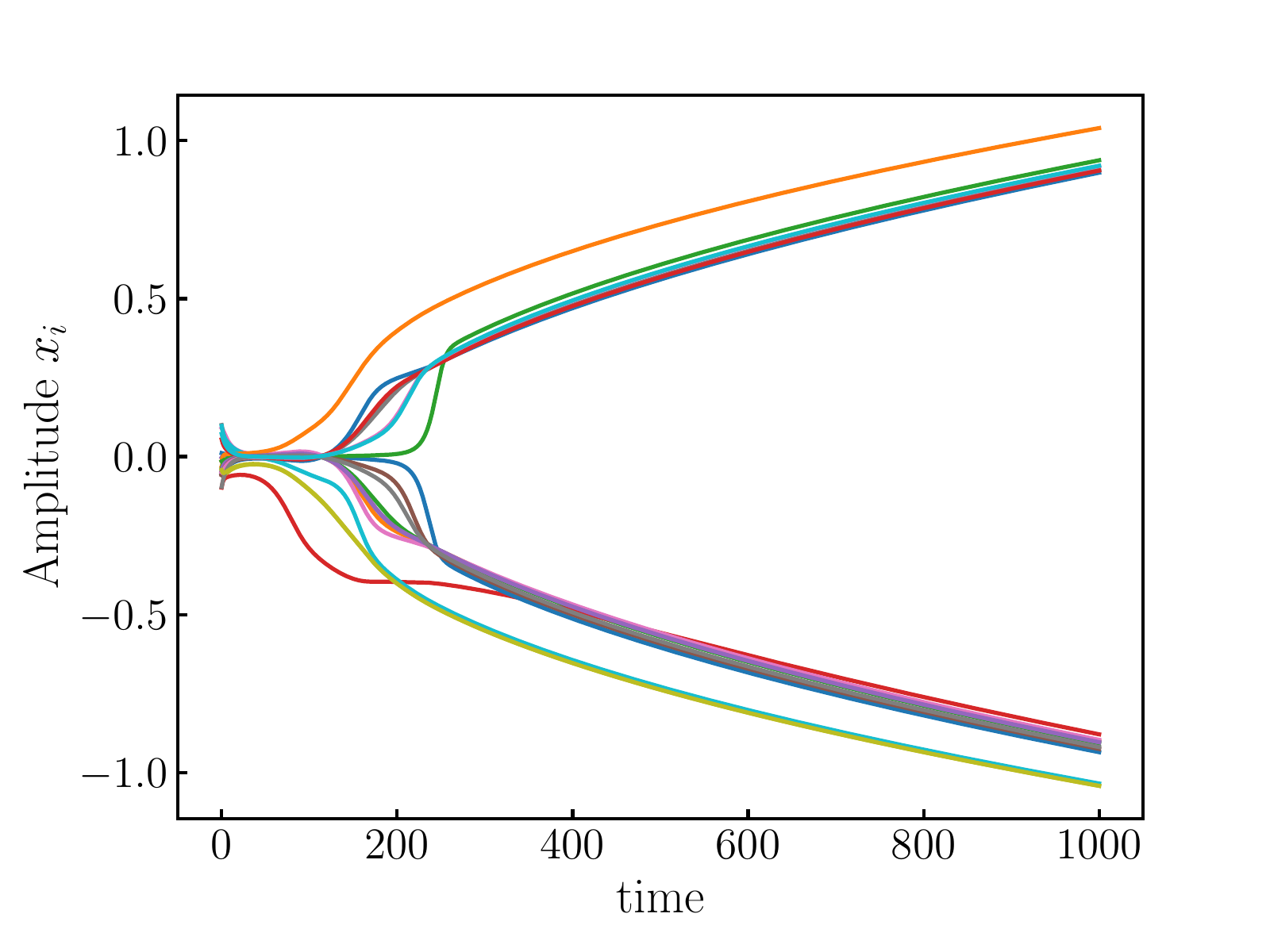}
    \end{minipage} &
    \hspace{-60pt}
    \begin{minipage}[t]{0.45\hsize}
      \centering
      \includegraphics[keepaspectratio, scale=0.4]{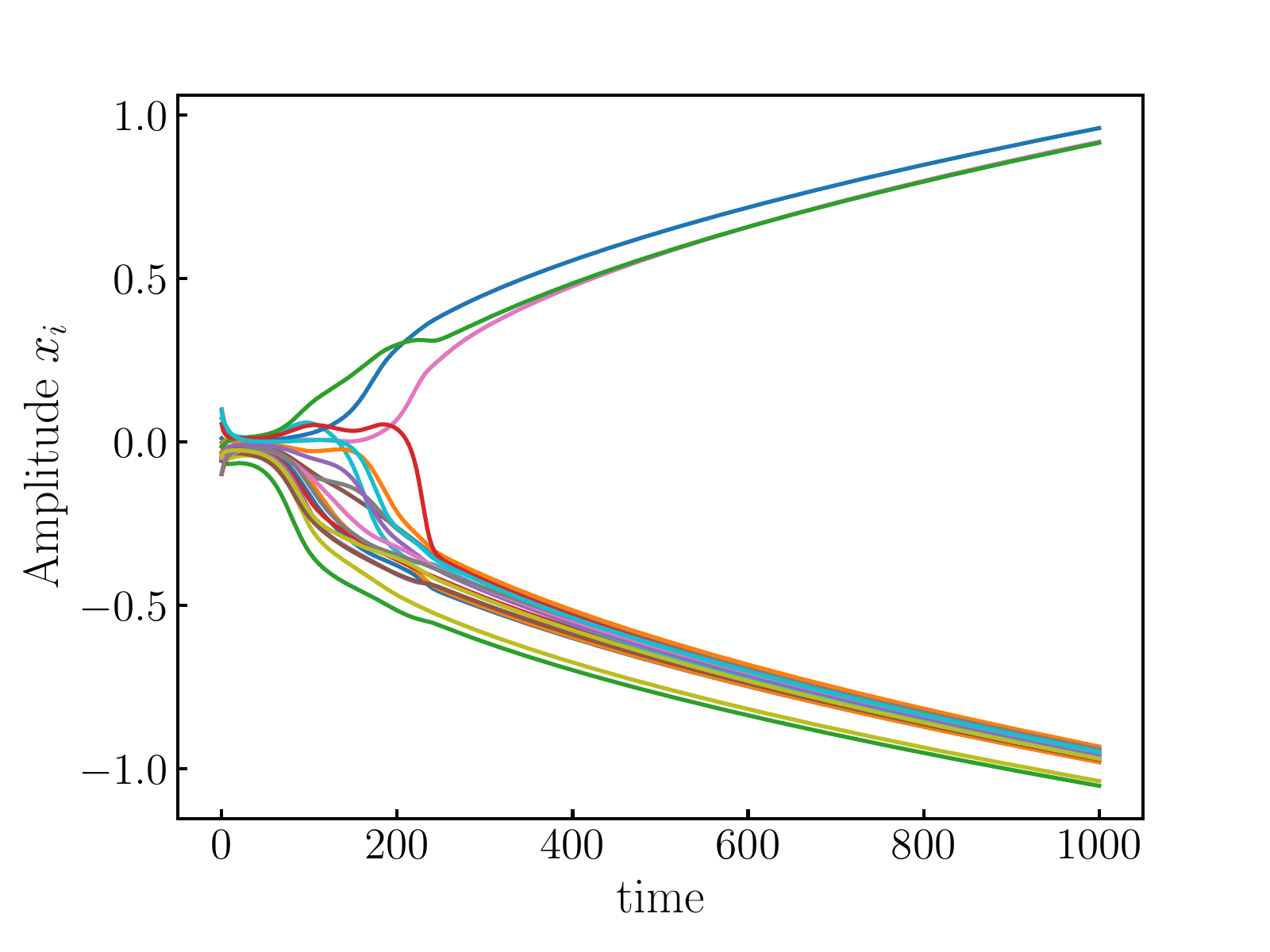}
    \end{minipage} &
    \hspace{-60pt}
    \begin{minipage}[t]{0.45\hsize}
      \centering
      \includegraphics[keepaspectratio, scale=0.4]{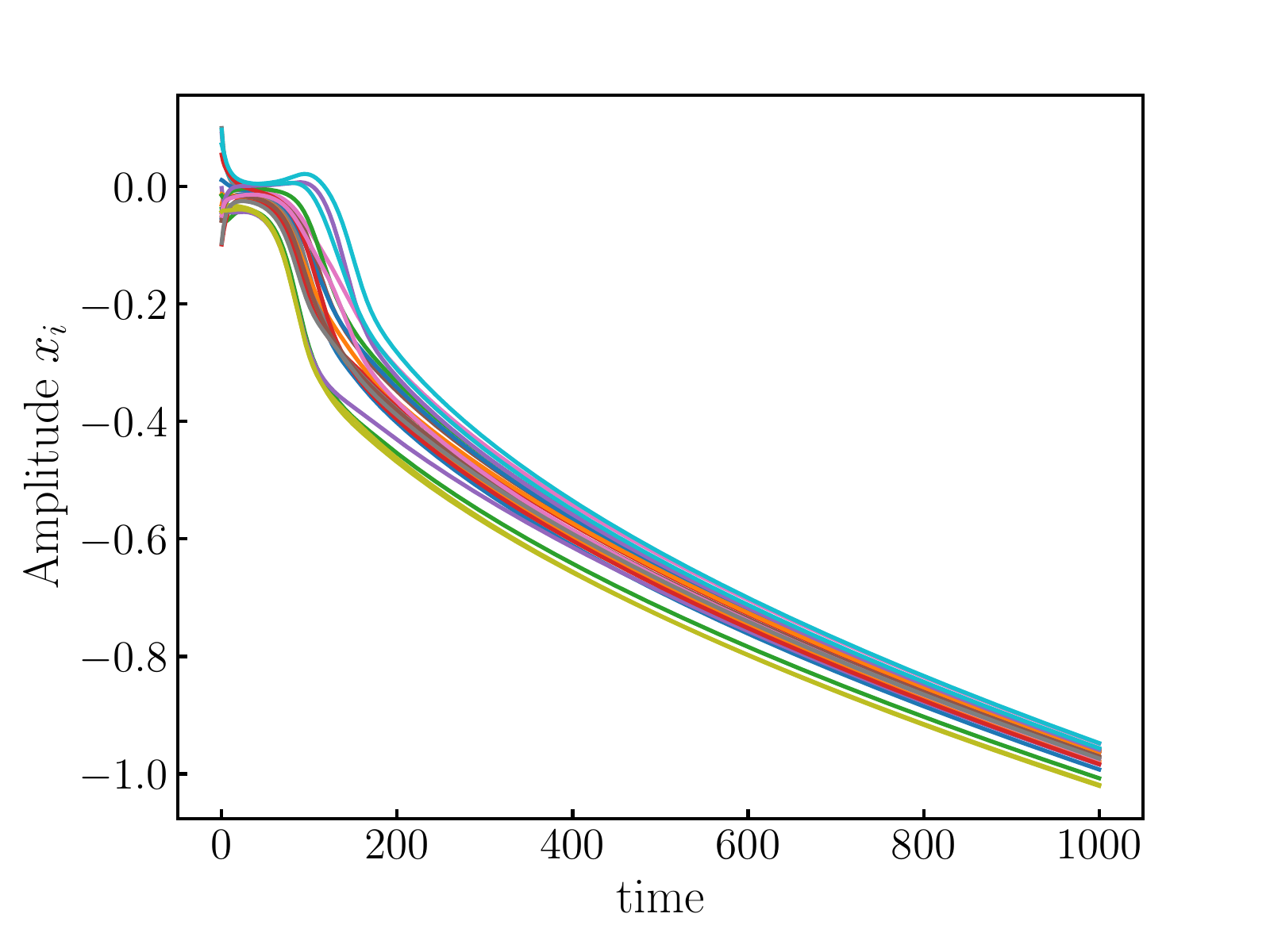}
    \end{minipage} \\

    \hspace{-45pt}
    \begin{minipage}[t]{0.45\hsize}
      \centering
      \includegraphics[keepaspectratio, scale=0.4]{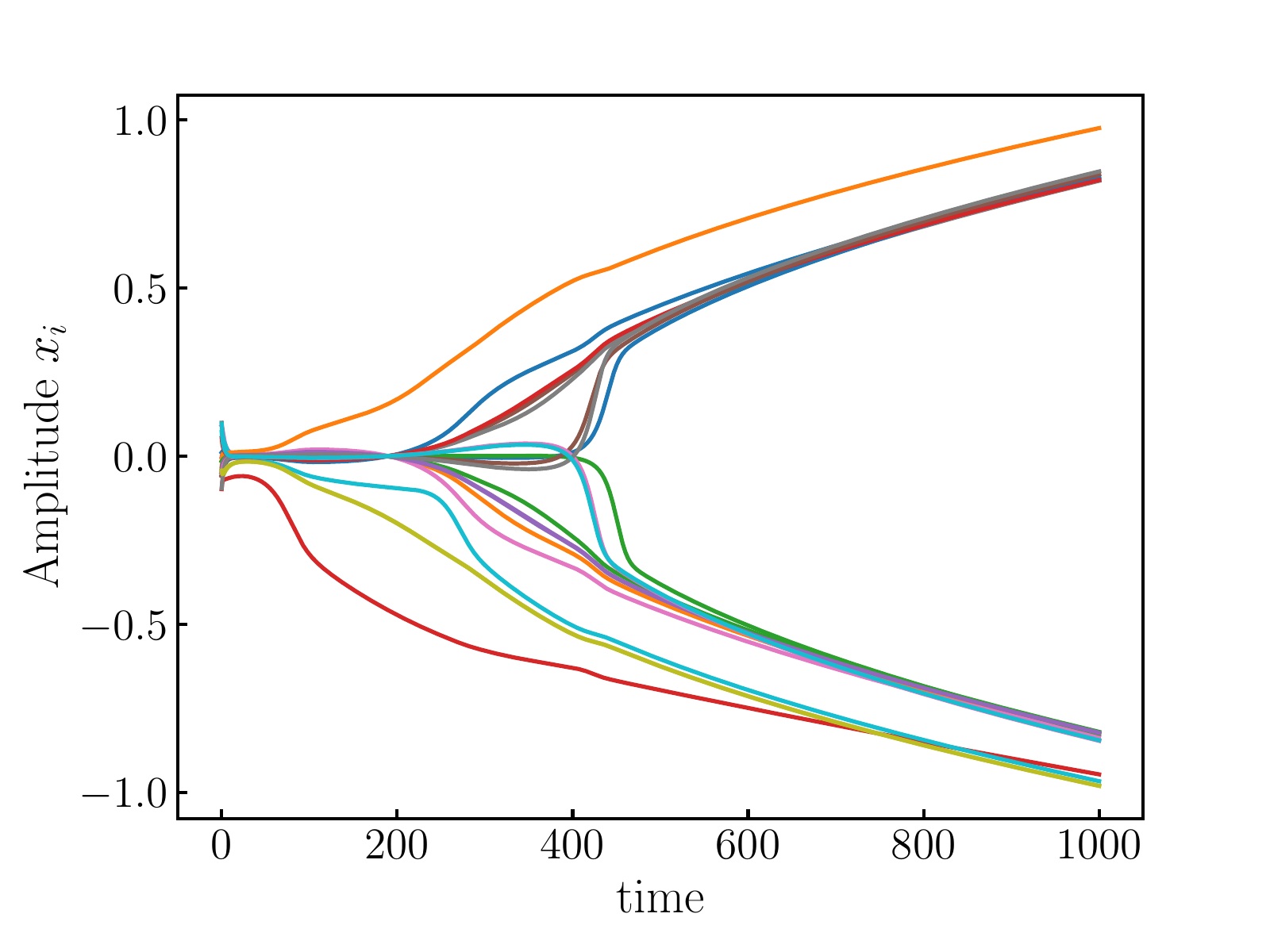}
    \end{minipage} &
    \hspace{-60pt}
    \begin{minipage}[t]{0.45\hsize}
      \centering
      \includegraphics[keepaspectratio, scale=0.4]{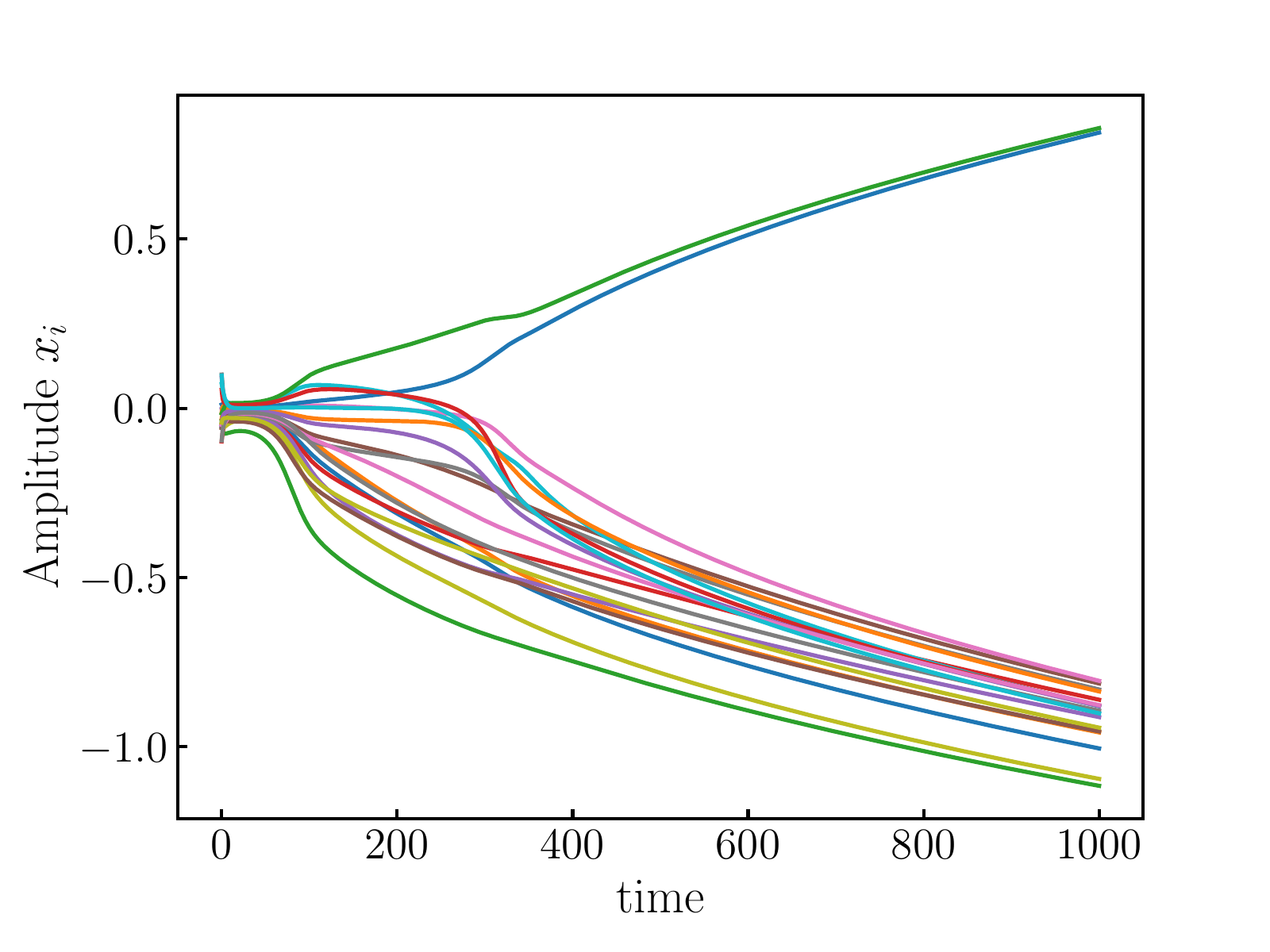}
    \end{minipage} &
    \hspace{-60pt}
    \begin{minipage}[t]{0.45\hsize}
      \centering
      \includegraphics[keepaspectratio, scale=0.4]{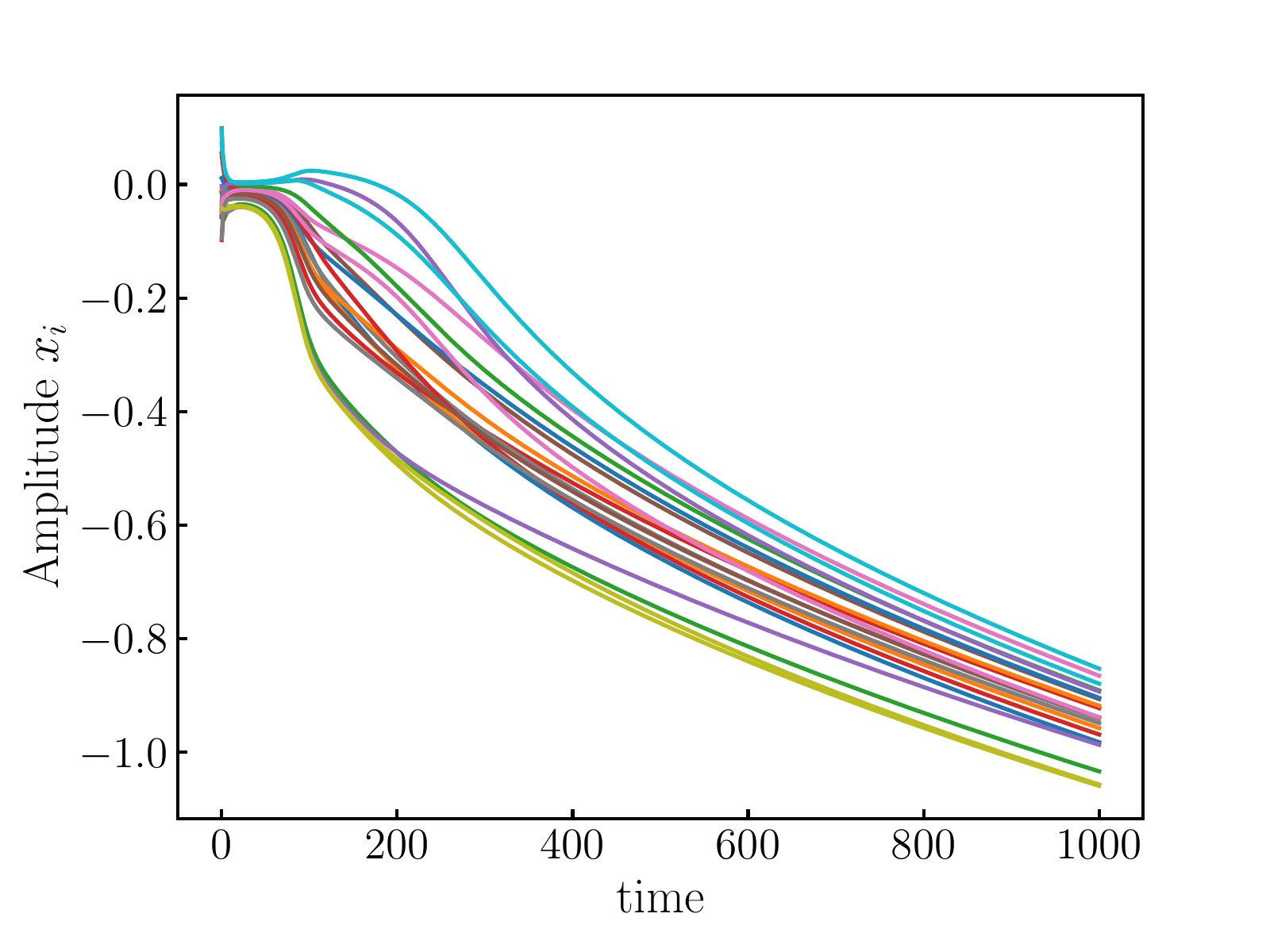}
    \end{minipage} \\
 
    \hspace{-45pt}
    \begin{minipage}[t]{0.45\hsize}
      \centering
      \includegraphics[keepaspectratio, scale=0.4]{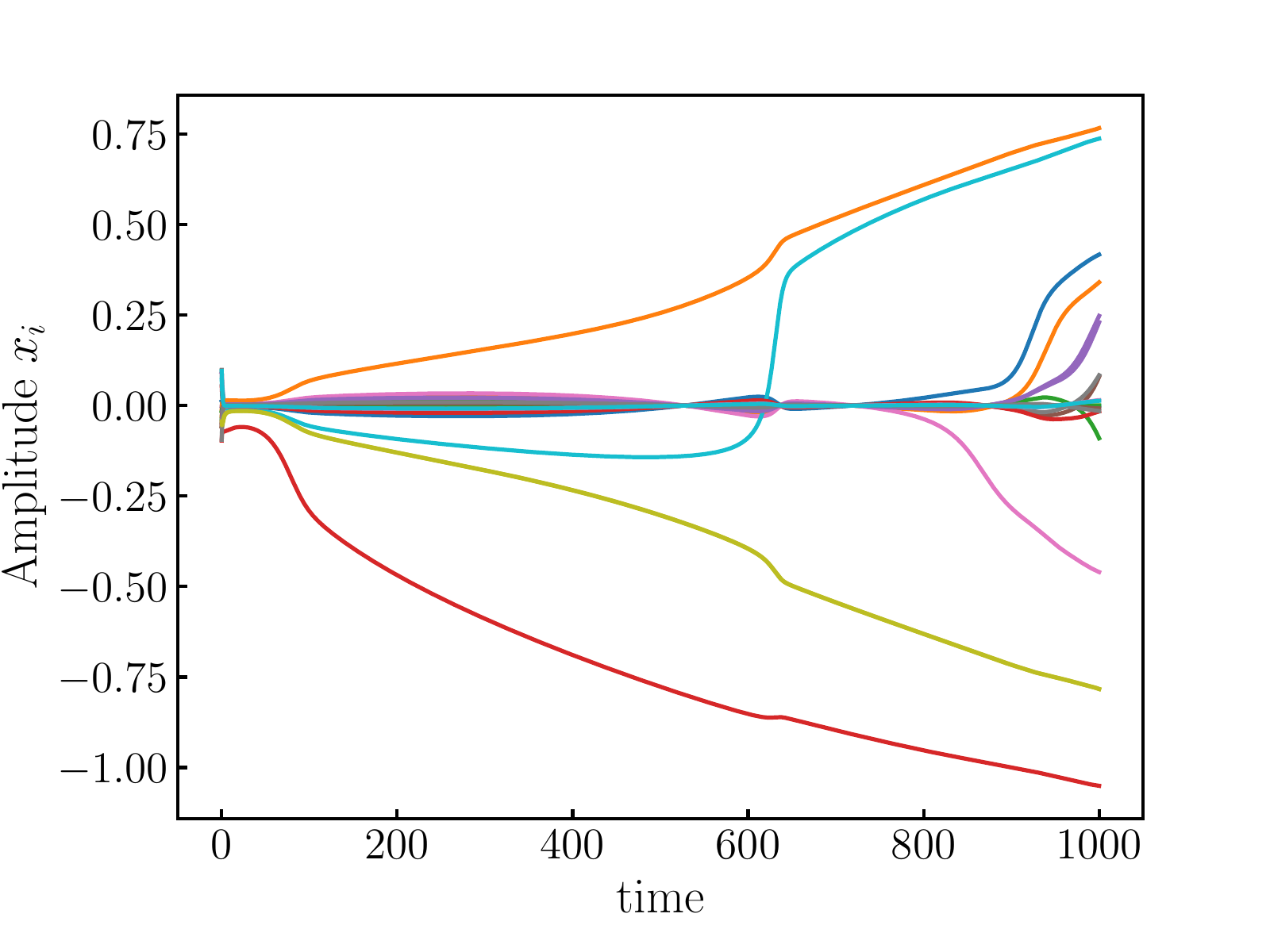}
    \end{minipage} &
    \hspace{-60pt}
    \begin{minipage}[t]{0.45\hsize}
      \centering
      \includegraphics[keepaspectratio, scale=0.4]{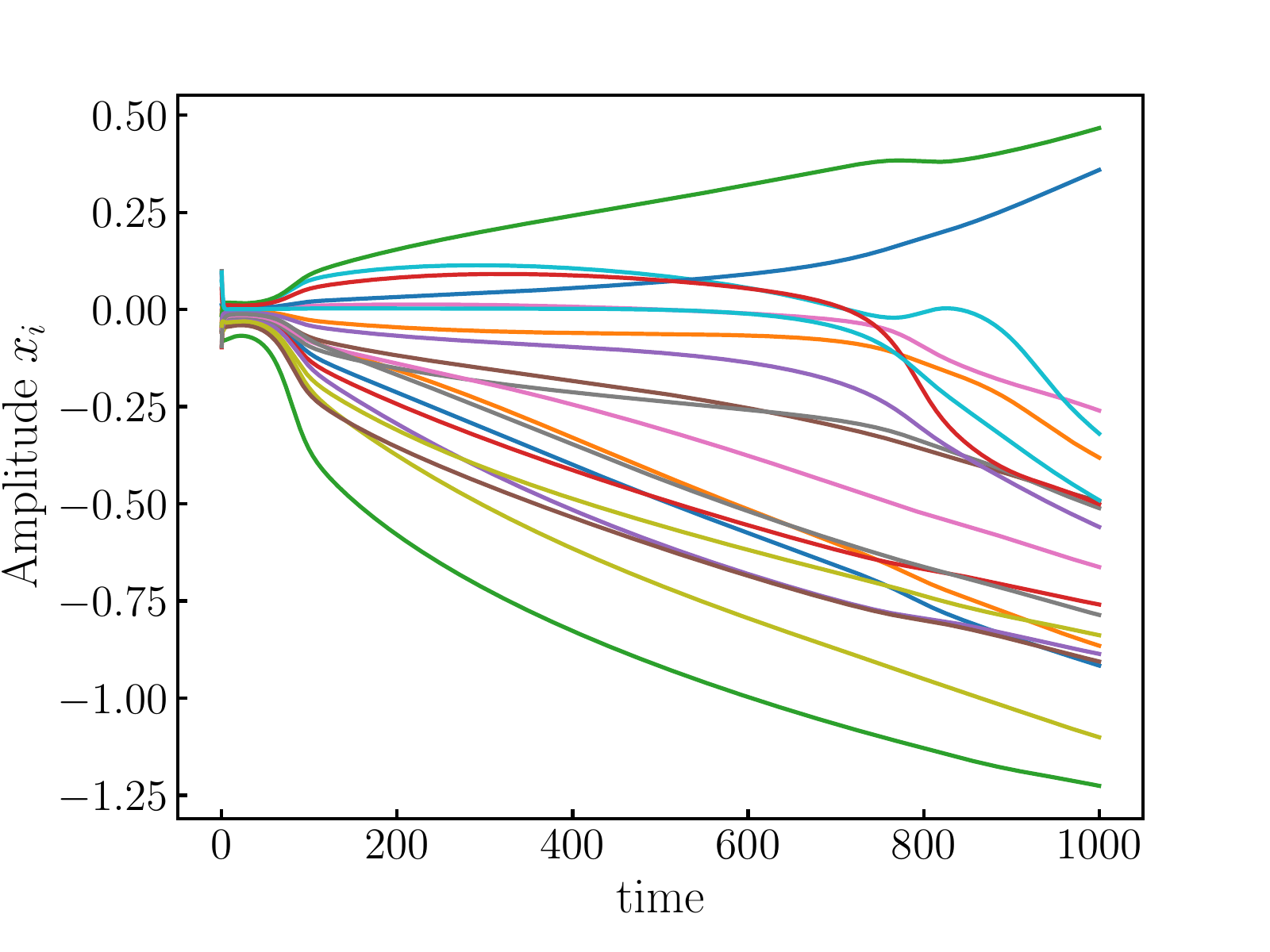}
    \end{minipage} &
    \hspace{-60pt}
    \begin{minipage}[t]{0.45\hsize}
      \centering
      \includegraphics[keepaspectratio, scale=0.4]{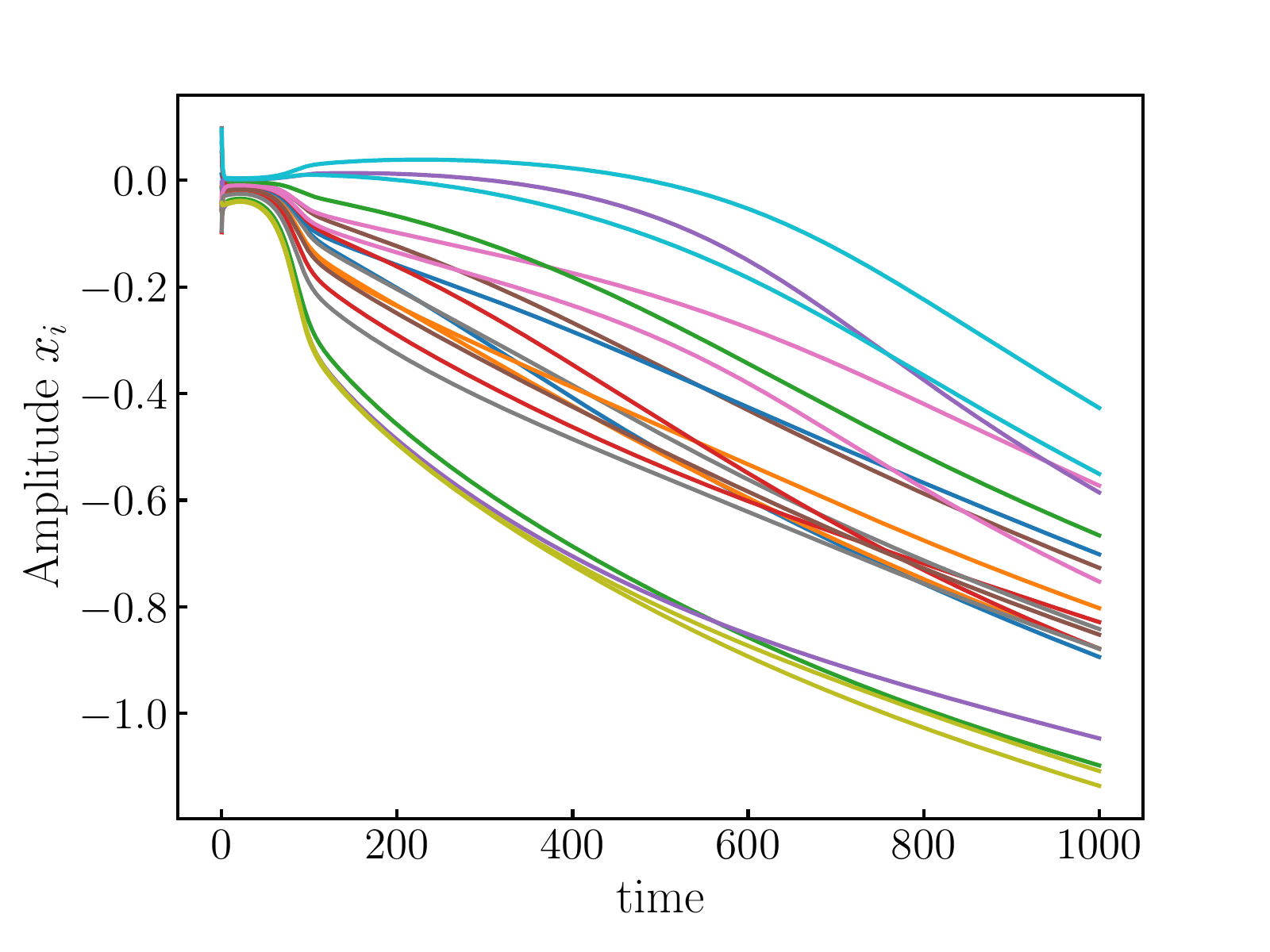}
    \end{minipage} 
  \end{tabular}
  \caption{Time evolution of the amplitudes $x_i(t)$ until $t=1000$ in $\beta=0.03, 0.1, 0.3$, and $1$ from the top in each row, for $\alpha=0.05, 0.55$, and $0.95$ from left to right in each column.}
  \label{fig:various_beta}
\end{figure}

Fig.~\ref{fig:various_beta_optimizations} also reports the number of stable fixed points, success probability of the optimizations in our model, and comparisons of success probability between our model and mean field CIM without noises
for $\beta=0.1$ and $0.03$ in each row.
In $\beta\to0$ limit, $x_i$ does not interact with each other.
It is expected that the differences between $\alpha$ will vanish and the results of optimizations will be close to a random sampling result.
In fact, the success probability in all $\alpha$ decreases for smaller $\beta$.
We note that the comparisons of $\mathcal{N}_{\text{SP}}$ between different $\beta$ do not give a precise relation for the success probability.
This is because the typical time scale of relaxations to local minima varies in different $\beta$. 
It can be seen that our observations for the relation between $\mathcal{N}_{\text{SP}}$ and the success probability are actually established only in each $\beta$.

\begin{figure}[h]
  \begin{tabular}{lll}
      \hspace{-45pt}
      \begin{minipage}[t]{0.45\hsize}
        \centering
        \includegraphics[keepaspectratio, scale=0.4]{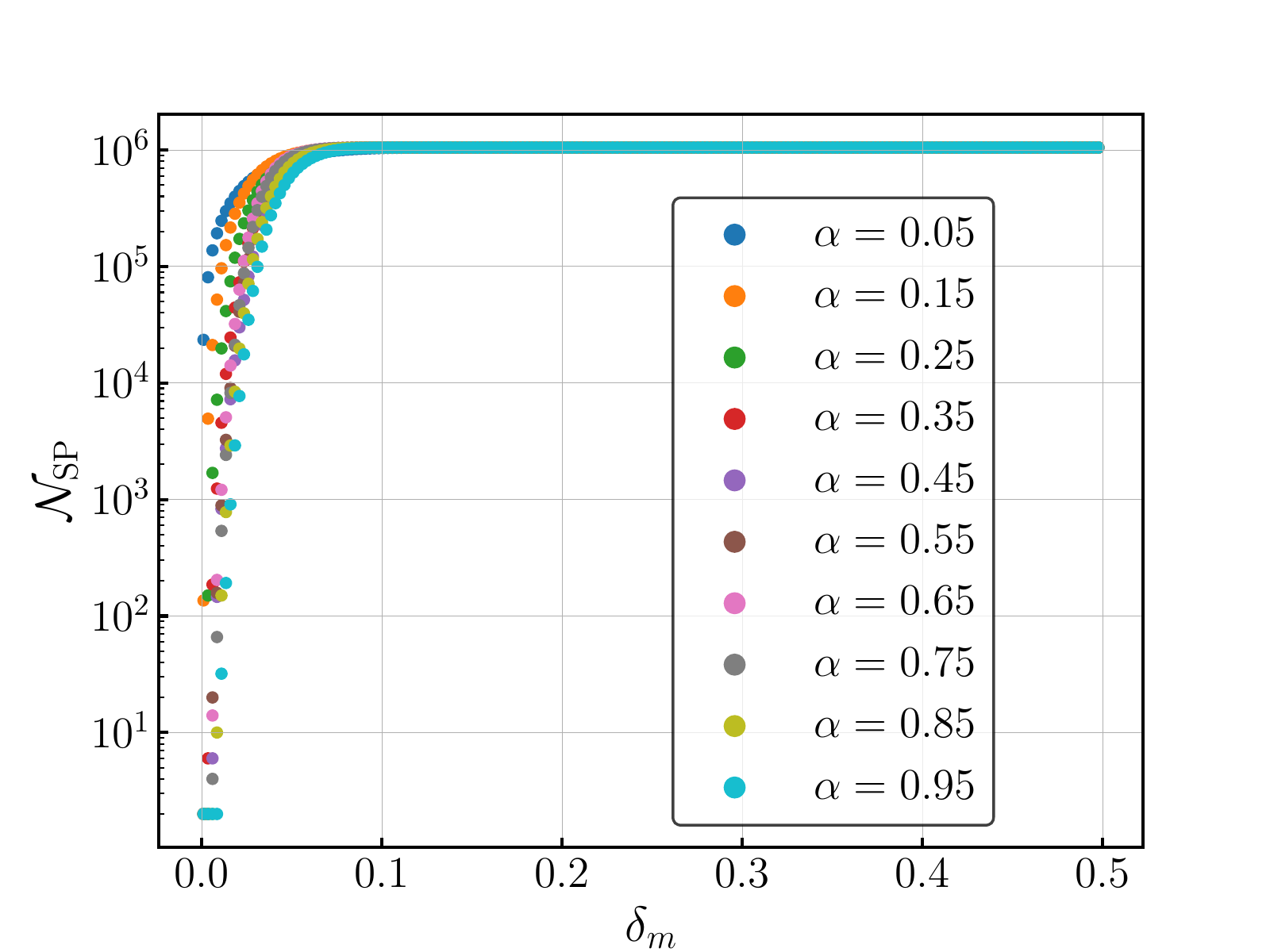}
      \end{minipage} &
      \hspace{-60pt}
      \begin{minipage}[t]{0.45\hsize}
        \centering
        \includegraphics[keepaspectratio, scale=0.4]{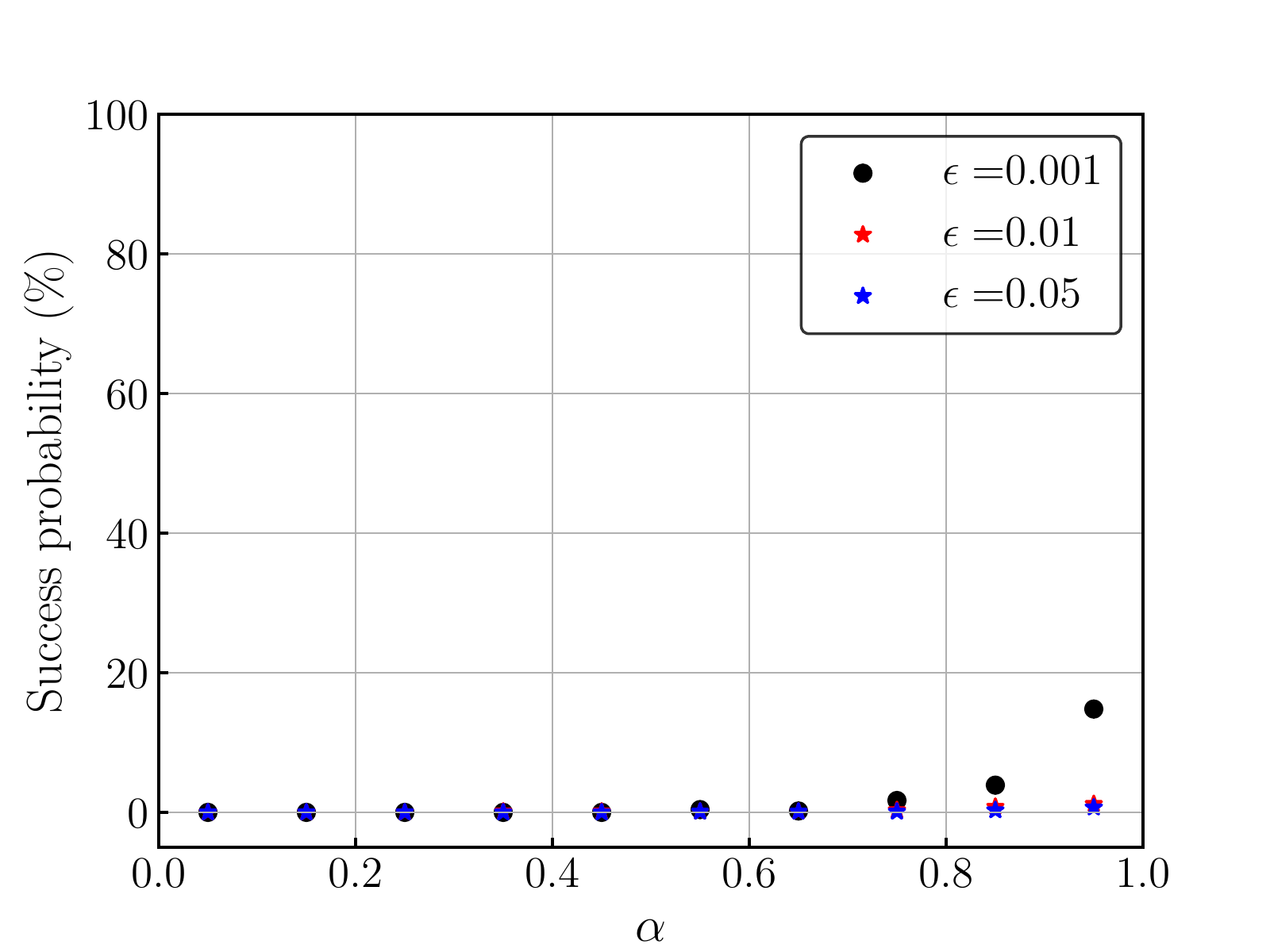}
      \end{minipage} &
      \hspace{-60pt}
      \begin{minipage}[t]{0.45\hsize}
        \centering
        \includegraphics[keepaspectratio, scale=0.4]{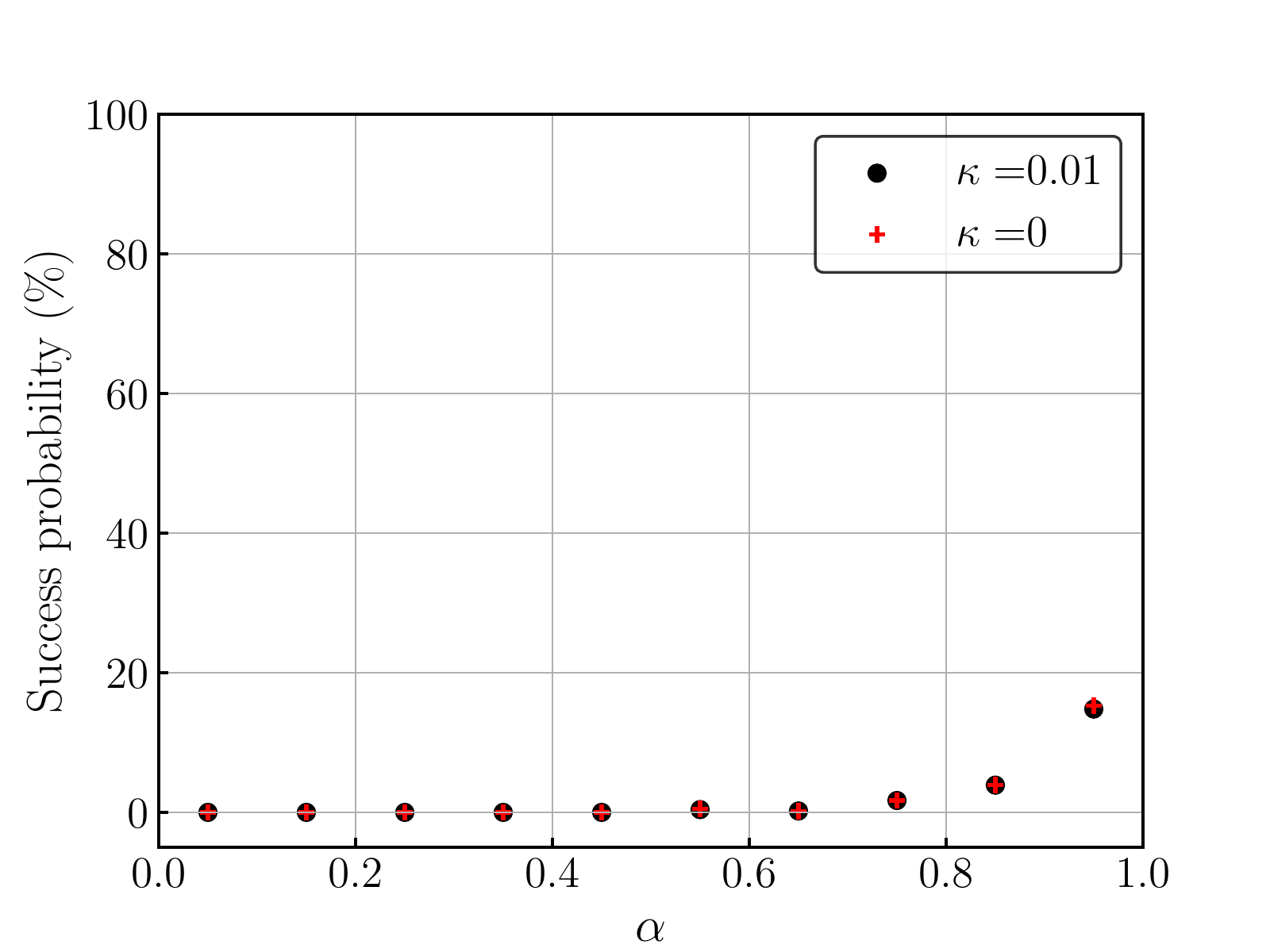}
      \end{minipage} \\    
    \hspace{-45pt}
    \begin{minipage}[t]{0.45\hsize}
      \centering
      \includegraphics[keepaspectratio, scale=0.4]{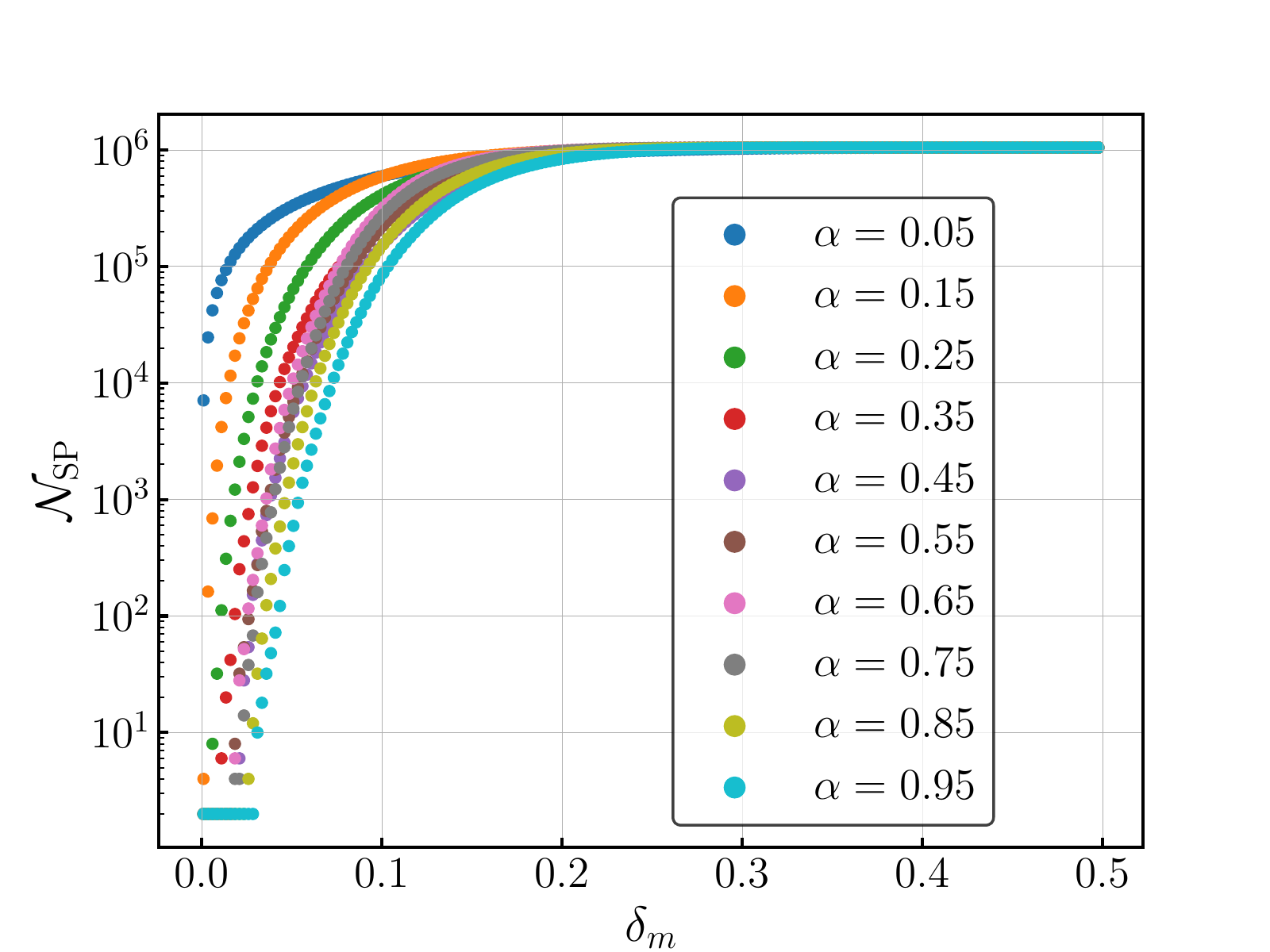}
    \end{minipage} &
    \hspace{-60pt}
    \begin{minipage}[t]{0.45\hsize}
      \centering
      \includegraphics[keepaspectratio, scale=0.4]{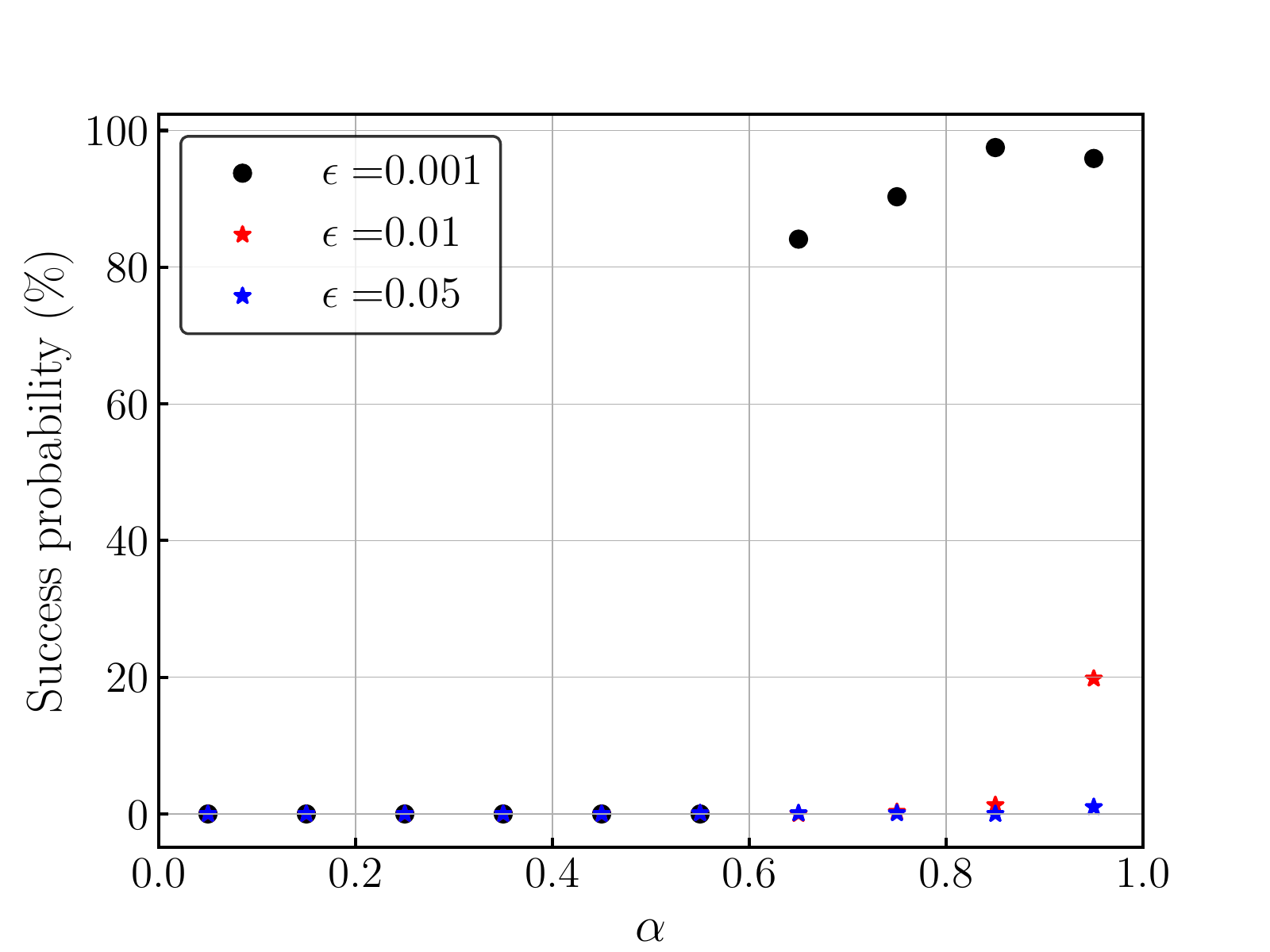}
    \end{minipage} &
    \hspace{-60pt}
    \begin{minipage}[t]{0.45\hsize}
      \centering
      \includegraphics[keepaspectratio, scale=0.4]{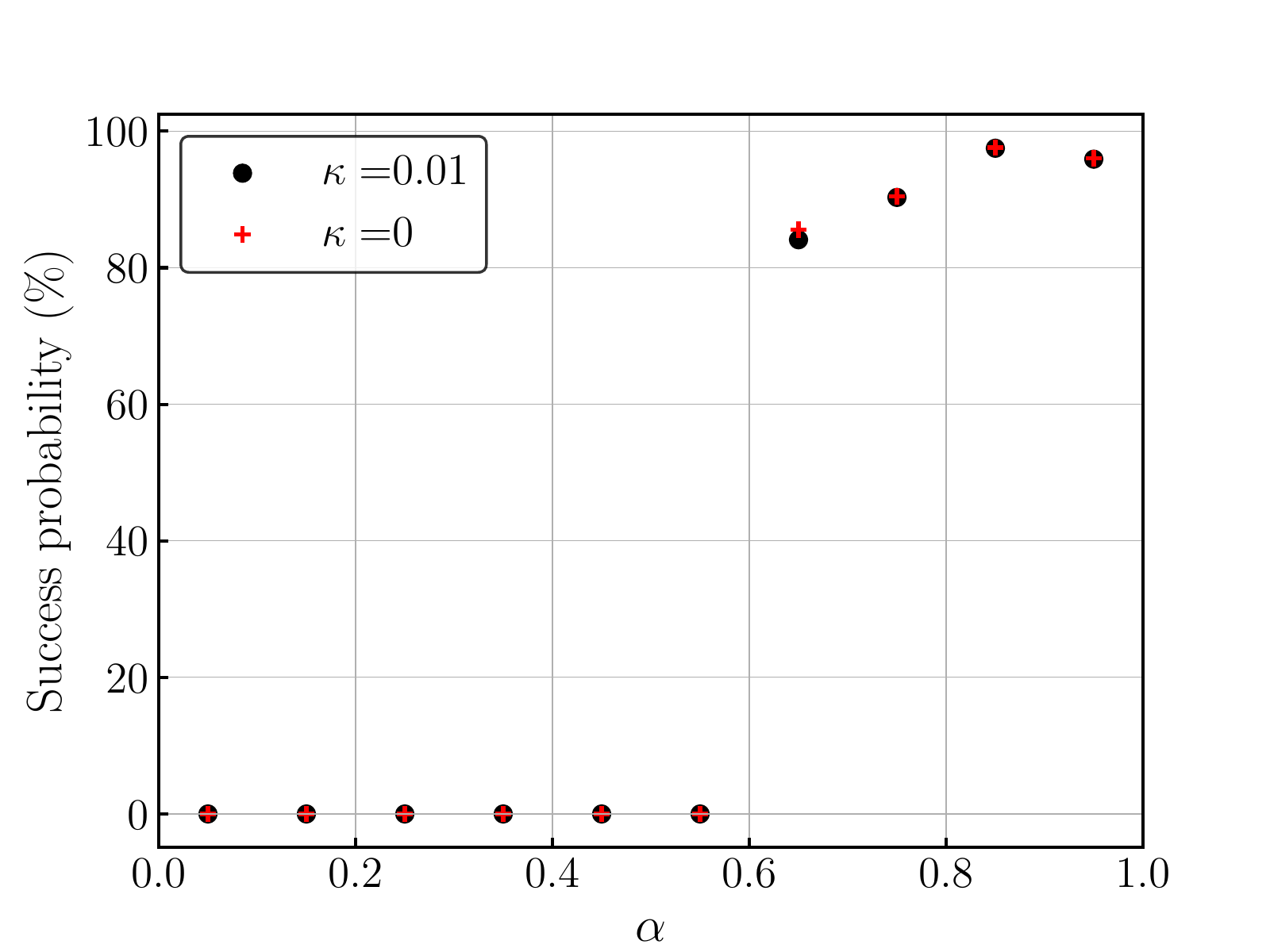}
    \end{minipage} 
  \end{tabular}
  \caption{The three figures in each row correspond to the results in Figs.~\ref{fig:nsp}~(Number of stable fixed points during the optimization:left),~\ref{fig:lin}~(Success probability of the optimizations for $\alpha=0.05,0.15,\ldots,0.95$:center), and \ref{fig:sp_normal}~(Success probability of the optimizations for $\alpha=0.05,0.15\ldots,0.95$ with $\kappa=0$ and $0.01$:right) in the main text. The results are obtained for $\beta=0.03$ (top) and $0.1$ (bottom) in each row.}
  \label{fig:various_beta_optimizations}
\end{figure}

\section{Numerical results for the normalized WPE with different random seeds}\label{other_seeds}
\setcounter{figure}{0}
\renewcommand{\thefigure}{\Alph{section}.\arabic{figure}}

To discuss dependencies of our results on the random seed, we here report the results corresponding to Figs.~\ref{fig:nsp},~\ref{fig:lin}, and \ref{fig:sp_normal} for other random seeds.
We perform the numerical calculations with the same parameters except random seeds.
The results are shown in Fig.~\ref{fig:test}.
Our considerations discussed in the main text are also valid in these cases.
\begin{figure}[h]
    \begin{tabular}{lll}
        \hspace{-45pt}
        \begin{minipage}[t]{0.45\hsize}
          \centering
          \includegraphics[keepaspectratio, scale=0.4]{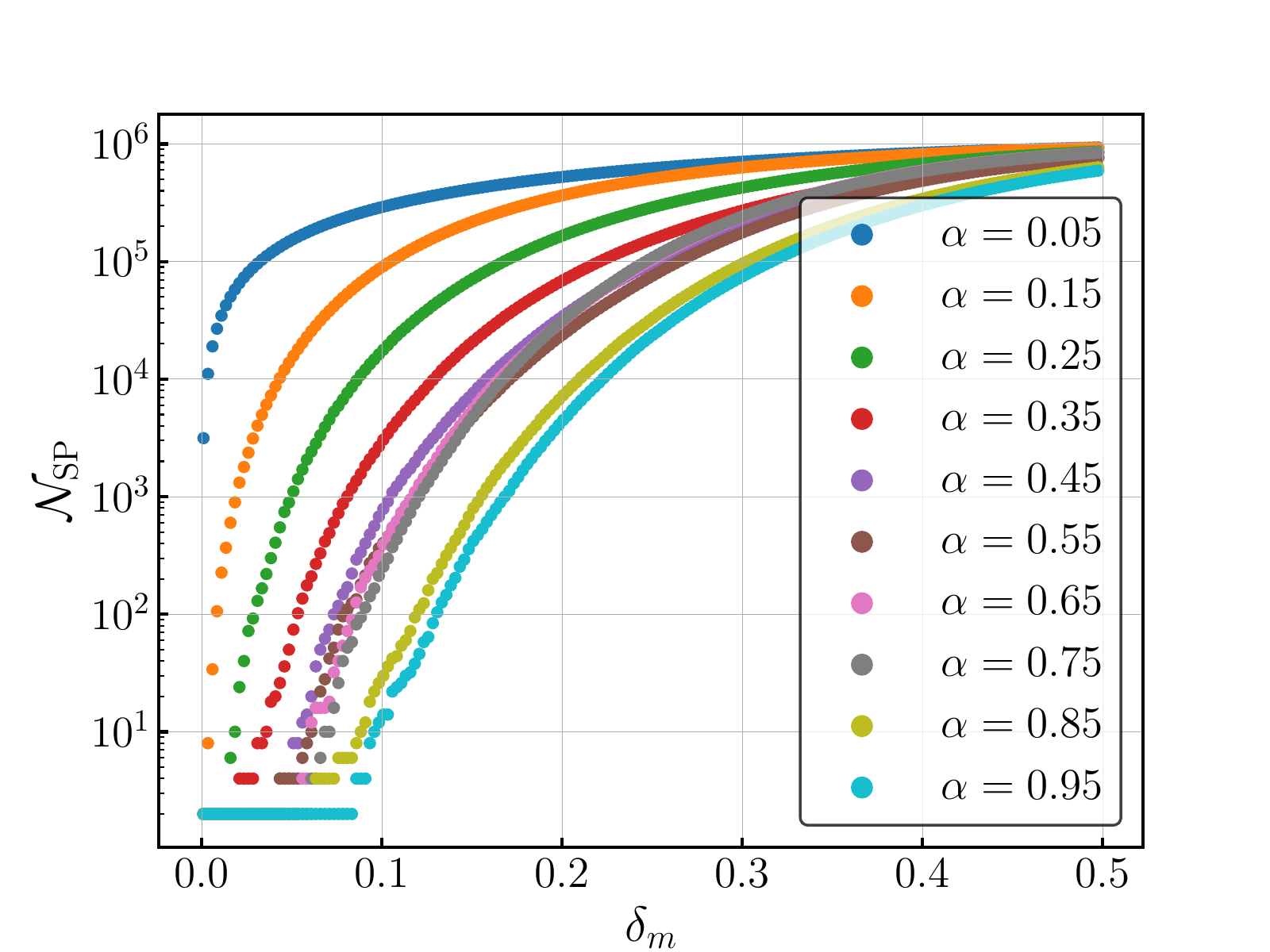}
          \label{fig:seed2_nsp}
        \end{minipage} &
        \hspace{-60pt}
        \begin{minipage}[t]{0.45\hsize}
          \centering
          \includegraphics[keepaspectratio, scale=0.4]{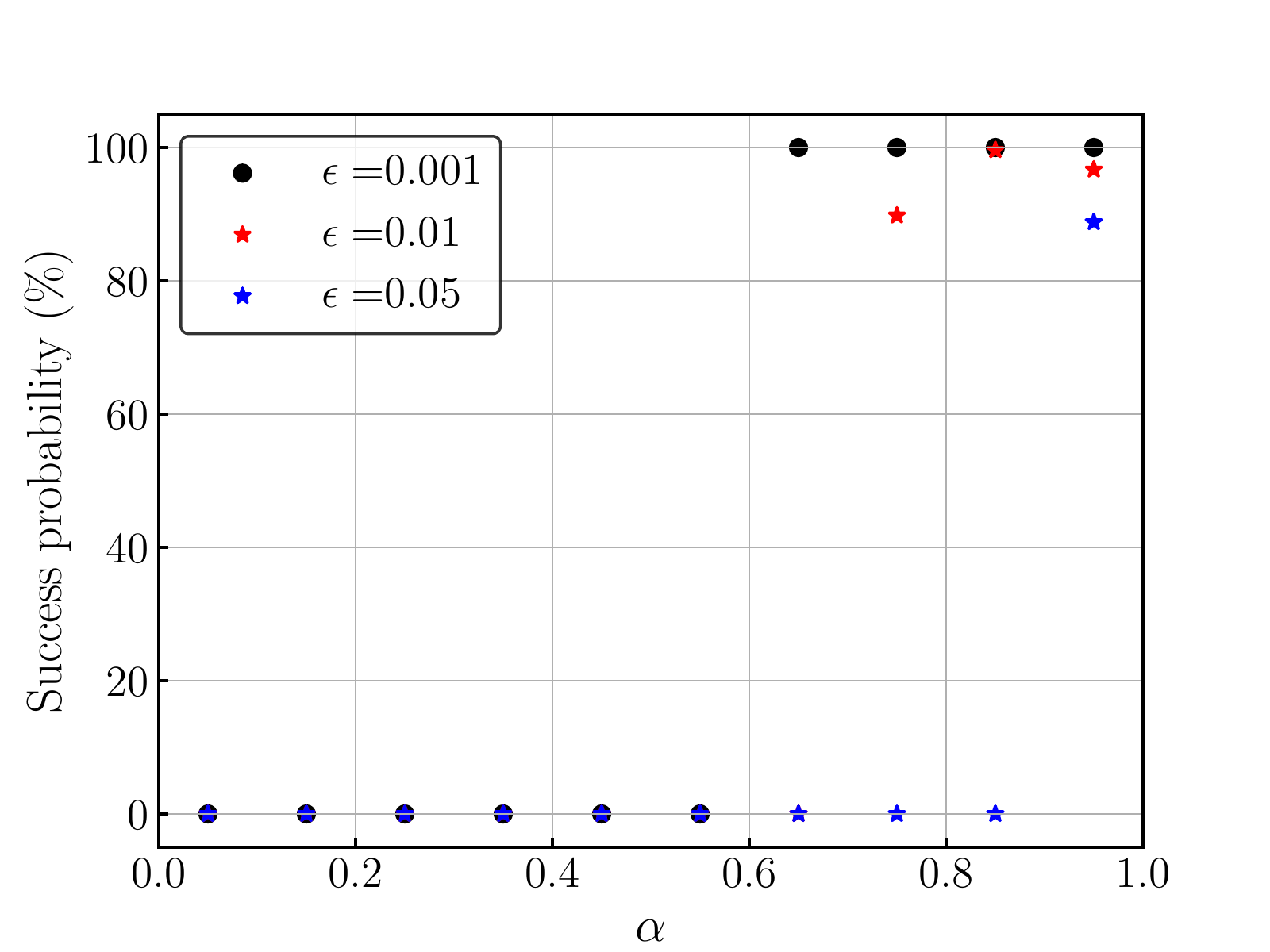}
          \label{fig:seed2_lin}
        \end{minipage} &
        \hspace{-60pt}
        \begin{minipage}[t]{0.45\hsize}
          \centering
          \includegraphics[keepaspectratio, scale=0.4]{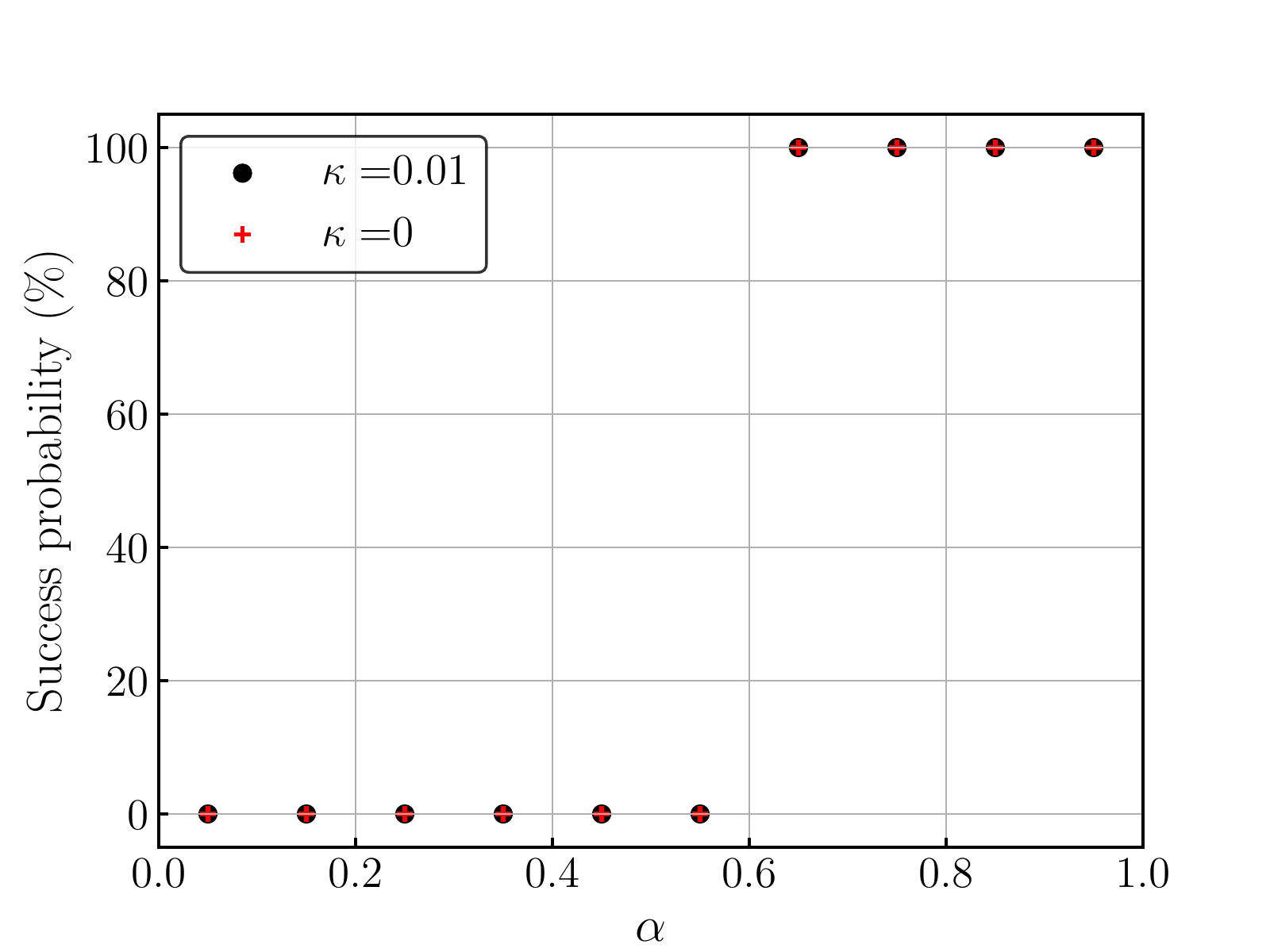}
          \label{fig:seed2_sp_normal}
        \end{minipage} \\      
      \hspace{-45pt}
      \begin{minipage}[t]{0.45\hsize}
        \centering
        \includegraphics[keepaspectratio, scale=0.4]{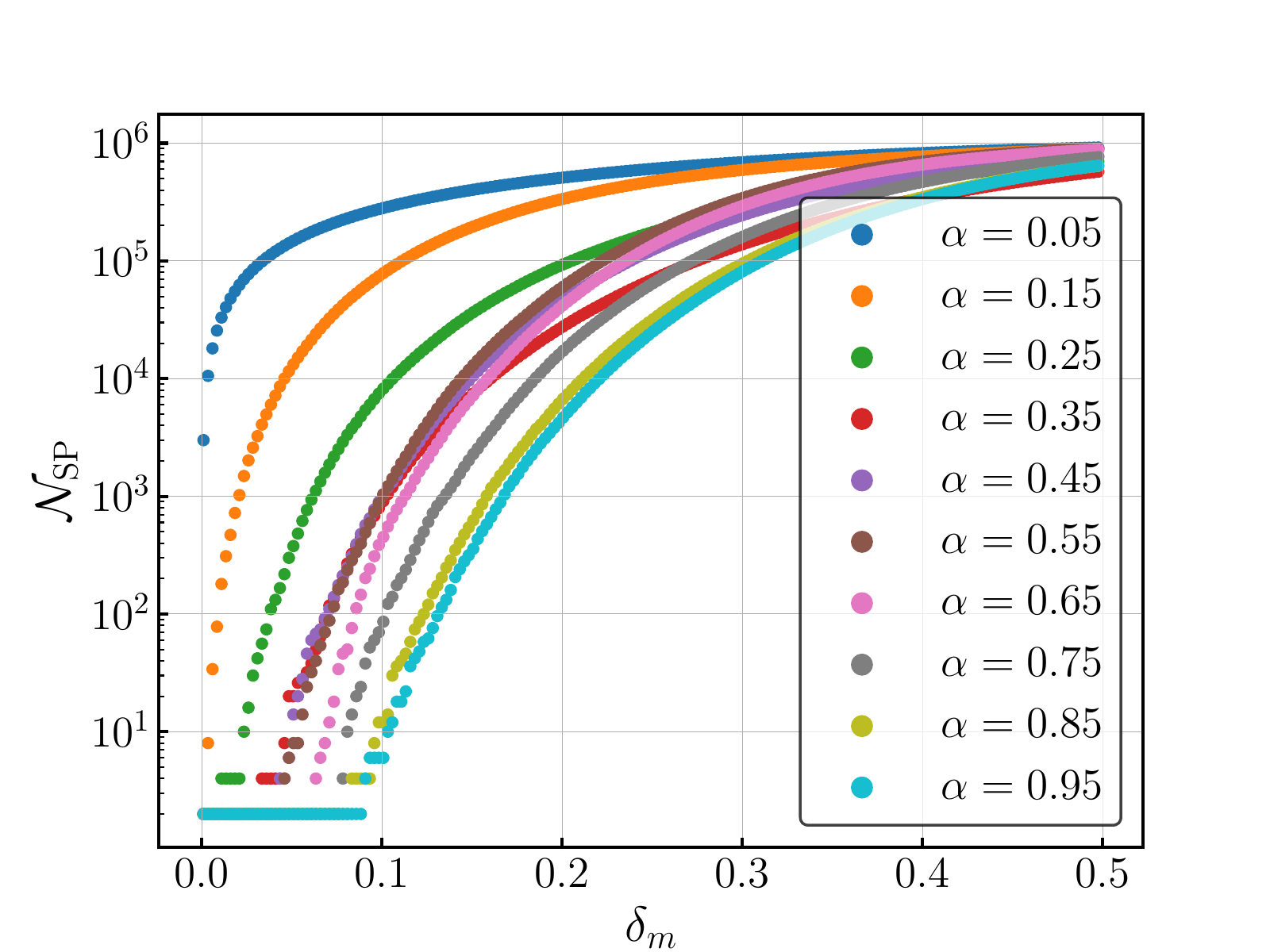}
        \label{fig:seed3_nsp}
      \end{minipage} &
      \hspace{-60pt}
      \begin{minipage}[t]{0.45\hsize}
        \centering
        \includegraphics[keepaspectratio, scale=0.4]{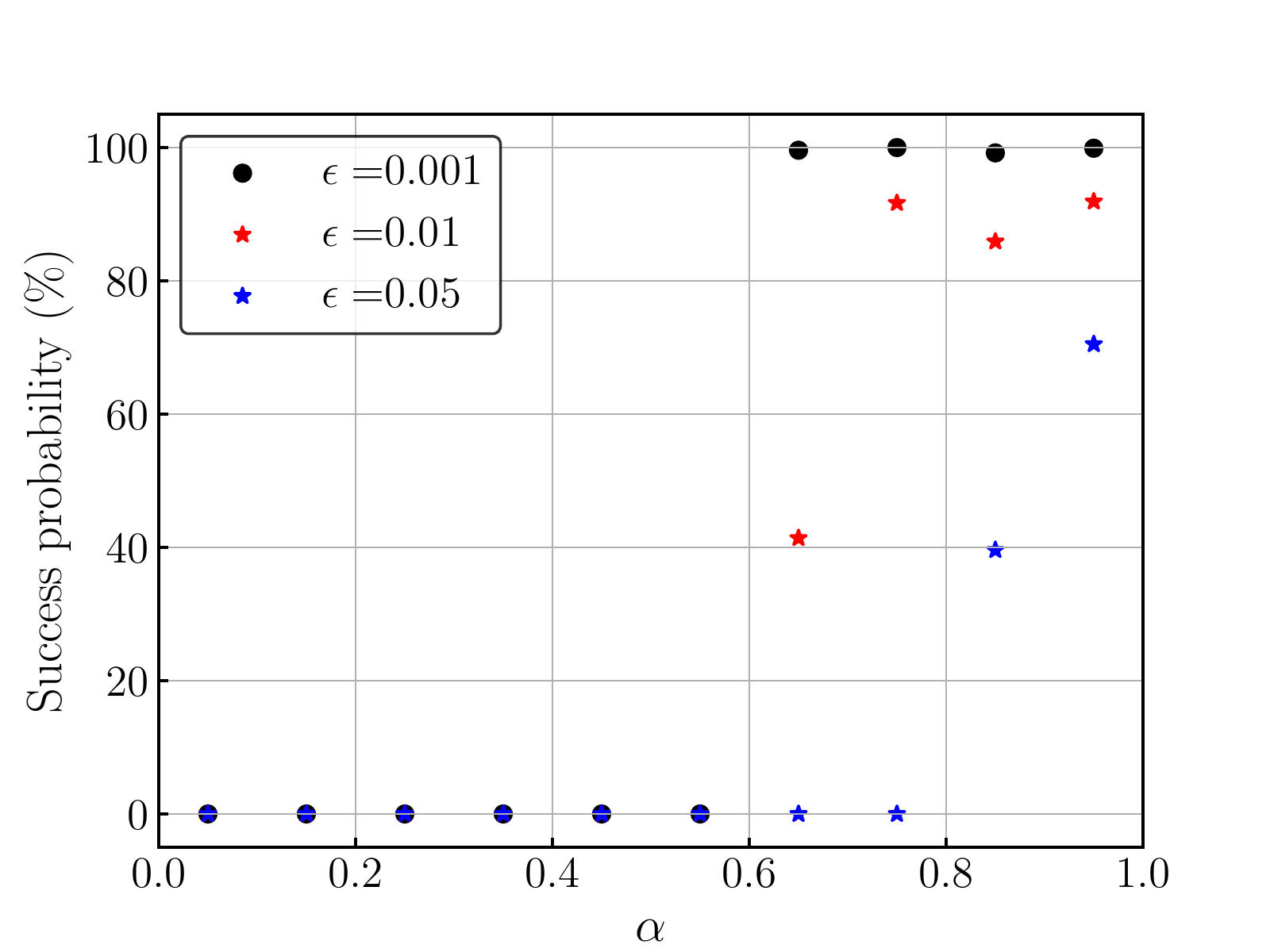}
        \label{fig:seed3_lin}
      \end{minipage} &
      \hspace{-60pt}
      \begin{minipage}[t]{0.45\hsize}
        \centering
        \includegraphics[keepaspectratio, scale=0.4]{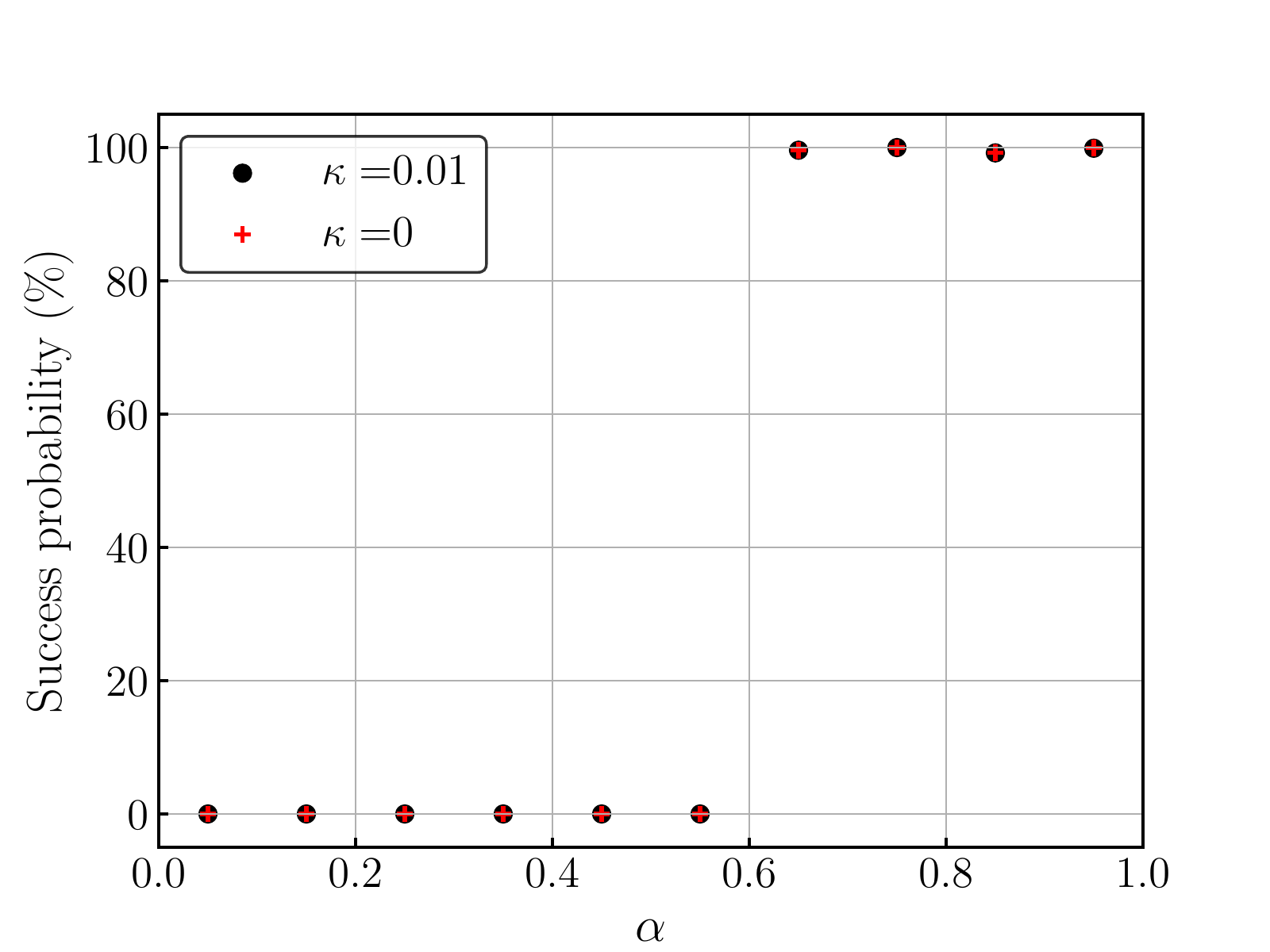}
        \label{fig:seed3_sp_normal}
      \end{minipage} \\
      \hspace{-45pt}
      \begin{minipage}[t]{0.45\hsize}
        \centering
        \includegraphics[keepaspectratio, scale=0.4]{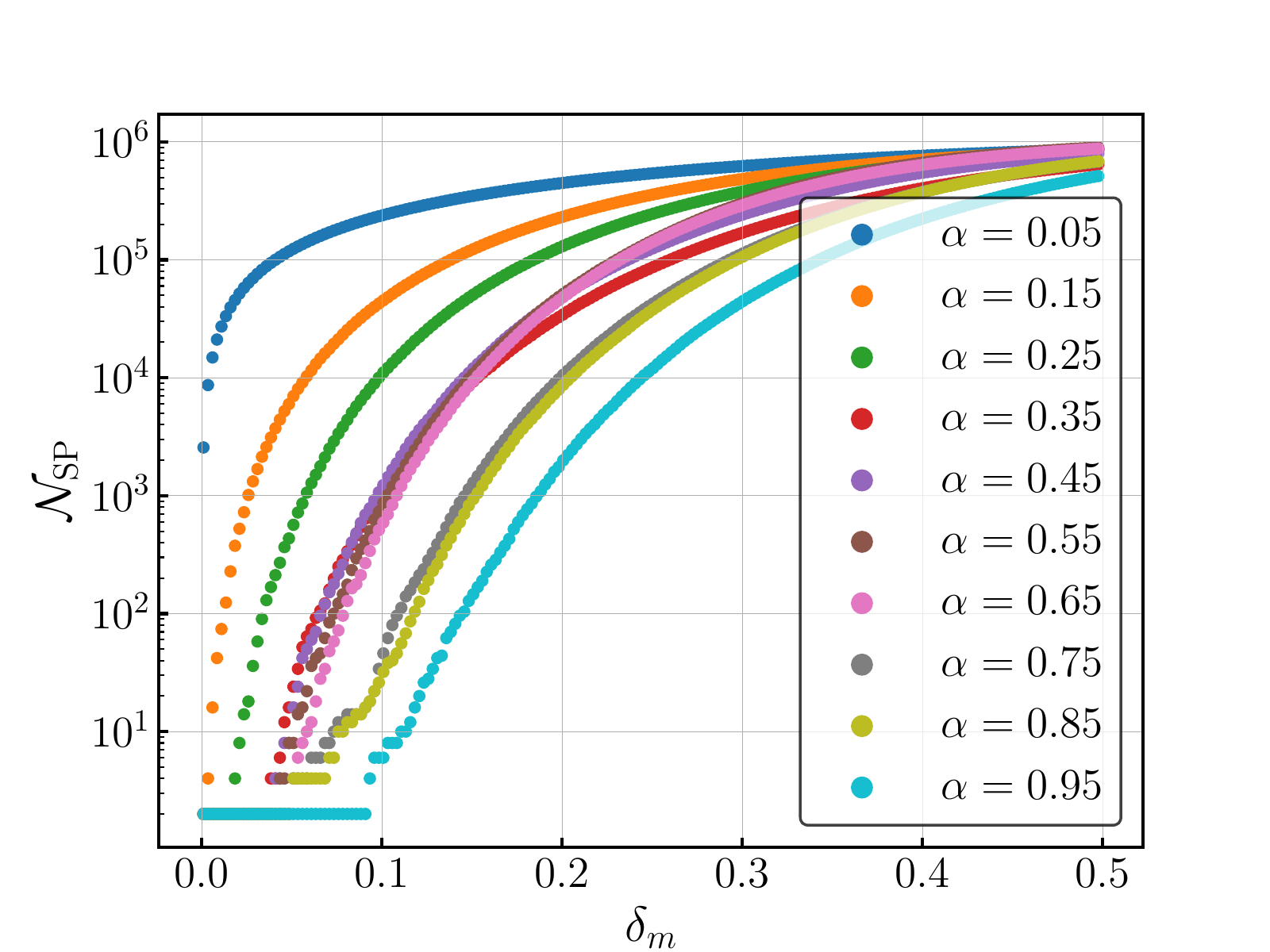}
        \label{fig:seed4_nsp}
      \end{minipage} &
      \hspace{-60pt}
      \begin{minipage}[t]{0.45\hsize}
        \centering
        \includegraphics[keepaspectratio, scale=0.4]{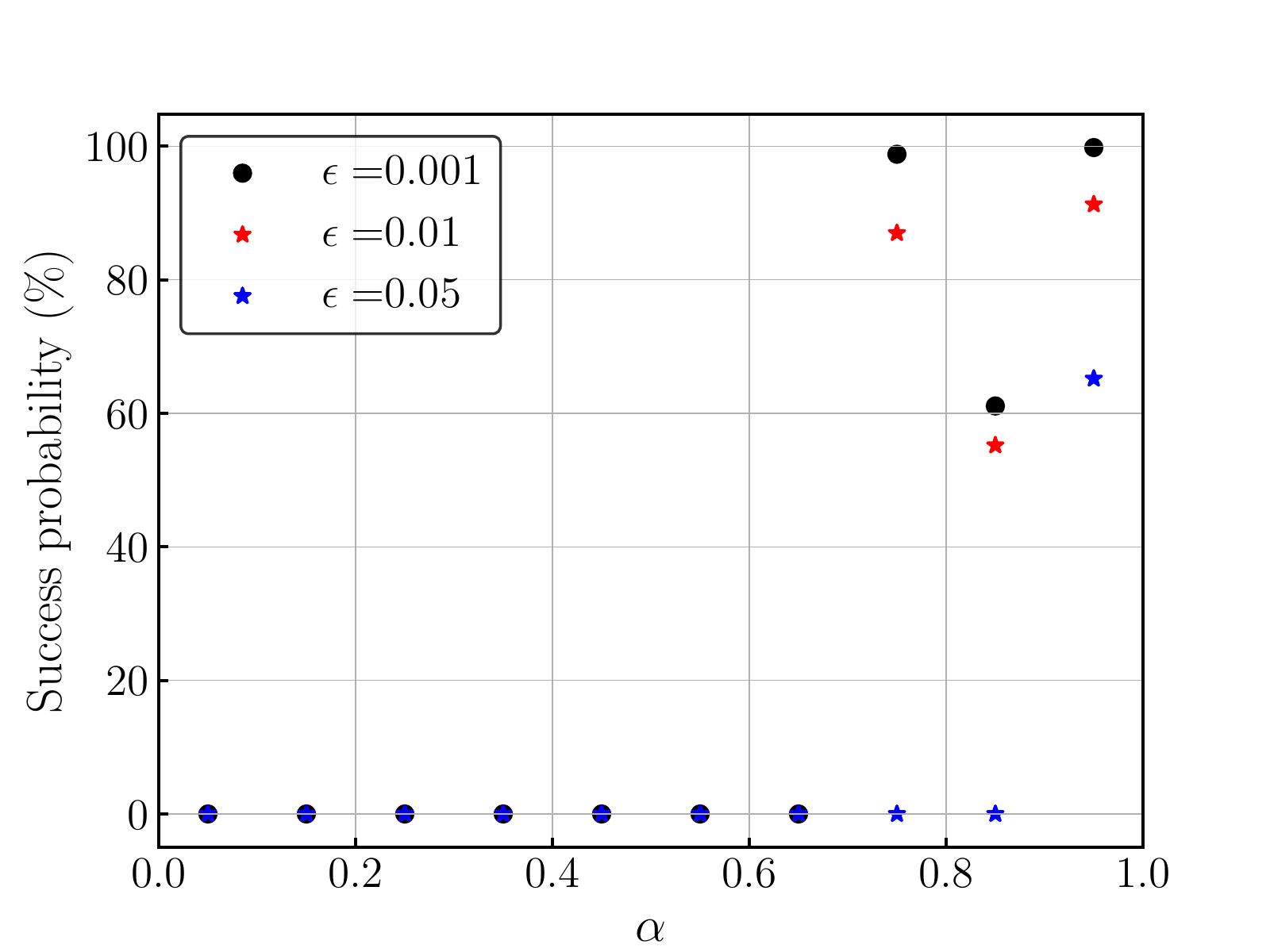}
        \label{fig:seed4_lin}
      \end{minipage} &
      \hspace{-60pt}
      \begin{minipage}[t]{0.45\hsize}
        \centering
        \includegraphics[keepaspectratio, scale=0.4]{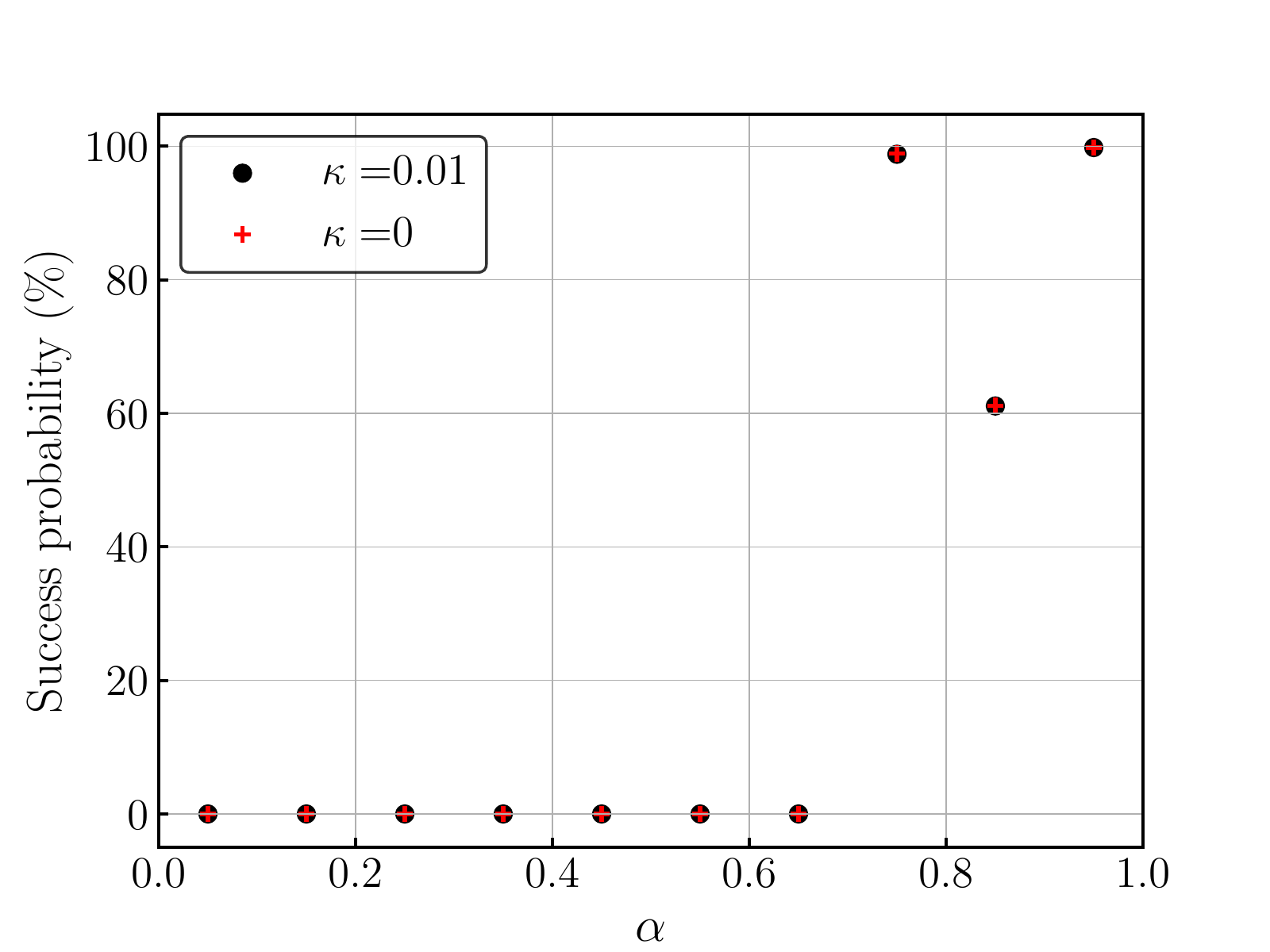}
        \label{fig:seed4_sp_normal}
      \end{minipage} \\   
      \hspace{-45pt}
      \begin{minipage}[t]{0.45\hsize}
        \centering
        \includegraphics[keepaspectratio, scale=0.4]{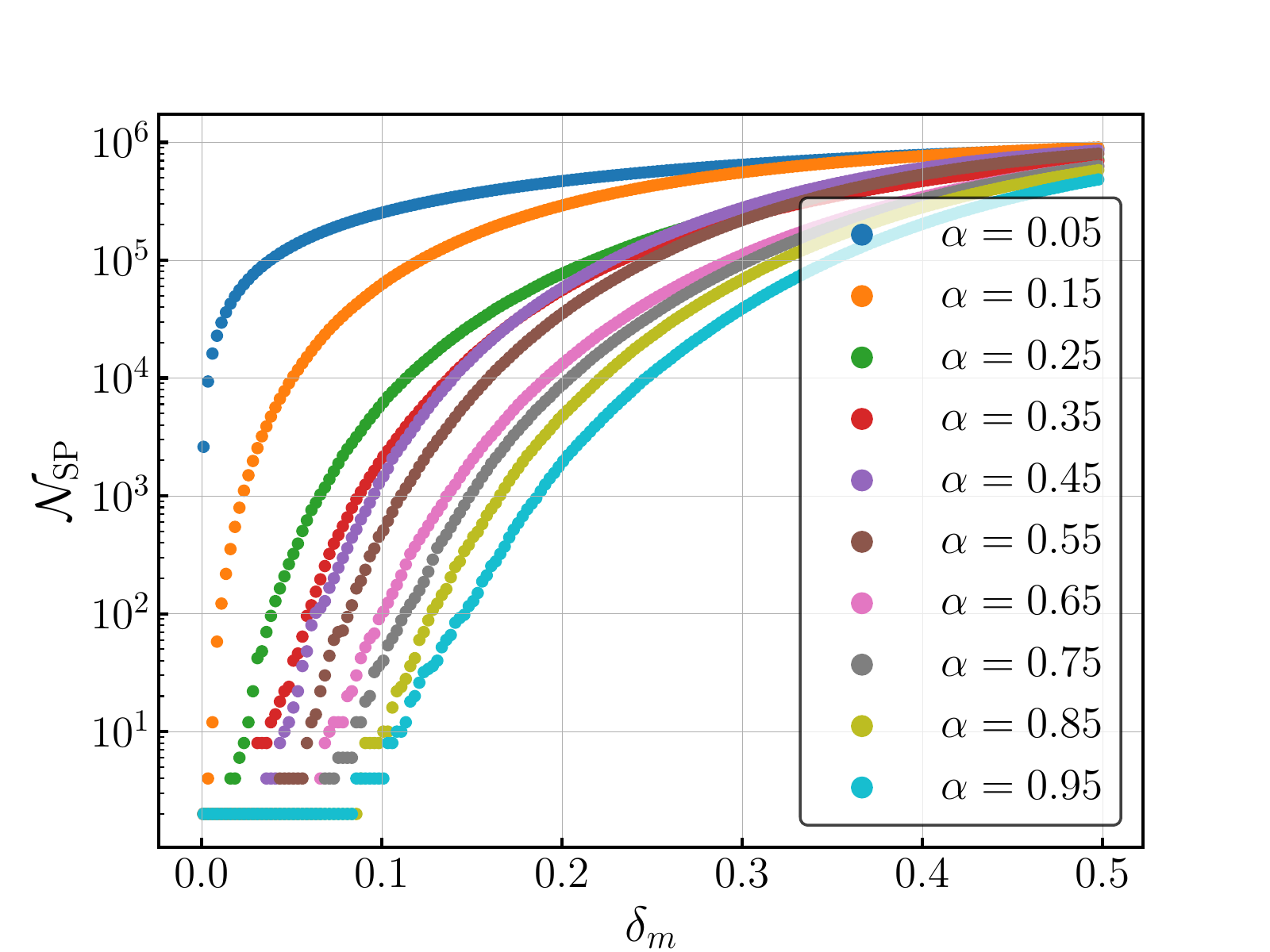}
        \label{fig:seed5_nsp}
      \end{minipage} &
      \hspace{-60pt}
      \begin{minipage}[t]{0.45\hsize}
        \centering
        \includegraphics[keepaspectratio, scale=0.4]{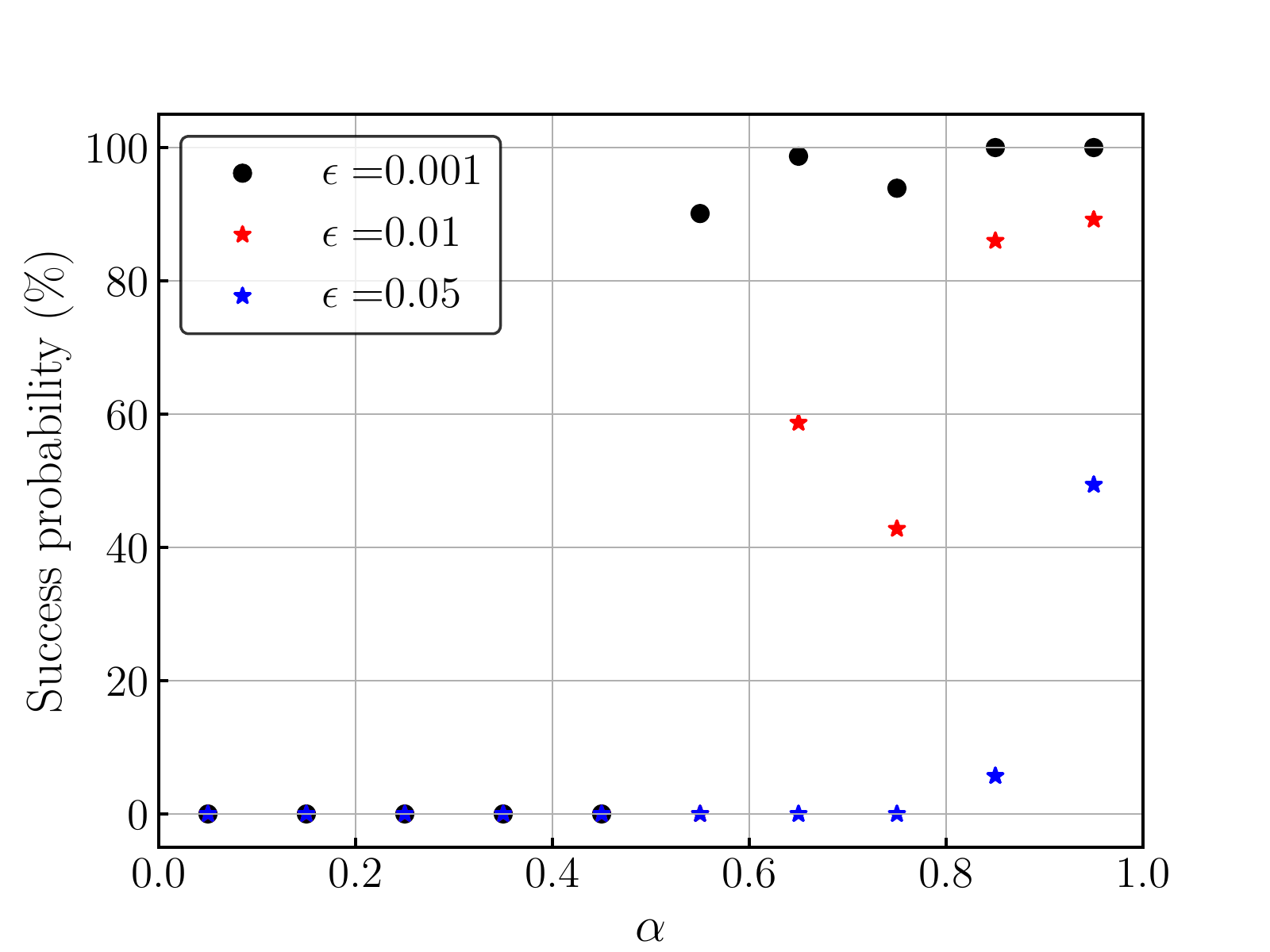}
        \label{fig:seed5_lin}
      \end{minipage} &
      \hspace{-60pt}
      \begin{minipage}[t]{0.45\hsize}
        \centering
        \includegraphics[keepaspectratio, scale=0.4]{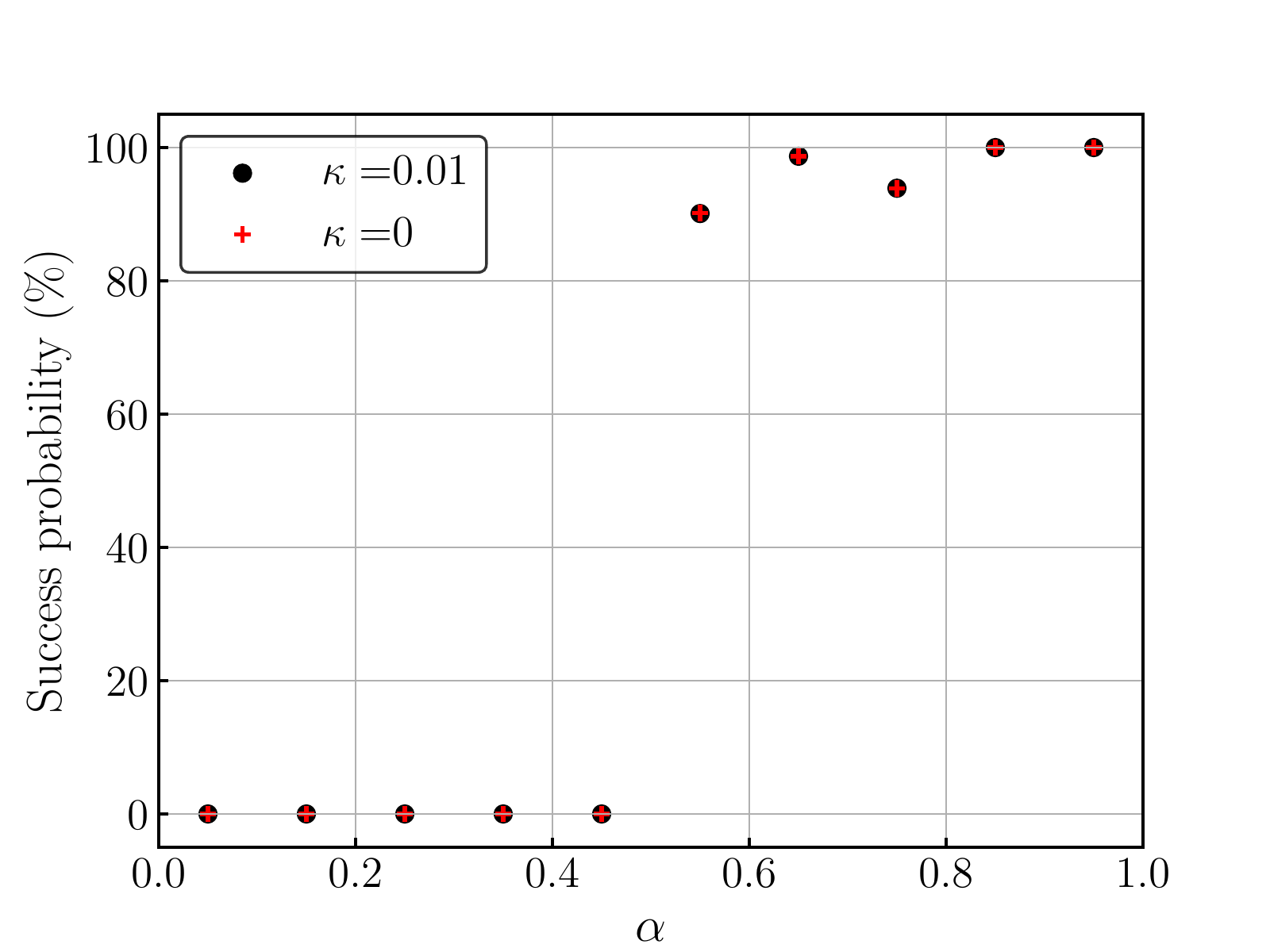}
        \label{fig:seed5_sp_normal}
      \end{minipage} 
    \end{tabular}
    \caption{The three figures in each row correspond to the results in Figs.~\ref{fig:nsp}~(Number of stable fixed points during the optimization for $\beta=0.5$:left),~\ref{fig:lin}~(Success probability of the optimizations for $\alpha=0.05,0.15,\ldots,0.95$:center), and \ref{fig:sp_normal}~(Success probability of the optimizations for $\alpha=0.05,0.15\ldots,0.95$ with $\kappa=0$ and $0.01$:right) in the main text for other random seed. The results are obtained by using a different random seed in each row. }
    \label{fig:test}
  \end{figure}

\end{widetext}

\end{document}